\documentclass[a4paper,11pt,titlepage]{article}
\makeatletter
\def\maxwidth{ \ifdim\Gin@nat@width>\linewidth
    \linewidth
  \else
    \Gin@nat@width
  \fi
}
\makeatother

\usepackage[pagebackref]{hyperref}
\renewcommand*{\backref}[1]{}
\renewcommand*{\backrefalt}[4]{{[\ifcase #1 Not cited.\or Cited on page~#2.\else Cited on pages #2.\fi ]}}

\usepackage[round, authoryear]{natbib}
\usepackage{lmodern,ifthen,amsmath,amssymb,csquotes,mathtools,tikz,fixmath,bm,doi,float,chngcntr,etoc}
\etocsettocstyle{}{}
\mathtoolsset{showonlyrefs}
\usepackage{xcolor,graphicx,booktabs,subfig,authblk}

\newcommand{\en}[3]{#1 #3 #2}
\newcommand{\EN}[3]{\left#1 #3 \right#2}
\newcommand{\enbc}[1]{\{ #1 \}}
\newcommand{\ENBC}[1]{\left\{ #1 \right\}}
\def\beq{\begin{equation}}
\def\eeq{\end{equation}}
\def\edgecovsymbol{x}
\def\edgecovspace{\mathcal{X}}
\def\edgecovv{\bm{\edgecovsymbol}}
\def\andedgecov{,\edgecovv}
\def\andnetedgecov{,\edgecovv_\sampidx}
\def\netcovsymbol{z}
\newcommand{\netcov}[1][]{\netcovsymbol\ifthenelse{\equal{#1}{}}{}{_{#1}}}
\def\netcovm{Z_{\sampidx}}
\def\netcovv{\bm{\netcovsymbol}}
\def\netcovnet{\netcovv_{\sampidx}}
\def\netsymbol{y}
\def\net{\bm{\netsymbol}}
\def\nets{\vec{\net}}
\def\rnetsymbol{Y}
\def\rnet{\bm{\rnetsymbol}}
\def\rnets{\vec{\rnet}}
\def\statsym{g}
\def\statv{\bm{\statsym}}
\newcommand{\stat}[1][]{\statsym\ifthenelse{\equal{#1}{}}{}{_{#1}}}
\def\targetsym{t}
\newcommand{\target}[1][]{\targetsym\ifthenelse{\equal{#1}{}}{}{_{#1}}}
\def\meantarget{\tau}
\def\vartarget{\Psi}
\def\targetv{\bm{\targetsym}}
\def\paramsymbol{\theta}
\def\lparamsymbol{\beta}
\newcommand{\param}[1][]{\paramsymbol\ifthenelse{\equal{#1}{}}{}{_{#1}}}
\def\lparamv{\bm{\lparamsymbol}}
\newcommand{\lparam}[1][]{\lparamsymbol\ifthenelse{\equal{#1}{}}{}{_{#1}}}
\def\paramv{\bm{\paramsymbol}}
\def\andparamv{;\paramv}
\def\cnmapsymbol{\eta}
\def\cnmapv{\bm{\cnmapsymbol}}

\def\obs{^\text{obs}}
\DeclareMathOperator{\observe}{obs}

\def\lmle{\hat{\lparamv}}
\def\0{\bm{0}}
\def\nparam{q}
\def\nstat{p}
\def\pop{N}
\def\actors{\pop}
\def\nactors{n}
\def\sampidx{s}
\def\samp{S}
\def\meanstat{\mu}
\def\meanstatv{\bm{\mu}}
\def\varstatm{\bm{\varSigma}}
\def\infom{\mathcal{I}}

\DeclareMathOperator{\ilogit}{logit^{-1}}

\DeclareMathOperator{\M}{Pr}

\DeclareMathOperator{\Binomial}{Binomial}

\DeclareMathOperator{\E}{\mathbb{E}}
\DeclareMathOperator{\Var}{\mathbb{V}\kern-0.072em ar}
\DeclareMathOperator{\vecf}{vec}

\DeclareMathOperator{\MVN}{MVN}

\DeclareMathOperator{\score}{sc}
\def\lk{\mathcal{L}}
\def\dyadspace{\mathbb{Y}}
\def\netspace{\mathcal{Y}}

\def\ncsymbol{\kappa}
\def\nc{\ncsymbol}

\def\ypnets{\net'\in\netspace}
\def\Yy{\rnet=\net}
\def\t{^\top}
\newcommand{\tsbinom}[2]{{\textstyle\binom{#1}{#2}}}
\def\inv{^{-1}}
\newcommand{\smallstackrel}[2]{\stackrel{\smash{\scriptscriptstyle #1}}{#2}}
\def\defeq{\smallstackrel{\text{def}}{=}}
\def\simiid{\smallstackrel{\text{i.i.d.}}{\sim}}
\def\simind{\smallstackrel{\text{ind}}{\sim}}

\def\Enc{\ensuremath{E_{\overline{12}}}}
\def\Ewc{\ensuremath{E_{12\vphantom{\overline{12}}}}}

\newcommand{\I}[1]{\mathbb{I}{\enbc{#1}}}
\def\reals{\mathbb{R}}
\def\Ident{I}
\def\kron{\otimes}

\newcommand{\redge}[1]{\rnetsymbol_{#1}}
\newcommand{\edge}[1]{\netsymbol_{#1}}
\def\rnet{\bm{\rnetsymbol}}
\def\net{\bm{\netsymbol}}

\def\pval{\ensuremath{P\text{-val.}}}
\def\pvals{\ensuremath{P\text{-vals.}}}
\def\pv{\ensuremath{P}}

\newcommand{\innerprod}[2]{{#1}\cdot{#2}}

\providecommand{\abs}[1]{\lvert#1\rvert}

\newcommand{\pkg}[1]{\texttt{#1}}

\newcommand{\Model}[1]{\emph{Model~#1}}

\DeclareMathOperator{\ERGM}{ERGM}

\newcommand\seq[2]{#1,\dotsc,#2}
\newcommand\inseq[2]{\in\enbc{\seq{#1}{#2}}}

\def\checkmark{\tikz\fill[scale=0.4](0,.35) -- (.25,0) -- (1,.7) -- (.25,.15) -- cycle;}

\def\ergmspec{{\netspace, \edgecovv, \statv}}

\newcommand\twostct[1][\nactors]{\sum_{i=1}^{#1} \tsbinom{\abs{\net_i}}{2}}
\newcommand\trict[1][\nactors]{\sum_{1\le i<j<k\le{#1}} \edge{i,j}\edge{j,k}\edge{k,i}}

\def\Mmin{0m}
\def\Mind{1d}
\def\Mnns{1n}
\def\MIp{1a}
\def\MIIp{2a}

\newcommand{\subdataleg}{ (Subsets: \textcolor[HTML]{E69F00}{$H$}\ \textcolor[HTML]{0072B2}{$\Ewc{}$}\ \textcolor[HTML]{CC79A7}{$\Enc{}$})}

\usepackage{upquote}
\begin{document}
\thispagestyle{empty}
\pagenumbering{gobble}

\def\acknowledge{Krivitsky wishes to thank the UCL Big Data Institute, Elsevier, the University of Wollongong Faculty of Engineering and Information Sciences, and the University of Hasselt for funding travel related to this project, Prof Raymond Chambers for helpful discussions, Prof Michael Schweinberger for invaluable feedback on drafts of this paper, and the Editor, the Associate Editors, and two anonymous Reviewers, whose comments have led to great improvements to our work. This work received funding from the European Research Council (ERC) under the European Union’s Horizon 2020 research and innovation program (PC and NH, grant number 682540---TransMID project, NH  grant number 101003688---EpiPose project), from US Army Research Office (PK, award W911NF-21-1-0335 (79034-NS)), and from US National Institutes of Health (PK, award R01 AI138783).
Computations were performed on the Katana computing cluster, supported by Research Technology Services at the University of New South Wales.}

\def\papertitle{A Tale of Two Datasets:\\ Representativeness and Generalisability of Inference for Samples of Networks}
\def\authorblock{\author[1]{Pavel N. Krivitsky\thanks{\href{mailto:p.krivitsky@unsw.edu.au}{\texttt{p.krivitsky@unsw.edu.au}}}}
\author[2]{Pietro Coletti}
\author[2,3]{Niel Hens}
\affil[1]{Department of Statistics and UNSW Data Science Hub

  School of Mathematics and Statistics

  University of New South Wales

  Sydney, Australia

~}
\affil[2]{I-BioStat

  Data Science Institute

  Hasselt University

  Hasselt, Belgium

~}
\affil[3]{Centre for Health Economics and Modelling Infectious Diseases

  Vaccine and Infectious Disease Institute

  University of Antwerp

  Antwerp, Belgium}
}

\renewcommand{\thefootnote}{\fnsymbol{footnote}}
\title{\papertitle}
\authorblock
\date{}
\maketitle

\begin{abstract}
\thispagestyle{empty}
\pagenumbering{gobble}

The last two decades have seen considerable progress in foundational aspects of statistical network analysis, but the path from theory to application is not straightforward. Two large, heterogeneous samples of small networks of within-household contacts in Belgium were collected using two different but complementary sampling designs: one smaller but with all contacts in each household observed, the other larger and more representative but recording contacts of only one person per household. We wish to combine their strengths to learn the social forces that shape household contact formation and facilitate simulation for prediction of disease spread, while generalising to the population of households in the region.

To accomplish this, we describe a flexible framework for specifying multi-network models in the exponential family class and identify the requirements for inference and prediction under this framework to be consistent, identifiable, and generalisable, even when data are incomplete; explore how these requirements may be violated in practice; and develop a suite of quantitative and graphical diagnostics for detecting violations and suggesting improvements to candidate models. We report on the effects of network size, geography, and household roles on household contact patterns (activity, heterogeneity in activity, and triadic closure).

~

\end{abstract}

\setcounter{tocdepth}{1}
\tableofcontents
\thispagestyle{empty}
\pagenumbering{gobble}

\renewcommand{\thefootnote}{\arabic{footnote}}

\clearpage
 \pagenumbering{arabic}

\section{Introduction}

Networks of human interaction provide invaluable insights into epidemiology of directly transmitted infectious disease, and there is a great deal of interest in translating network data into epidemic models \citep[for a review]{KeeEa05}. It is common to focus on epidemiologically important settings such as households \citep{GoSa17h,GriGoe15h} and schools \citep[for example]{MaFo15}, and such data are often used as they are for simulating disease spread and evaluating the impact of intervention strategies \citep[for example]{CeSa21}. However, observation of larger and broader epidemiologically relevant networks is limited by time, resources, and considerations such as privacy, so it is often indirect or incomplete in a variety of ways, therefore requiring statistical models to learn network structure from the available data and reconstruct (simulate) networks consistent with it \citep{KrMo17i}.

Exponential-Family Random Graph Models (ERGMs), also called $p^*$ models \citep[among many]{WaPa96l,LuKo12e,ScKr20e}, are a popular framework for specifying probability models for networks, postulating an exponential family on the sample space of graphs. Most applications of ERG modelling concern a single, completely observed population network, but methods for incomplete or indirect observation exist \citep[among others]{HaGi10m,KrMo17i}. At the same time, questions of ERGM asymptotics and inference---particularly under varying network size---have been debated in the literature \citep{ScKr20e}.

Increasingly, networks are collected in samples, however. Examples include social networks, such as multiple classrooms \citep{Lu03g,StSc19m}, multiple households \citep{GoSa17h,GriGoe15h}, and multiple persons' social support networks \citep{ErHa11u}; but also other types of networks such as connections among brain regions for multiple subjects \citep[Sec.~8]{ScKr20e}.

Though simpler mathematically, inference from samples of independent, non-overlapping networks is no less substantively challenging. In practice, the straightforward ``i.i.d.''\ inference scenario (e.g., brain networks) is relatively rare, and it is far more common---particularly for social and contact networks---to observe multiple non-overlapping settings, similar to each other in nature and with the same notion of a relationship, but varying in size, composition, and exogenous influences. This variation is important because size and composition of networks can have profound effects on their structure \citep{KrHa11a}. Furthermore, the selection of networks to be observed may itself be a complex process, and the selected networks themselves may be incompletely observed, requiring network model inference to be integrated with survey sampling inference.

Samples of networks also offer new opportunities. With just one network, methods to diagnose how well the model fits---and how well the inference generalises---are limited to working within that network: \citet{HuGo08g} proposed a form of lack-of-fit testing that compares the observed network's relational features not explicitly in the model to the distribution of those features simulated from the fitted model, and \citet{KoWa18o} leveraged missing data techniques to compute an analogue of Cook's distance for each actor (the effect of observing each actor's observed relations on parameter estimates). On the other hand, models for independent (if heterogeneous) samples of networks can be diagnosed using familiar techniques developed for regression---provided those techniques can be adapted to networks, including partially observed networks.

A variety of techniques exist for ERG modelling of samples of networks. One popular approach is meta-analysis, pooling individual networks' estimates \citep{Lu03g}. This approach is impractical for large samples of small networks, because the model may be nonidentifiable on each network individually \citep{VeSl21e}. More recent is multilevel (hierarchical) modelling: \citet{ZiVa06m} developed it for the related $p^2$ model, and \citet{SlKo16m} for a Bayesian ERGM with random effects. \citet{VeSl21e} described exact maximum likelihood inference for samples of very small networks. Also, when modelling a time series of networks, \emph{transitions} between successive networks are typically treated as conditionally independent \citep[for example]{LeCr18t}. 

However, assessing an ERGM's goodness-of-fit for a sample of networks has tended to be limited to comparing distributions of observed network statistics to expected \citep[for example]{SlKo16m,StSc19m} and replicating diagnostics of \citeauthor{HuGo08g} for each network \citep{VeSl21e}. Little attention has been paid to methods appropriate for partially observed networks, for large samples of networks, and to identifying precisely how the model is misspecified.

Here, we consider two samples of within-household contact networks: one more complete but restricted to households with a young child, the other larger and more representative but with only one member's relations observed in each household, and both heterogeneous in household sizes and compositions. We wish to fit a probability model to these samples to pool their information and combine their strengths, which will allow us to learn about the social forces affecting the formation of their contacts (i.e., inference) and predict their unobserved relations or other households in the population (i.e., prediction and simulation). More generally, we seek to answer three questions:
\begin{enumerate}
\item What are we estimating when we jointly fit a model to multiple networks?
\item What do we need to assume to combine information from multiple networks?
\item How do we test these assumptions?
\end{enumerate}

In Section~\ref{sec:data}, we begin to address Question 1 by describing the household contact datasets and applying the principles of model-based survey sampling inference to make explicit assumptions associated with inference from samples of networks that were previously left implicit. In Section~\ref{sec:model}, we review ERGM inference for missing data, describe a parametrisation for jointly modelling an ensemble of networks, and discuss its inferential properties---and the requirements for valid inference, addressing Question 2. We then consider in Section~\ref{sec:diagnostics} the different ways in which these requirements may be violated and combine missing data theory with classic generalised linear model (GLM) diagnostics to produce tools for diagnosing lack of fit in the proposed framework, addressing Question 3; and in addition propose fast model selection techniques for ERGMs for ensembles of networks. Finally, in Section~\ref{sec:application}, we apply these techniques to select and diagnose models for our data, and report our substantive findings.

Further details, discussion, and results are provided in Appendices~\ref{app:data}--\ref{app:extra-results}, referenced throughout this article. When appendix figures and tables are referenced, they are prefixed by the appendix, e.g., Figure~B5 is Figure~5 from Appendix~B.

\section{\label{sec:data}Data and Inferential Questions}

\subsection{\label{sec:data-design}Study Designs}

Two paper-based surveys \citep{HoaCo21c,GoSa17h} were conducted in Belgium in 2010--2011, using similar survey instruments but differing in sampling design. In both surveys, recruited by random-digit dialling, respondents (or their guardians) reported their and their household members' demographic information and recorded their contacts over the course of one day, including the contacts' ages and genders. Approximate duration and frequency of each contact was also recorded, but here, we focus our attention on presence or absence of contacts involving skin-to-skin touching.

The first major difference between the surveys is that whereas in the \emph{egocentric} (\emph{$E$}) survey \citep{HoaCo21c}, only one member in each household (the \emph{ego}) was enrolled; in the other \citep{GoSa17h}, the whole \emph{household} (\emph{$H$}) was. This within-household sampling design impacts profoundly the information available about the households: while all contacts are known for the networks in the $H$ dataset (Figure~\ref{fig:H-dataset}), only contacts incident on the one respondent in the household are known for the $E$ dataset (Figure~\ref{fig:E-dataset}), though, importantly \citep{KrMo17i}, enough information (discussed in Appendix~\ref{app:data-design}) was collected to identify these contacts uniquely within the household for almost all households.
\begin{figure}
  \centering
  \subfloat[\label{fig:H-dataset}$H$ dataset: Contacts among household members for household \#11.]{
    \centering
    \small
    \begin{tabular}{|rl|c|c|c|c|}
      \hline
      &                  & \textbf{1} & \textbf{2} & \textbf{3} & \textbf{4} \\
      \hline
      Female, 40 & \textbf{1} &   & 1 & 1 & 1 \\
      \hline
      Male, 41 & \textbf{2} & 1 &   & 1 & 1 \\
      \hline
      Male, 13 & \textbf{3} & 1 & 1 &   & 0 \\
      \hline
      Female, 11 & \textbf{4} & 1 & 1 & 0 &   \\
      \hline
    \end{tabular}
  }
  \qquad
  \subfloat[\label{fig:E-dataset}$E$ dataset: Report of within-household contacts by ego \#8.]{
    \centering
    \small
    \begin{tabular}{|rl|c|c|c|c|}
      \hline
      &                  & \textbf{1} & \textbf{2} & \textbf{3} & \textbf{4} \\
      \hline
      Male, 26 & \textbf{1} &   & 0 & 0 & 1 \\
      \hline
      Female, 54 & \textbf{2} & 0 &   & ? & ? \\
      \hline
      Male, 57 & \textbf{3} & 0 & ? &   & ? \\
      \hline
      Female, 23 & \textbf{4} & 1 & ? & ? &   \\
      \hline
    \end{tabular}    
  }
  \caption{Example observation units from the two datasets. Household composition is observed for both, but whereas every contact in the $H$ households is observed, in the $E$ households contacts not involving the ego are missing by design. \label{fig:datasets}}
\end{figure}

The second major difference is that the $H$ survey was restricted to households with a child aged at most 12, whereas the $E$ survey was not. For convenience, we define \Ewc{} to be the set of households in $E$ with at least one such child---that potentially \emph{could have been} in the $H$ dataset---and \Enc{} to be those without any, that could not. (Throughout, we will use ``presence of a child'' and similar wording to refer to this specific criterion.)

More incidentally, the surveys differed slightly in their geographical localisation: both surveys included households in the Flemish (Dutch-speaking) areas of Belgium, but only the $H$ survey included households from the (majority-French-speaking) Brussels-Capital region. Also, both surveys' designs called for fine-grained stratification by age, but the surveyor was not able to adhere to it exactly; and in the $H$ survey, households for which any members' contacts were not successfully recorded were dropped altogether.

\subsection{\label{sec:data-description}Descriptive Statistics}

After the preprocessing discussed in Appendix~\ref{app:data-design}, dataset $H$ comprises 317 households of size 2--7 for a total of 1262 members/respondents. Requiring less effort per household to collect, $E$ comprises 1463 respondents whose households (ranging in size 2--8) have a total of 4780 members with 52\% of the households' relationship states observed. 

In $H$, individuals in their mid 20s are underrepresented; $E$ is more representative in this respect, though individuals living alone or in shared housing (disproportionately young adults and seniors) are still excluded. Both datasets' households are on average gender-balanced, but among $E$'s respondents women aged 25--55 are overrepresented and adolescents of both genders underrepresented relative to $E$ households' composition.
Most households ($H$: 71\%, $E$: 75\%) were observed on a weekday; 11\% of the $H$ households are in Brussels. $\Ewc$ constitutes 35\% of $E$.

With respect to social structure, the networks are, on average, dense ($H$: 93\%, $\Ewc$: 90\%, $\Enc$: 67\%), with $H$'s networks being more dense on average (vs.\ \Ewc: $\pv \ensuremath{=0.021}$; vs.\ \Enc: $\ensuremath{<0.001}$), and those of \Ewc{} more dense than those of \Enc{} ($\pv \ensuremath{<0.001}$). $H$'s networks exhibit high triadic closure (global clustering coefficient $\frac{3\times \text{\# triangles}}{\text{\#2-stars}}$ averaging $92\%$); it cannot be estimated on the partially observed $E$ dataset.

More information can be found in Appendix~\ref{app:data-summaries}.

\subsection{Implications for Inference}

$E$ dataset is representative but consists of egocentric, incomplete networks; $H$ dataset is very selective but of complete networks. $E$ generalises better to the population of Flanders. $H$ allows higher-order (e.g., triadic) effects to be estimated, and includes Brussels. Combining information from multiple surveys with different strengths is not uncommon, and a variety of approaches can be taken \citep{ElRa18c}.

These data and the substantive problem are particularly amenable to a model-based approach: the ERGM framework seamlessly integrates exogenous (e.g., age) and endogenous (e.g., friend-of-a-friend) effects likely to be relevant. The model-based approach is also feasible: unlike some egocentric data  \citep{KrMo17i} each respondent's contacts in $E$ could be identified uniquely within the household, so model-based inference of \citet{HaGi10m} is possible.

For the purposes of prediction (e.g., given the distribution of household compositions, how would an infection brought home from school spread?) we require the analysis to generalise to the population of households in Flanders and Brussels. The missing information principle \citep{OrWo72m,BrCh94m} suggests that if the model is accurate enough, it can be generalised to the population despite the heterogeneous and biased sample. More precisely, we require that the sampling process be \emph{ignorable} or, if viewed as a missing data process, \emph{missing at random}: the unobserved relationship states must be conditionally independent of the selection process given the model and what is observed \citep{Ru76i,HaGi10m}. In our case, this also means that the model must render the dataset from which the network had come ignorable, which entails accounting for network size, composition, geography, and other relevant effects.

For the purposes of inference (e.g., do mothers have more contact with their children than fathers?), we in addition require consistency and a sampling distribution for our model parameters' estimators. Fortunately, we can treat these networks as an independent sample: the probability that any member of any of the households in either sample has interacted with a member of one of the other households in either sample is low, and such an interaction is unlikely to affect within-household information in a systematic way in the first place. However, there are further nuances, discussed in Section~\ref{sec:nonident}.

\section{Model Specification and Inference\label{sec:model}}
In modelling an independent sample of networks, we represent two levels of effects: 1) the exogenous and endogenous social forces affecting each network's relations; and 2) the effects of a network's exogenous properties such as size, composition, and sampling stratum membership on those social forces. For example, does the presence of a child (a network composition property) affect the contacts between adult men and women in the household (an exogenous relation effect)? Is triadic closure (an endogenous relation effect) stronger or weaker in due to household size (a network property)? We discuss these levels in turn.

\subsection[ERGMs for Completely and Partially Observed Networks]{Exponential-Family Random Graph Models for Completely and Partially Observed Networks\label{sec:ergm}}

We refer the reader to the text book by \citet{LuKo12e} and a review by \citet{ScKr20e} for detailed discussions of ERGMs' formulation, interpretation, and inference. For our purposes, let $\actors=\enbc{1, 2, \dotsc, \nactors}$, for $n\ge 2$, be the set of
actors whose relations are of interest. Since physical contacts are inherently two-way, we will focus on undirected graphs: the set of potential relations of interest $\dyadspace\subseteq\enbc{\enbc{i,j}\in \actors\times\actors:i\ne j}$ is a subset of the set of \emph{dyads}---distinct unordered pairs of actors. Then, the set of possible graphs of interest $\netspace\subseteq 2^{\dyadspace}$ (the set of all possible subsets of $\dyadspace$). We use $\net\in\netspace$ for the graph data structure, and $\edge{i,j}\in\enbc{0,1}$ as indicator of $i$ and $j$ being connected in $\net$ (with $\edge{i,j}\equiv\edge{j,i}$).

An ERGM is specified by its sample space $\netspace$, a collection $\edgecovv\in\edgecovspace$ of quantitative and categorical exogenous attributes of actors (e.g., age and gender) or dyads (e.g., distance) used as predictors, and a (sufficient by construction) statistic $\statv:\netspace\times\edgecovspace\mapsto\reals^p$. This statistic op\-er\-a\-tion\-a\-li\-ses the hypothesised social forces affecting the network's relations. With free model parameters $\paramv\in\reals^p$,
a random graph $\rnet\sim \ERGM_\ergmspec(\paramv)$ if
\begin{equation}\M_\ergmspec ({\Yy};\paramv)=\exp\enbc{\innerprod{{\paramv}}{ \statv(\net\andedgecov)}}/{\nc_\ergmspec(\paramv)},\ {\net\in\netspace},\end{equation} where
$\nc_\ergmspec(\paramv)=\sum_{\ypnets} \exp\enbc{\innerprod{\paramv}{\statv(\net'\andedgecov)}}$
is the normalising constant. For the sake of brevity, we will omit specification elements \enquote{$\netspace$}, \enquote{$\edgecovv$}, \enquote{$\statv$}, and \enquote{$\paramv$} where unambiguous. \renewcommand{\andparamv}{}\renewcommand{\andedgecov}{}

Network statistics that we will use in this work include the edge count $\abs{\net}$ to model propensity to have relations; edge counts within or between exogenous groups of actors to model homophily and other types of mixing; and endogenous effects: count of 2-stars $\stat_{\text{2-star}}(\net)=\twostct$ (where $\abs{\net_i}$ is the degree---number of ties incident on actor $i$) to model degree heterogeneity and count of triangles $\stat_{\text{triangles}}(\net)=\trict$ to model triadic closure. Ordinarily, we would not use the latter two because of their well-known tendency to induce badly behaved \enquote{degenerate} models in large networks and instead use less degeneracy-prone---perhaps \emph{curved} (Appendix~\ref{app:curved})---effects \citep[Sec.~3.1 for context and history]{ScKr20e}. However, this application's networks are very small and thus largely unaffected, so we use them for their simplicity.

With respect to this ERGM, we may take expectations $\E\en(){\cdot\andparamv\andedgecov}$ and variances $\Var\en(){\cdot\andparamv\andedgecov}$, including those of the sufficient statistic: let $\meanstatv(\paramv)\defeq\E\enbc{{\statv(\rnet\andedgecov)}\andparamv\andedgecov}$ and $\varstatm(\paramv)\defeq\Var\enbc{\statv(\rnet\andparamv)\andedgecov\andparamv}$.

Given an observed network $\net$, an ERGM is typically estimated by maximum likelihood, with
$l(\paramv)\defeq \log \M (\Yy;\paramv\andedgecov)$ and
Fisher information
${\infom(\paramv)} = {-l''(\paramv)}  = \varstatm(\paramv)$.
For most interesting models, the normalising constant $\nc(\paramv)$ is intractable, and estimation requires MCMC-based techniques \citep[Sec.~1.2.1 for references]{ScKr20e}.

If the network is incompletely observed, likelihood estimation proceeds as follows \citep{HaGi10m}: to the unobserved true population network $\net$, an observation process $\observe(\cdot)$ (deterministic or conditioned-on) is applied, producing
the observed data structure $\net\obs\defeq\observe(\net)$, with $\edge{i,j}\obs\in\enbc{0,1,\text{NA}}$ representing observed-absent, observed-present, and unobserved potential relations, respectively.
For the $E$ dataset, $\observe(\net)$ is such that
\[
  \edge{i,j}\obs \equiv
  \begin{cases}
    \edge{i,j} & \text{if $i=1 \lor j=1$}, \\
    \text{NA}  & \text{otherwise}.
  \end{cases}
\]

Let $\netspace(\net\obs)\defeq\enbc{\net'\in\netspace:\observe(\net')=\net\obs}$: all complete networks that could have produced $\net\obs$ or, equivalently, all possible imputations of unobserved relations of $\net\obs$; and define conditional expectation
\begin{align*}\meanstatv(\paramv\mid\net\obs)&\defeq\E\enbc{\statv(\rnet\andedgecov)\mid{\rnet\in\netspace(\net\obs)}\andparamv\andedgecov}\end{align*}
and, analogously, conditional covariance $\varstatm(\paramv\mid\net\obs)$.
Then, under noninformative sampling and/or missingness at random, the face-value log-likelihood is $l(\paramv)= \log {\textstyle\sum_{\net\in\netspace(\net\obs)}}\M (\Yy;\paramv\andedgecov)$ and \emph{observed} information is \citep{OrWo72m,Su74m,HaGi10m}
\begin{align}
\infom\obs(\paramv) &= \varstatm(\paramv)-\varstatm(\paramv\mid\net\obs).\label{eq:ergm-obs-info}
\end{align}
Unlike the completely observed case, \eqref{eq:ergm-obs-info} is not the Fisher information, because it depends on the data $\net\obs$. The Fisher information, then, also takes the expectation over the possible values of $\net\obs$ under the model:
\begin{subequations}\label{eq:ergm-info}
\begin{align}
  \infom(\paramv) &= \varstatm(\paramv)-\E_{\rnet}\en[]{\varstatm\enbc{\paramv\mid\observe(\rnet)}}\label{eq:ergm-info-vardiff}\\
                  &=\Var_{\rnet}\en[]{\meanstatv\enbc{\paramv\mid\observe(\rnet)}}.\label{eq:ergm-info-varcond}
\end{align}
\end{subequations}

\subsection{Multivariate Linear Models for ERGM Parameters\label{sec:mlergm}}
Now, consider a sample of networks indexed ${s=1,...,\samp}$, that we wish to model jointly, incorporating network-level effects. There is no unique way to do so; the following approach---drawing on multivariate linear regression models and on seemingly unrelated regression models---has the advantages of familiarity, interpretability, and good inferential properties.

Let $\netcovnet\in\reals^\nparam$ be a row vector of network-level covariates of interest and $\lparamv\in\reals^{\nparam\times\nstat}$ the parameter \emph{matrix}. Set network-level parameters $\paramv_{\sampidx}\defeq(\netcovnet\lparamv)\t$. Then, jointly, \[(\rnet_1,\rnet_2,\dotsc,\rnet_{\samp})\sim\ERGM_{\netcovv,\vec{\netspace},\vec{\edgecovv},\vec{\statv}}(\lparamv)\]
if 
$\rnet_{{\sampidx}}\simind\ERGM_{\netspace_{\sampidx},\edgecovv_{\sampidx},\statv_{\sampidx}}(\paramv_{{\sampidx}})$. Thus, the components of the network model specification (sample space, sufficient statistic, and any covariates---respectively, $\netspace_{\sampidx}$ $\statv_{\sampidx}$, and $\edgecovv_{\sampidx}$, collected into $\samp$-vectors $\vec{\netspace}$, $\vec{\statv}$, $\vec{\edgecovv}$) may vary arbitrarily between networks, but their parameter vectors $\paramv_{{\sampidx}}$ are parametrised in turn, with elements of $\lparamv$ determining, in a manner analogous to the linear predictor of a GLM, how network-level covariates affect the ERGM parameters.

Although here we treat $\nparam$ and $\nstat$ as the same for all networks, we show in Appendix~\ref{app:parametrisation} that there is no loss of generality as long as selected elements of $\lparamv$ can be fixed at 0. Also, this framework can be viewed as a special case (for $\varSigma=\0$) of the model of \citet{SlKo16m}, whose prior can be expressed $\paramv_{\sampidx}\simiid \MVN\enbc{(\netcovnet\lparamv)\t, \varSigma}$.

\paragraph{Example: Network size effects} For a given type of social setting (e.g., classroom, household), bigger networks will typically have lower density ($\abs{\net}/\enbc{\nactors(\nactors-1)/2}$ for undirected networks), with mean degree ($\abs{\net}/\enbc{\nactors/2}$) being close to invariant to size. This includes our data (e.g., Figure~\ref{fig:size-vs-dens}); but the ``default'' ERGM behaviour is to preserve network density \citep{KrHa11a} so that mean degree grows in proportion to $\nactors$. \citeauthor{KrHa11a} proposed to adjust this behaviour by an offset term of the form $-\log(\nactors) \abs{\net}$: other things being equal, the odds of a relation in a network of size $\nactors$ would be scaled by $\nactors\inv$, stabilising the mean degree. But, their result is asymptotic, reliant on sparsity, and only adjusts lower-order properties (density, mixing, and, fortuitously, degree distribution).

\citet{BuAl15f} proposed that the effect of network size on density could be estimated from a sample of networks, with $\log(\nactors)$ above multiplied by a free parameter $\gamma$, rather than by $-1$, making the mean degree approximately proportional to $\nactors^{\gamma+1}$. Here, we can accomplish this by setting $\netcov[\sampidx,k]=\log(\nactors_\sampidx)$ and $\stat[\sampidx,l](\net_\sampidx)=\abs{\net_\sampidx}$ for some indices $k$ and $l$; then $\gamma\equiv\lparam[k,l]$. Considering that our networks are small and dense, we can model a nonlinear network size effect by adding a quadratic covariate $\netcov[\sampidx,k+1]=\log^2(\nactors_\sampidx)$, with $\lparam[k+1,l]$ then becoming its coefficient. (The resulting design matrix is given in Appendix~\ref{app:parametrisation}.) Alternatively, orthogonal polynomial contrasts, a spline, or dummy variables could be used.

\subsection{Inference\label{sec:inference}}
\renewcommand{\andnetedgecov}{}
We now describe this framework's inferential properties. (Corresponding results for curved ERGMs are given in Appendix~\ref{app:curved}.)
Let $\Ident_d$ be an identity matrix of dimension $d$; let $\kron$ be the Kronecker product; and let $\netcovm\defeq\Ident_{\nstat}\kron\netcovnet\in \reals^{p\times pq}$. Then, we can reexpress $\paramv_{\sampidx}= \lparamv\t\netcovnet\t \equiv {\netcovm \vecf(\lparamv)}$, for an exponential family with a complete-data likelihood
\begin{align}
      \lk(\lparamv)&={\exp\ENBC{\innerprod{ \vecf(\lparamv)}{\sum_{\sampidx=1}^\samp\netcovm\t\statv_{\sampidx}(\net_{\sampidx}\andnetedgecov)}}}\big/\prod_{\sampidx=1}^\samp \nc_{\netspace_{\sampidx},\edgecovv_{\sampidx},\statv_{{\sampidx}}}\enbc{(\netcovnet\lparamv)\t}.\label{eq:mlergm-ef}
\end{align}
Let $\meanstatv_{\sampidx}(\lparamv\mid\net\obs_{\sampidx})\defeq\E\enbc{\statv_{\sampidx}(\rnet_{\sampidx})\mid{\rnet_{\sampidx}\in\netspace(\net\obs_{\sampidx})}; (\netcovnet\lparamv)\t}$ and analogously for $\meanstatv_{\sampidx}(\lparamv)$, $\varstatm_{\sampidx}(\lparamv\mid\net\obs_{\sampidx})$, and $\varstatm_{\sampidx}(\lparamv)$. Then, its partially observed Fisher information is
\begin{align}
\infom(\vecf\lparamv) &= \sum_{\sampidx=1}^\samp \netcovm\t\Var_{\rnet_{\sampidx}}\en[]{\meanstatv\enbc{\lparamv\mid\observe(\rnet_{\sampidx})}}\netcovm. \label{eq:mlergm-info}\end{align}
For those networks in the sample that are completely observed, $\meanstatv_{\sampidx}(\lparamv\andnetedgecov\mid\net_\sampidx\obs)\equiv \statv_{\sampidx}(\net_\sampidx)$ and $\Var_{\rnet_{\sampidx}}\en[]{\meanstatv\enbc{\lparamv\mid\observe(\rnet_{\sampidx})}}\equiv\varstatm_{\sampidx}(\lparamv)$.
Since this is an independent sample of networks, consistency and asymptotic normality of $\hat{\lparamv}$ in $\samp$ can be shown \citep{Su74m}, provided the sampling process is noninformative and $\infom(\vecf\lparamv)$ is nonsingular asymptotically, which requires the model to be identifiable.

\section{Diagnosing Multivariate Linear ERGMs\label{sec:diagnostics}}
Whether or not the estimation can be consistent and the inference be generalised to a broader population of households depends on the model being identifiable given available data and on its goodness-of-fit---both of which must take into account that at least some of the networks in the sample are partially observed. Here, we discuss likely causes and diagnostics for nonidentifiability, develop a generalisation of residual diagnostics to partially observed networks, and consider a variety of ways in which a model may fit the data poorly and how to diagnose this.

\subsection{Causes and Diagnostics for Nonidentifiability\label{sec:nonident}}

The key condition for consistency by \citet{Su74m} is that $\infom(\vecf\lparamv)$ must be nonsingular. Substantively, there is a number of reasons this condition might not be satisfied.
\paragraph{Nonidentifiable model specification} A model may erroneously contain a relationship type or other network feature that is not possible in any potentially sampled network. For a trivial example, counting the number of connections between adults and children is not meaningful in a survey of households without children, nor is counting 2-stars in households of size 2. Similarly, given the large selection of potential network features, and a large selection of potential network-level covariates, it is easy to inadvertently specify a model that is not full-rank. An example of this is network size as a covariate in a sampling process that observes networks of only one distinct size; or a quadratic network size effect if only two distinct sizes are observed. Then the minuend of \eqref{eq:ergm-info-vardiff} (i.e., $\varstatm(\lparamv)$), respectively, has zeros on the diagonal or linear dependence, and the model is not identified even under complete observation.

This form of nonidentifiability can usually be detected during estimation by examining the variance--covariance matrices of simulated sufficient statistics. 

\paragraph{Network observation process not informative of the model} If the sampling process entails partially observed networks, some observation processes may render some otherwise identifiable model specifications nonidentifiable.

\textbf{Example~1} Consider an undirected network with actors partitioned into groups $A$ and $B$. A 3-parameter model whose statistic comprises the counts of all edges, of edges within group $A$, and of edges between members of $A$ and members of $B$ is identifiable, and its $\varstatm(\paramv)$ is full-rank. But, if only relationships incident on members of group $A$ ($A$--$A$ and $A$--$B$) are observed, while $B$--$B$ relations are missing by design, then the elements of $\meanstatv\en(){\paramv\mid\net\obs}$ are affinely dependent, making $\infom(\paramv)$ singular. (See Appendix~\ref{app:nonident-1}.)

\textbf{Example~2} For an i.i.d.\ sample of $\samp$ 3-node undirected networks, it is possible to estimate a 3-parameter model with a sufficient statistic comprising edges, 2-stars, and triangles; but not if any one of the 3 possible relations is unobserved in each network: a direct enumeration of the sample space in Appendix~\ref{app:nonident-2} shows that \eqref{eq:ergm-info-varcond} is singular.

This form of nonidentifiability is more insidious. Its main symptom is that intermediate estimates of the difference in \eqref{eq:ergm-info-vardiff} are not positive definite; but the algorithm of \citet{HaGi10m} obtains this difference by subtracting the two simulated variance--covariance matrices, and for data with high missingness fraction and models with many parameters in particular, a false positive can result from Monte Carlo error.

\subsection{Residual Diagnostics for Partially Observed Networks\label{sec:pearson-calc}}
Traditional model diagnostics---whether for linear regression or for ERGMs \citep{HuGo08g}---work by comparing the observed data points to those predicted by the fitted model. The approach of \citeauthor{HuGo08g} in particular is to simulate networks from the fitted model, and compare the statistics of the simulated networks---particularly those statistics \emph{not} in the original model---to their observed values. If the observed value falls outside of the range of the simulated, lack of fit is indicated. However, most of the networks in $E$ are partially observed, and this means that there is no ``true'' observed value for a network feature. We therefore derive equivalent diagnostics for partially observed networks.

For notational convenience, let $\nets=[\net_\sampidx]_{\sampidx=1}^\samp$ refer to a vector of completely observed networks. Consider a real-valued function $\target(\nets)$ that evaluates a particular network feature of interest, either cumulatively over all of the networks or for a specific network. Analogously to $\meanstatv(\cdot)$  and $\varstatm(\cdot)$ in Section~\ref{sec:ergm}, let $\meantarget(\lparamv)\defeq\E\enbc{{\target(\rnets)};\lparamv}$ and $\vartarget(\lparamv)\defeq\Var\enbc{{\target(\rnets)};\lparamv}$, and likewise for the conditional expectations.

\begin{subequations}\label{eq:pearson-resid}
We can form a standardised (Pearson) residual for $\target(\nets)$ by evaluating \beq R_{\target}=\enbc{\target(\nets)-\meantarget(\lmle)} \big / {\sqrt{\vartarget\en(){\lmle}}},\label{eq:pearson-resid-full}\eeq
with the expectation and the variance estimated by simulating from the fitted model. Under the true model, this residual \noeqref{eq:pearson-resid-full} would, by construction, have mean 0 and variance close to 1; this also facilitates outlier detection.

If the networks are not completely observed, $\target(\nets)$ cannot be evaluated directly, and it is natural to replace it with its empirical best predictor \citep{HuHa08e,StSc19m,KrHu23e}, \[\meantarget\en(){\lmle\mid\nets\obs}\defeq\E\enbc{\target(\rnets)\mid{\rnets\in\netspace(\nets\obs)}},\] where $\nets\obs$ is defined analogously to $\nets$.
Then, \beq R_{\target}=\enbc{\meantarget\en(){\lmle\mid\nets\obs}-\meantarget\en(){\lmle}} \big / {\sqrt{\Var_{\rnets}\en[]{\meantarget\enbc{\lmle\mid\observe(\rnets)}}}},\label{eq:pearson-resid-miss}\eeq
\end{subequations}
Estimating the variance in the divisor in \eqref{eq:pearson-resid-miss} is not trivial. We discuss it in Appendix~\ref{app:pearson-var}.

\subsection{Causes and Diagnostics for Lack-of-Fit}

\paragraph{Within-network}

It may be the case that the within-network model fits poorly. Network statistics used for diagnostics by \citet{HuGo08g} include the full degree distribution, counts of shared partners (i.e., for a given pair of connected actors, how many common connections do they have?), and the distribution of geodesic distances. All of these can be used as $\target(\cdot)$, but it may be impractical for two reasons. Firstly, family networks are relatively small and very dense. This makes the statistics typically used less than informative. Secondly, the sheer number of networks in the dataset means that diagnosing each network individually is infeasible, but, at the same time, pooling their within-network diagnostics is likely to wash out any effects because of their heterogeneity.

Nonetheless, even if a statistic is suboptimal and difficult to interpret, for a model that fits well, $R_{\target}$ will still have mean 0 and variance close to 1.

\paragraph{Between-network\label{sec:bw-net-diag}}

It may be the case that the model for the network-level parameters ($\paramv_\sampidx$) as a function of global parameters ($\lparamv$) fits poorly: in particular, it may fail to account for network size and composition effects.
At network level, the model has a form similar to that of a GLM. We can thus use the developments of Section~\ref{sec:pearson-calc} directly to make familiar diagnostic plots: for some statistic $\target_\sampidx(\nets)\defeq \target(\net_\sampidx)$  (e.g., density), we can plot residuals $R_{\target_\sampidx}$ for $\sampidx=1,\dotsc,\samp$ against their respective $\meantarget_\sampidx(\lmle)\defeq \E\enbc{\target_\sampidx(\rnets)}$ (the fitted values) or against a candidate predictor $\netcov_{\sampidx,\text{new}}$. Or, we can use $\sqrt{\abs{R_{\target_\sampidx}}}$ instead for a scale--location plot, analogously to the standard diagnostic plots in \citet{R}.

We can also use residuals to test lack-of-fit hypotheses and assess potential explanatory power of $\netcov_{\sampidx,\text{new}}$---without the computationally costly ERGM fitting---by regressing $\target_\sampidx(\nets)-\meantarget_\sampidx(\lmle)$ on $\netcov_{\sampidx,\text{new}}$, weighted by their inverse-variance ($\Var\inv\enbc{{\target_\sampidx(\rnets)}}$). Individual networks are independent, so the residuals should be nearly independent as well. 

\paragraph{Between-dataset\label{sec:bw-data-diag}}

If we wish for the fitted model to generalise and render the sampling designs ignorable, the model must account for differences in datasets without incorporating dataset effects directly. This can be done via a hypothesis test, such as a simulation score test, along the lines of that described by \citet{Kr12e} in the context of valued ERGMs, by testing the significance of an explicit dataset effect without refitting the model. Details are given in Appendix~\ref{app:score}.

\paragraph{Non-systematic heterogeneity\label{sec:non-syst-diag}}
Lastly, even if there is no systematic bias in the model, there may be between-network heterogeneity due to unobserved factors. The above-described Pearson residuals incidentally provide us with a way to tell whether there is any heterogeneity left to explain: if there is none, $R_{\target_\sampidx}$ in \eqref{eq:pearson-resid} will, by construction, have mean 0 and variance around 1.

\section{Application\label{sec:application}}

We now return to the data we had introduced in Section~\ref{sec:data}, discuss model specification, and report model diagnostics and results. As one reads this section, it may be helpful to refer to Appendix~\ref{app:parametrisation} for how these effects are represented in the framework described.

We have implemented the methodology described in an extension to the \pkg{ergm} package \citep{HuHa08e,KrHu23e} for the \citet{R} statistical environment. To make this methodology accessible to a broad audience, we have published our implementation in an R package, \pkg{ergm.multi}. The most recent versions of the packages can be found on the Comprehensive R Archive Network \citep{R} or the Statnet Project software repositories (\url{https://statnet.org}).

\subsection{Initial Model}

A model used to join these two datasets must be substantively meaningful and interpretable. It must account for within-network conditional dependence among the relations. It must make the network size, composition, and dataset effects ignorable to enable generalisable inference. And, it must do so without requiring more information than is available in the data. We therefore dedicate a great deal of attention to formulating and justifying each of the model's elements.

Here, we develop the initial model, \Model{0}, which we will then refine using diagnostics.

\paragraph{Household roles\label{sec:mixing-eff}}

Our data do not record family relations (e.g., who is married to whom and who is whose child), so we must infer household roles from age and gender. In doing so, there is a tension between interpretability and accuracy: family roles are most conveniently modelled with discrete age categories; but outside of a few critical ages defined exogenously (e.g., school attendance, legal adulthood, and retirement), age effects are likely to be continuous, best modelled semiparametrically (e.g., with splines). Our compromise is to use relatively fine-grained edge categories.

A first classification was done according to age: \emph{young child} (under 6), \emph{preadolescent} (6--12), \emph{adolescent} (13--18), \emph{young adult} (19--24), \emph{older adult} (25--60), and \emph{senior} (over 60). (The age cut at 12 was chosen specifically to account for the design boundary.) In order to investigate gender-specific interactions \citep{GoSa17h} we subdivided older adults into \emph{older female adults} and \emph{older male adults}. A total of 7 categories results.

We then modelled mixing by counting the contacts between pairs of these categories---essentially cells of a symmetric $7\times 7$ contingency table. Our data about some of these cells are limited, both because some age groups are underrepresented for design reasons discussed above and in Section~\ref{sec:data-description}, and because some age combinations, such as young children and seniors, are rarely found together in a household (Figure~\ref{fig:mixmat_samp_size} for pairwise counts). Thus, guided by substantive interest, sample size, and design effects, some of the cells were combined for modelling. For example, in modelling contacts with seniors, we combine young children with preadolescents and adolescents with young adults because of their very small sample sizes; but we do not combine all four cells because the combined cell would then cross the age-12 boundary. Similarly, despite a small sample size, young adults with young children were retained as a separate count, because their chances of being parent and child are relatively high. The final parametrisation is visualised in Figure~\ref{fig:mixmat}.

\paragraph{Endogenous effects\label{sec:endogenous-eff}}

To model actor heterogeneity and triadic closure, we use 2-star and triangle counts, defined in Section~\ref{sec:ergm}. An additional caveat is that the $E$ dataset, by virtue of only containing relations incident on one individual per household, does not contain information about triadic closure. (See Section~\ref{sec:nonident} Example~2 and Appendix~\ref{app:nonident-2}.) We thus assume that net of all other effects, the effect of triadic closure on a household of a given size that does not have a child is the same as the effect of triadic closure on a household of that size that does have a child. It is not possible to test this assumption with the available data.

\paragraph{Network size effects\label{sec:netsize-eff}}

The effects of network size on our networks is not trivial: for example, in the analysis of the $H$ dataset by \citet{GoSa17h}, three different density and two different triadic parameters were used, depending on household size. In \Model{0}, we use the polynomial effects of $\log \nactors_\sampidx$ described in Section~\ref{sec:mlergm} on edge, 2-star, and triangle counts. This also further guards against ERGM degeneracy, by allowing 2-star and triangle coefficients to decrease with network size.

\paragraph{Other network-level effects\label{sec:other-eff}}

Some of the surveys were conducted on a weekend and others on a weekday (Table~\ref{tab:HE-misc}). Past literature \citep[for example]{GoSa17h} suggests that contact patterns may differ depending on the day.

Contact patterns may differ systematically between families that live in detached housing and families that live in apartments. This potential effect has received limited attention in the literature to date. Our data do not include housing type but do include postal codes. The population densities in those postal codes can then be used as a proxy for housing type. We use this potential predictor to illustrate the technique proposed in Section~\ref{sec:bw-net-diag}  of regressing the network-level residuals on potential network-level predictors.
Alternatively, we might ask whether or not the post code belongs to any of Belgium's larger cities.

These properties may be predictive of edge, 2-star, or triangle counts. For the sake of parsimony, \Model{0} initially incorporates only weekend effect on density, and diagnostics are used to suggest additional effects.

\paragraph{Design effects\label{sec:design-eff}}

Last but not least, if we wish to generalise our inference to the population of households, our model must make any design effects ignorable. The substantively motivated effects described above already control for some of those. For example, recall that households in $H$ were omitted if there was nonresponse from even a single member. To the extent that the nonresponse rate is a function of household size (i.e., the bigger the household, the more likely there is at least one nonrespondent), a model that accurately controls for network size will reduce the informativeness of this nonresponse. Similarly, although both surveys' selection was strongly affected by household members' ages, particularly children, granular modelling of age mixing effects---particularly for young children and preadolescents---already reduces this design effect's informativeness.

It is, however, also possible that interactions among adult household members, and other structural features, are affected by a child's presence---something that \citeauthor{GoSa17h}'s data could not be used to test. We can adjust for this by including the presence of a child in a household as a network-level covariate for density overall, for the endogenous effects, or for mixing. Given the wide range of possibilities, we do not incorporate any such effects into \Model{0}, instead using residual diagnostics on it to select them.

To account for only the $H$ dataset containing Brussels households, we add an indicator of a Brussels post code as a network-level covariate for density. This completes \Model{0}.

\subsection{Diagnostics\label{sec:example-diagnostics}}

We now apply the techniques from Section~\ref{sec:diagnostics} to the proposed models. To validate our diagnostic techniques, we also fit a number of reduced models: in Appendix~\ref{app:misspec}, we demonstrate how our techniques can identify their deficiencies.  In the following, recall our partitioning of $E$ into \Ewc{} (at least one child at most 12, potentially in $H$) and \Enc{} (no such children).

\paragraph{Effects of a child in a household} As discussed, we use diagnostics for \Model{0} to select which network features depend on the presence of a child: we calculate the Pearson residuals for counts of edges, 2-stars, triangles, and every pair of actor categories (excluding young children, preadolescents, and seniors), breaking them down by subset $H$, \Ewc, and \Enc. Then, category pairs with extreme residuals that have the same sign for $H$ and $\Ewc$ but the opposite sign for $\Enc$ may suggest a relevant child effect.

Three mixing cells have residuals with this sign pattern (Table~\ref{tab:pearson-mix-m0}): contacts between older female and male adults ($H$:~$2.6$, \Ewc:~$1.8$, \Enc:~$-1.8$), two male adults ($-1.6$, $-1.8$, $0.6$), and two female adults ($0.6$, $0$, $-0.1$). The latter two are likely spurious and cannot be used in any case due to small sample size (per Appendix~\ref{app:data-samp-size}).

Adding the effect of absence of a child on the coefficient for contacts between older female and male adults yields \Model{1}; none of its global or mixing residuals (Table~\ref{tab:pearson-mix-m1}) exceed 2 in magnitude. We focus on \Model{1} going forward, but, for illustrative purposes, we also report \Model{\MIp}, with the absence-of-child effect on edge count instead.

\paragraph{Additional substantive models\label{sec:additional-eff}} As proposed in Section~\ref{sec:other-eff}, we regressed (per Section~\ref{sec:bw-net-diag}) edge, 2-star, and triangle count residuals of \Model{1} on a number of candidate predictors, with full results given in Table~\ref{tab:reg-res-m1}. The linear effect of log-population-density is the most promising ($\pv\ensuremath{=0.046}$), yielding \Model{2}, and, for illustrative purposes, we also report \Model{\MIIp}, adding the effect of the household being in a major city ($\pv\ensuremath{=0.30}$) instead.
Specifications for all models are summarised in Table~\ref{tab:coef-fit}, and their complete results and diagnostics are provided in Appendix~\ref{app:extra-results}.

\paragraph{Unaccounted-for between-dataset differences}

\begin{figure}
\begin{center}

\subfloat[Edge residuals vs.\ fitted\label{fig:resid-1a}]{

{\centering \includegraphics[width=0.801173333333333\maxwidth]{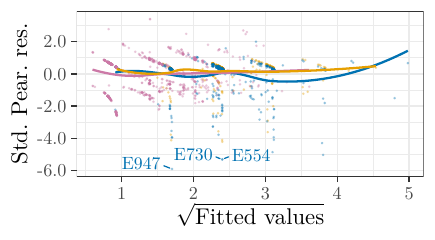} 

}

}\subfloat[Edge scale--location plot\label{fig:resid-1b}]{

{\centering \includegraphics[width=0.801173333333333\maxwidth]{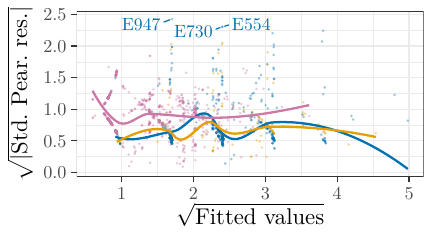} 

}

}\newline
\subfloat[Edge residuals vs.\ network size\label{fig:resid-2}]{

{\centering \includegraphics[width=0.801173333333333\maxwidth]{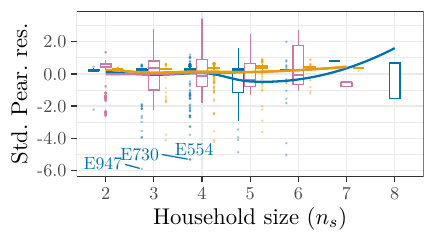} 

}

}\subfloat[Triangle residuals vs.\ fitted ($H$ only)\label{fig:resid-3}]{

{\centering \includegraphics[width=0.801173333333333\maxwidth]{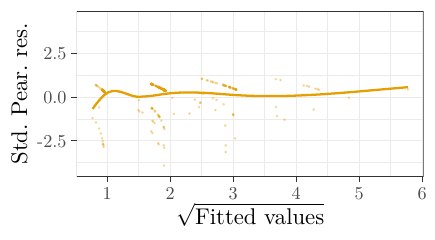} 

}

}\newline
\subfloat[Residuals for the total number of within-household contacts (edges) of all actors of a given age; symbol size represents the number of actors\label{fig:resid-4}]{

{\centering \includegraphics[width=\maxwidth]{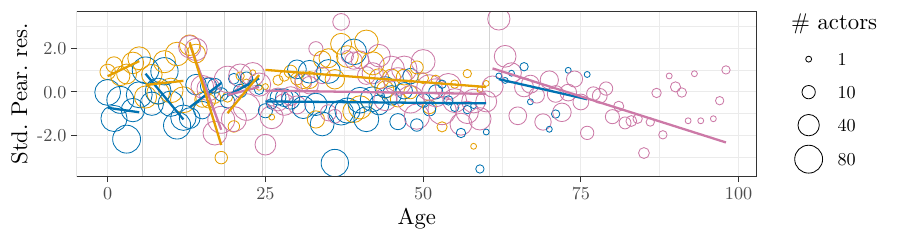} 

}

}
\end{center}
\caption{Selected Pearson residual plots of network statistics for \Model{1}, modelled after the diagnostic plots produced by R's \citeyearpar{R} built-in \pkg{stats} package for GLMs. Outliers are identified by their dataset and index within dataset.\subdataleg\label{fig:resid}}
\end{figure}

Selected residual plots are provided in Figure~\ref{fig:resid}. We also provide smoothing curves for each subset individually: these curves diverging would indicate that the model had failed to account for some systematic difference between the datasets. Panel~\subref{fig:resid-3} (triangle counts) excludes $E$'s networks, because those contain almost no triadic information.

The network residuals for both edges and triangles (Panels~\subref{fig:resid-1a}--\subref{fig:resid-3}) are skewed downward and exhibit a striped pattern. This is to be expected regardless of model fit: the underlying network statistics are small counts close to their exogenous upper bounds. There do not appear to be any clear patterns beyond that, and the edge residuals for $H$, \Ewc{}, and \Enc{} coincide on average (Panel~\subref{fig:resid-1a}). This suggests that the model fuses the two datasets well. The scales of the residuals (Panel~\subref{fig:resid-1b}) do not exhibit unambiguous patterns either, except for the residuals of \Enc{} having consistently higher variances than others---whereas the more similar $H$ and \Ewc{} networks have similar residual variances.

The dataset hypothesis tests described above and in Appendix~\ref{app:score} yield $P$-values \ensuremath{0.068}, \ensuremath{0.11}, and  \ensuremath{0.18} for the edge, the 2-star, and their omnibus test, respectively (with details given in Table~\ref{tab:dataset-m1}). Thus, at the conventional significance level, we do not detect unaccounted-for differences between datasets for these features. (In contrast, the respective $P$-values for \Model{0}  (Table~\ref{tab:dataset-m0}) are \ensuremath{0.019}, \ensuremath{0.06}, and \ensuremath{0.062}, which is more suggestive.)

\paragraph{Outliers} Our residual plots reveal some households inconsistent with typical behaviour. For example, the $E$ households \#947, \#730, and \#554 highlighted in Panels~\subref{fig:resid-1a} and~\subref{fig:resid-2} are families of three or four with two older adults (male and female) and a young child (the ``respondent''), but no within-household contacts with the child on the day of the survey. 

\paragraph{Network size effects}

Edge residuals against the network size are shown in Panel~\subref{fig:resid-2}. A model that fails to account for network size would display a linear or curved pattern in the residuals. We see no evidence of such a pattern.

We confirm this with lack-of-fit tests, regressing edge, 2-star, and triangle count residuals on network size treated as categorical (i.e., a dummy variable for each size but one). Lack of fit would be indicated by statistical significance of these regressions; we do not find it for any of the three network statistics (weighted ANOVA omnibus \pvals{} \ensuremath{0.20}, \ensuremath{0.17}, and \ensuremath{0.22}, respectively, full results in Table~\ref{tab:nns-anova-m1}).

\paragraph{Non-systematic heterogeneity}
The standard deviations of Pearson residuals for edges, 2-stars, and triangles are 1.00, 0.99, and 0.98, respectively, all close to 1 as hoped. (Breakdown by data subsets in Table~\ref{tab:pearson-sd-m1}.)

\paragraph{Continuous age effects}
Panel~\ref{fig:resid-4} shows the Pearson residuals of the total number of within-household contacts of individuals of each age. We see some indications that discretising ages into categories has an impact. For example, a senior's propensity to interact appears to drop off as they age, which the cutoff at 60 does not capture. The $H$ household density tends to be slightly underpredicted relative to comparable households in \Ewc, though the differences do not appear to be particularly strong or consistent.

\subsection{Results \label{sec:results}}

\paragraph{Model comparison}

Table~\ref{tab:coef} gives the parameter estimates for \Model{1} and \Model{2}, suggested by our residual diagnostics. (Those for \Model{\MIp} and \Model{\MIIp} are in Tables~\ref{tab:coef-m1p} and~\ref{tab:coef-m2p}.) AIC (Table~\ref{tab:models-aod}) is indifferent between \Model{1} ($\text{AIC}=3696.6$) and \Model{2} ($3696.8$) within margin of MCMC error and prefers them over \Model{\MIIp} ($3697.8$) and \Model{\MIp} ($3713.6$), the latter only slightly preferred over \Model{0} ($3714.6$).
This is as predicted by the residual analyses discussed in Section~\ref{sec:additional-eff}.

\begin{table}
  \caption{\label{tab:coef}Parameter estimates for \Model{1} and \Model{2}.}
  \footnotesize
  \begin{center}

\begin{tabular}{lrr}
\toprule
\multicolumn{1}{l}{Relationship Effect} & \multicolumn{2}{c}{Coefficient (S.E.)} \\
\cline{2-3}
\quad{}$\times$ Network-Level Effect & \Model{1}$^{\hphantom{\star\star\star}}$ & \Model{2}$^{\hphantom{\star\star\star}}$\\
\midrule
edges $\times$ $\log(\nactors_{\sampidx})$ & $-14.28 \; (2.87)^{\star\star\star}$ & $-13.78 \; (2.98)^{\star\star\star}$\\
\quad $\times$ $\log^2(\nactors_{\sampidx})$ & $5.69 \; (1.29)^{\star\star\star}$ & $5.47 \; (1.34)^{\star\star\star}$\\
\quad   if Brussels post code & $0.08 \; (0.19)^{\phantom{\star}\phantom{\star}\phantom{\star}}$ & $-0.02 \; (0.20)^{\phantom{\star}\phantom{\star}\phantom{\star}}$\\
\quad $\times$ $\log(\text{pop.\ dens.\ in post code})$ &  & $0.04 \; (0.03)^{\phantom{\star}\phantom{\star}\phantom{\star}}$\\
\quad   if on weekend & $0.14 \; (0.06)^{\star\phantom{\star}\phantom{\star}}$ & $0.13 \; (0.06)^{\star\phantom{\star}\phantom{\star}}$\\
2-stars & $1.91 \; (0.78)^{\star\phantom{\star}\phantom{\star}}$ & $1.14 \; (0.82)^{\phantom{\star}\phantom{\star}\phantom{\star}}$\\
\quad $\times$ $\log(\nactors_{\sampidx})$ & $-2.15 \; (0.41)^{\star\star\star}$ & $-1.22 \; (0.44)^{\star\star\phantom{\star}}$\\
\quad $\times$ $\log^2(\nactors_{\sampidx})$ & $0.34 \; (0.11)^{\star\star\phantom{\star}}$ & $0.07 \; (0.11)^{\phantom{\star}\phantom{\star}\phantom{\star}}$\\
triangles & $5.55 \; (0.97)^{\star\star\star}$ & $7.30 \; (0.96)^{\star\star\star}$\\
\quad $\times$ $\log(\nactors_{\sampidx})$ & $-3.46 \; (1.39)^{\star\phantom{\star}\phantom{\star}}$ & $-5.65 \; (1.44)^{\star\star\star}$\\
\quad $\times$ $\log^2(\nactors_{\sampidx})$ & $0.93 \; (0.70)^{\phantom{\star}\phantom{\star}\phantom{\star}}$ & $1.60 \; (0.74)^{\star\phantom{\star}\phantom{\star}}$\\
Young Child with Young Child & $8.60 \; (1.49)^{\star\star\star}$ & $8.66 \; (1.54)^{\star\star\star}$\\
Young Child with Preadolescent & $9.10 \; (1.48)^{\star\star\star}$ & $9.15 \; (1.54)^{\star\star\star}$\\
Preadolescent with Preadolescent & $8.17 \; (1.45)^{\star\star\star}$ & $8.24 \; (1.51)^{\star\star\star}$\\
Adolescent with Adolescent & $7.70 \; (1.43)^{\star\star\star}$ & $7.75 \; (1.49)^{\star\star\star}$\\
Young Child with Young Adult & $9.64 \; (1.76)^{\star\star\star}$ & $9.67 \; (1.81)^{\star\star\star}$\\
Preadolescent with Young Adult & $7.25 \; (1.46)^{\star\star\star}$ & $7.28 \; (1.51)^{\star\star\star}$\\
Adolescent with Young Adult & $7.73 \; (1.45)^{\star\star\star}$ & $7.82 \; (1.51)^{\star\star\star}$\\
Young Adult with Young Adult & $7.66 \; (1.44)^{\star\star\star}$ & $7.70 \; (1.49)^{\star\star\star}$\\
Young Child with Older Female Adult & $10.26 \; (1.45)^{\star\star\star}$ & $10.32 \; (1.51)^{\star\star\star}$\\
Preadolescent with Older Female Adult & $9.67 \; (1.43)^{\star\star\star}$ & $9.73 \; (1.49)^{\star\star\star}$\\
Adolescent with Older Female Adult & $8.90 \; (1.43)^{\star\star\star}$ & $8.96 \; (1.48)^{\star\star\star}$\\
Older Female Adult with Older Female Adult & $7.45 \; (1.46)^{\star\star\star}$ & $7.50 \; (1.52)^{\star\star\star}$\\
Young Child with Older Male Adult & $9.09 \; (1.43)^{\star\star\star}$ & $9.14 \; (1.49)^{\star\star\star}$\\
Preadolescent with Older Male Adult & $8.76 \; (1.42)^{\star\star\star}$ & $8.83 \; (1.48)^{\star\star\star}$\\
Adolescent with Older Male Adult & $8.20 \; (1.42)^{\star\star\star}$ & $8.26 \; (1.48)^{\star\star\star}$\\
Older Female Adult with Older Male Adult & $10.11 \; (1.44)^{\star\star\star}$ & $10.17 \; (1.49)^{\star\star\star}$\\
\quad   if child absent & $-1.22 \; (0.30)^{\star\star\star}$ & $-1.20 \; (0.30)^{\star\star\star}$\\
Older Male Adult with Older Male Adult & $6.59 \; (1.45)^{\star\star\star}$ & $6.66 \; (1.50)^{\star\star\star}$\\
Older Female Adult with Senior & $8.12 \; (1.42)^{\star\star\star}$ & $8.20 \; (1.47)^{\star\star\star}$\\
Older Male Adult with Senior & $7.51 \; (1.45)^{\star\star\star}$ & $7.58 \; (1.50)^{\star\star\star}$\\
Senior with Senior & $7.82 \; (1.40)^{\star\star\star}$ & $7.89 \; (1.46)^{\star\star\star}$\\
Adolescent with Young Child or Preadolescent & $8.07 \; (1.43)^{\star\star\star}$ & $8.13 \; (1.48)^{\star\star\star}$\\
Young Adult with Older Adult & $8.02 \; (1.43)^{\star\star\star}$ & $8.07 \; (1.48)^{\star\star\star}$\\
Young Child or Preadolescent with Senior & $8.29 \; (1.52)^{\star\star\star}$ & $8.34 \; (1.57)^{\star\star\star}$\\
Adolescent or Young Adult with Senior & $9.93 \; (1.70)^{\star\star\star}$ & $10.01 \; (1.76)^{\star\star\star}$\\
\midrule
AIC & $ 3696.6  \; ( 0.2 )^{\makebox[0pt]{\tiny\dag}\hphantom{\star\star\star}}$ & $ 3696.8  \; ( 0.2 )^{\hphantom{\star\star\star}}$\\
BIC & $ 3926.9  \; ( 0.2 )^{\hphantom{\star\star\star}}$ & $ 3933.7  \; ( 0.2 )^{\hphantom{\star\star\star}}$\\
log-likehood & $ -1813.3  \; ( 0.1 )^{\hphantom{\star\star\star}}$ & $ -1812.4  \; ( 0.1 )^{\hphantom{\star\star\star}}$\\
\bottomrule
\end{tabular}

Significance: $^{\star\star\star}\le 0.001<^{\star\star}\le 0.01< ^\star \le 0.05$

$^{\dag}$Standard errors for AIC, BIC, and log-likelihood are due to MCMC error.
\end{center}
\end{table}

A test of population density effect in \Model{2} is not significant at conventional level ($\hat{\beta} = 0.04, \operatorname{SE} = 0.031,  \pv \ensuremath{=0.19}$),
so we do not find evidence of housing type having an effect---or regional population density is a poor proxy; we leave these questions for future work, except to suggest that type of housing should be considered for future data collection.

\paragraph{Substantive conclusions}

We discuss results primarily from \Model{1}, though \Model{2} yields the same conclusions. Only a few of the effects are interpretable in isolation. In particular, we can conclude with some confidence ($\hat{\beta} = 0.14, \operatorname{SE} = 0.056,  \pv \ensuremath{=0.015}$) that weekends have a positive effect on the number of contacts that are observed in the household, in line with prior literature \citep{GriGoe15h}. Presence of a child in a household is associated with a higher propensity of older male adults and older female adults (likely the parents) to interact with each other (if no child: $\hat{\beta} = -1.2, \operatorname{SE} = 0.30,  \pv \ensuremath{<0.001}$). We do not detect an effect of being in Brussels on network density ($\hat{\beta} = 0.1, \operatorname{SE} = 0.19,  \pv \ensuremath{=0.68}$). 

\begin{figure}

{\centering \subfloat[Edge (relative to $\nactors=2$)\label{fig:nseff-1}]{\includegraphics[width=0.8\maxwidth]{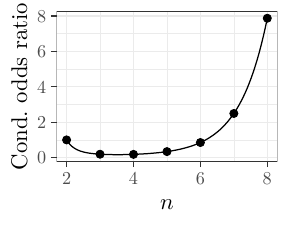} }
\subfloat[2-star\label{fig:nseff-2}]{\includegraphics[width=0.8\maxwidth]{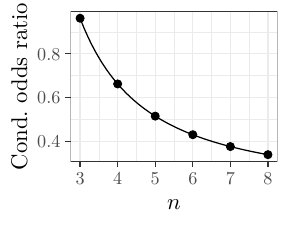} }
\subfloat[Triangle\label{fig:nseff-3}]{\includegraphics[width=0.8\maxwidth]{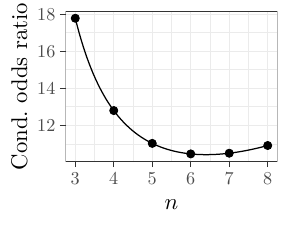} }

}

\caption[Estimated effects of network size on conditional odds of an instance of a graph feature]{Estimated effects of network size on conditional odds of an instance of a graph feature. Two-stars and triangles are only possible for $\nactors \ge 3$.}\label{fig:nseff}
\end{figure}

The estimated polynomial log-network-size effects are shown in Figure~\ref{fig:nseff}. Though they are difficult to interpret in isolation from each other, we observe that edge effect counterbalances 2-star and triangle effects, which also decrease with network size, guarding against ERGM degeneracy as hoped. Overall, there is strong evidence that network size effects are present ($\pv \ensuremath{<0.001}$), including quadratic ($\pv \ensuremath{<0.001}$), and including those on 2-stars and triangles ($\pv \ensuremath{<0.001}$). Both 2-star ($\pv \ensuremath{<0.001}$) and triangle ($\pv \ensuremath{<0.001}$) effects are significant in the presence of others. (Test details are given in Table~\ref{tab:omnibus-m1}.)

\begin{figure}

{\centering \includegraphics[width=\maxwidth]{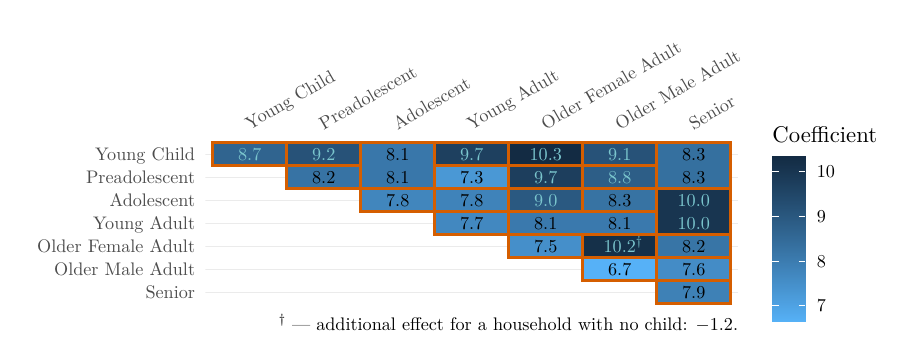} 

}

\caption[Parameter estimates for mixing by family role]{Parameter estimates for mixing by family role. Borders denote parametrisation. Because there is no ``intercept'' effect in the model, testing them against 0 is not meaningful.}\label{fig:mixmat}
\end{figure}

We report the parameter estimates for household member category mixing in a more intuitive layout in Figure~\ref{fig:mixmat}. Since, unlike \citet{GoSa17h}, we do not use a baseline category (``intercept''), they are not interpretable in isolation but only in contrast with each other. Thus, we conclude that older female adults (i.e.\ mothers) tend to interact more than older male adults (i.e.\ fathers) with young children ($\hat{\Delta} = 1.2, \operatorname{SE} = 0.47,  \pv \ensuremath{=0.013}$), preadolescents ($\hat{\Delta} = 0.9, \operatorname{SE} = 0.29,  \pv \ensuremath{=0.002}$), and adolescents ($\hat{\Delta} = 0.7, \operatorname{SE} = 0.25,  \pv \ensuremath{=0.006}$). We thus confirm and expand on similar findings by \citet{GoSa17h}. We also tested whether older female adults interacted with seniors more than older male adults did, whether because they are more likely to care for elderly parents or because in a marriage, the female spouse is typically younger than the male spouse, and there is some evidence for this ($\hat{\Delta} = 0.6, \operatorname{SE} = 0.31, \text{one-tailed }  \pv \ensuremath{=0.025}$). As with housing type, we recommend that future studies record specific familial relations for contacts.

\section{Conclusion}

Motivated by two collections of networks representing the same phenomena but collected using very different sampling designs, we combined their strengths, facilitating population-wide simulation of household networks. In the process, we identified the requirements of this procedure and developed generally applicable techniques for specifying and diagnosing models for large samples of networks, techniques that, through their relationship to GLMs, can be used by researchers from a wide variety of disciplines. The techniques we have developed do not rely on the networks being completely observed. To make this methodology accessible to a broad audience, we produced a user-friendly R package \texttt{ergm.multi}.

Our two surveys were conducted in Flanders and Brussels in 2010--2011. It is important to design and analyse household surveys in different settings with different inclusion criteria---but, ideally, compatible measurement instruments---to gain further insights on the contact patterns and the effects of endogenous factors such as triadic closure, exogenous individual attributes such as age, and exogenous household attributes such as size and type of residence. This work provides a foundation for identifying and testing these effects and for confirming the validity of the analysis---and opens the door to design of future cost-effective yet highly informative hybrid network studies.

A number of methodological research directions remain. In our work, we used Pearson residuals. Other types of residuals, such as deviance, tend to be better behaved and could, perhaps, be derived for this family of models. Similarly, Cook's distance may be possible to compute inexpensively for each network using the approach of \citet{KoWa18o}.

We did not find evidence of non-systematic heterogeneity of networks. Where such is present, it can be accounted for in a mixed effects framework \citep{SlKo16m} at an additional computational cost, or perhaps by constructing ERGM sufficient statistics to absorb the variation \citep{Bu17b,Kr12e}. Alternatively, quasi-likelihood and generalised estimating equation approaches may be extended to samples of networks.

ERGM computational and diagnostic techniques are agnostic to the structure of the sample space, so these approaches directly generalise to directed, temporal, valued, and multilayer network scenarios. For our two surveys in particular, physical contact was not the only relational measurement: the respondents were also asked about the approximate duration of interaction (close proximity) on an ordinal scale (time ranges). Along similar lines, the techniques for calculating standardised residuals under partially observed data may be applicable to other domains that involve modelling independent samples of units which are themselves partially observed.

\section*{Acknowledgements}

\acknowledge

\bibliography{bibliography}
\addcontentsline{toc}{section}{References}

\appendix

\counterwithin{figure}{section}
\counterwithin{table}{section}

\renewcommand{\thefigure}{\thesection\arabic{figure}}
\renewcommand{\thetable}{\thesection\arabic{table}}

\clearpage
\section{\label{app:data}Data}
\subsection{\label{app:data-design}Data Collection and Preprocessing}

In both surveys, households were recruited via random-digit dialling on mobile phones and landlines and asked to record their social contacts on a diary during one quota-based assigned day \citep{HoaCo21c}, 318 in the $H$ survey and 1759 for $E$ survey. For each contact the gender and the exact age  or the estimated age interval of each contacted person were recorded, together with contacts' features including location, duration and frequency/intimacy level. Two types of contacts were recorded, one requiring a two-way conversation of at least three words and one requiring skin-to-skin contact. Participants also filled in a background survey, including information about age and gender of their household members. Contacts with a household member were identified requiring the contact was with a member of the household and checking that the age and gender of the contacted person matched the ones of a household member.

\paragraph{Eligibility and missing data}
A number of households and individuals in the survey data were excluded from our analysis. Respondents living alone and whose postal codes were not in Belgium were excluded as not in the population of interest (though it is also possible that their postal codes were misrecorded). Others were excluded due to missing data or within-household contacts being not uniquely identifiable by age and gender. This latter factor would disproportionately affect same-sex couples, large households, and households with twins.

The causes of exclusion and numbers affected are given in Table~\ref{tab:HE-problems}. The percentage of the eligible respondents ultimately excluded for reasons that may be informative in ways not accounted for by the model (i.e., identifiability of contacts) is very small. There are, of course, other approaches, such as imputation, which may work better, and we leave them for future research.
\begin{table}[p]
  \caption{\label{tab:HE-problems} Household and respondent exclusion criteria and their impact. Units excluded are only counted once (towards the earlier of the criteria listed). Percentages are relative to the previous total.}
  \footnotesize
\begin{center}

\begin{tabular}{lrr}
\toprule
 & $H$ & $E$\\
\midrule
Total recruited & $342$ (100.0\%) & $1759$ (100.0\%)\\
$-$ Some responses not recorded & $-24$ (\hphantom{0}\hphantom{0}7.0\%) & $$ \hphantom{00.0\%}\\
\midrule
Total recorded & $318$ (\hphantom{0}93.0\%) & $1759$ (100.0\%)\\
$-$ Living alone & $$ \hphantom{00.0\%} & $-227$ (\hphantom{0}12.9\%)\\
$-$ Postal code not in Belgium & $$ \hphantom{00.0\%} & $  -4$ (\hphantom{0}\hphantom{0}0.2\%)\\
\midrule
Total eligible & $318$ (100.0\%) & $1528$ (\hphantom{0}86.9\%)\\
$-$ Day of week missing & $ -1$ (\hphantom{0}\hphantom{0}0.3\%) & $  -1$ (\hphantom{0}\hphantom{0}0.1\%)\\
$-$ Postal code missing & $$ \hphantom{00.0\%} & $  -7$ (\hphantom{0}\hphantom{0}0.5\%)\\
$-$ No household members listed & $$ \hphantom{00.0\%} & $  -7$ (\hphantom{0}\hphantom{0}0.5\%)\\
$-$ Household size not reported & $$ \hphantom{00.0\%} & $  -1$ (\hphantom{0}\hphantom{0}0.1\%)\\
$-$ Household size and composition mismatch & $$ \hphantom{00.0\%} & $ -19$ (\hphantom{0}\hphantom{0}1.2\%)\\
$-$ Household contacts not uniquely identifiable & $$ \hphantom{00.0\%} & $ -30$ (\hphantom{0}\hphantom{0}2.0\%)\\
\midrule
Total included in the analysis & $317$ (\hphantom{0}99.7\%) & $1463$ (\hphantom{0}95.7\%)\\
\bottomrule
\end{tabular}

\end{center}
\end{table}

\paragraph{Manual contact matching} To account for potential errors in data entry from paper survey to digital form, we allowed for imperfect matching. If no exact match was found, we also considered the following contacts:
\begin{enumerate}
\item Exactly matched the age of a household member but not their gender.
\item Exactly matched the gender of a household member and but had a discrepancy in age, either off by 1 (to account for ageing between recruitment and survey completion) or a 1-digit discrepancy particularly when the digits were visually similar. For example, a report of a contact with a 43-year-old male would match a 48-year-old male household member if there were no better matches.
\end{enumerate}
This error-correction was performed blindly with respect to the analysis. Its effect was to increase the number of matched contacts in $E$ households from 2860 to 2925.

\paragraph{Ageing effects}

Two $H$ households (\#39 and \#51) were found not to contain any children 12 or under but contained children aged 13. This is likely to be due to their birthdays occurring in the period between the household's recruitment into the study and the day of the contact diary. Since these households were selected based on the presence of a child, we treated them no differently from other $H$ households in our diagnostics, though we modelled those actors' relations as we would ordinary 13-year-olds for consistency.

\subsection{\label{app:data-summaries}Additional Data Summaries}

Data summaries most critical to the analysis are included in Section~\ref{sec:data-description} in the body of the article. This appendix provides additional summaries and visualisations.

\begin{figure}[p]
  \subfloat[$H$ dataset: 317 households with a total of 1262 members/respondents]{\label{subfig:H-hists}

{\centering \includegraphics[width=\maxwidth]{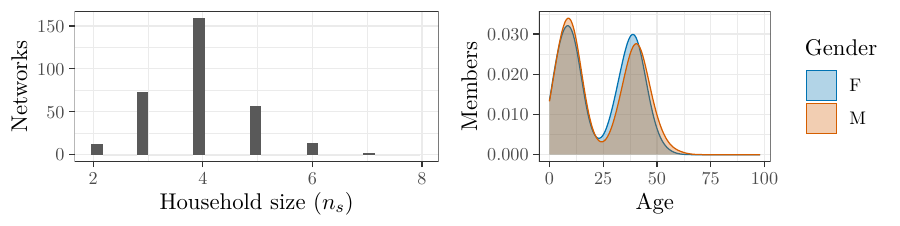} 

}

}

  \subfloat[$E$ dataset: 1463 households/respondents with a total of 4780 members]{\label{subfig:E-hists}

{\centering \includegraphics[width=\maxwidth]{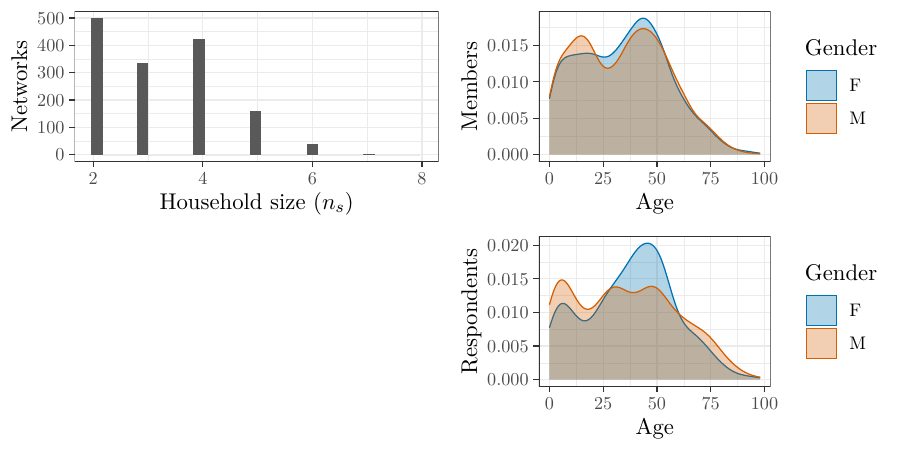} 

}

}
\caption{\label{fig:HE-hists}Household size distribution (left), household members' age and gender distribution (right), and respondents' age and gender distribution (bottom, $E$ only).}
\end{figure}

Figure~\ref{fig:HE-hists} shows the distributions of household sizes and ages and genders of household members in the two datasets. Differences between the two are due to design, with individuals aged around 25 years certainly underrepresented in the $H$ dataset (Panel~\subref{subfig:H-hists}), as they are unlikely to be either parents or children of parents of children 12 or under. The members of the households in the $E$ dataset (Panel~\subref{subfig:E-hists}) are more representative of the population (though not completely so, as individuals living alone ($\nactors_\sampidx=1$) are excluded); but survey respondents (Panel~\subref{subfig:E-hists}, bottom) have a different age and gender distribution from $E$ household members in general, with females aged 25--55 years overrepresented and adolescents of both genders underrepresented.

\begin{figure}

{\centering \includegraphics[width=\maxwidth]{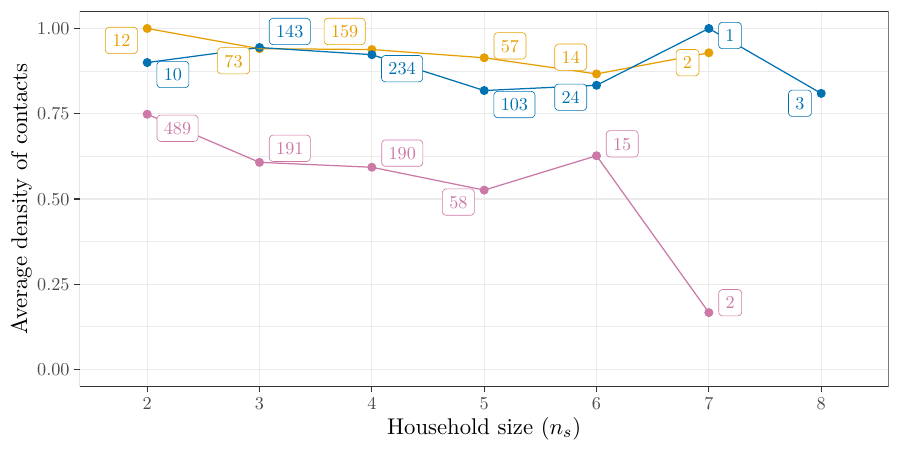} 

}

\caption[Mean household density of contacts (fraction of observable relations that are present) for each household size and data subset, and sample sizes for each household size and subset combination]{Mean household density of contacts (fraction of observable relations that are present) for each household size and data subset, and sample sizes for each household size and subset combination. \subdataleg}\label{fig:size-vs-dens}
\end{figure}

Densities of contacts in the households as a function of household size and subset ($H$, $\Ewc$, $\Enc$, respectively, $H$ dataset, households in $E$ dataset with a child at most 12, and $E$ dataset households without) are shown in Figure~\ref{fig:size-vs-dens}, along with sample sizes of each combination. $E$ households without a child differ systematically from those with a child ($H$ or $E$), and tend to be concentrated among smaller household sizes, whereas those with a child tend to have a size of around 4. Overall, all subsets show a negative relationship between household size and its density of contacts. This is as expected \citep{KrHa11a}.

More household and structural summaries are provided in Table~\ref{tab:HE-misc}.

\begin{table}[p]
  \caption{\label{tab:HE-misc} Miscellaneous dataset composition information. In the $H$ dataset, every household member is a ``respondent''. Clustering coefficient is only meaningful for $\nactors_\sampidx\ge 3$ and cannot be computed for the $E$ dataset households.}
  \footnotesize
\begin{center}

\begin{tabular}{lrr}
\toprule
 & $H$ & $E$\\
\midrule
Household members & 1262 (100\%) & 4780 (100\%)\\
\qquad Female & 633 (\hphantom{0}50\%) & 2342 (\hphantom{0}49\%)\\
\qquad Male & 629 (\hphantom{0}50\%) & 2438 (\hphantom{0}51\%)\\
Respondents &  & 1463 (100\%)\\
\qquad Female &  & 768 (\hphantom{0}52\%)\\
\qquad Male &  & 695 (\hphantom{0}48\%)\\
Dyads (observed) & 2007 (100\%) & 6387 (\phantom{0}52\%)\\
\addlinespace
Child present in household & 317 (100\%) & 518 (\hphantom{0}35\%)\\
Household observed on weekend & 91 (\hphantom{0}29\%) & 364 (\hphantom{0}25\%)\\
Household located in Brussels & 36 (\hphantom{0}11\%) & 0 (\hphantom{0}\hphantom{0}0\%)\\
\addlinespace
Mean (std. dev.) net. density & 0.93 (0.15) & 0.75 (0.38)\\
Mean (std. dev.) net. clustering coef. & 0.92 (0.21) & \\
\bottomrule
\end{tabular}

\end{center}
\end{table}

\subsection{\label{app:data-samp-size}Sample Size Considerations}

\begin{figure}[p]

{\centering \subfloat[Total pairs of individuals observed: the optimistic estimate\label{fig:mixmat_samp_size-1}]{\includegraphics[width=\maxwidth]{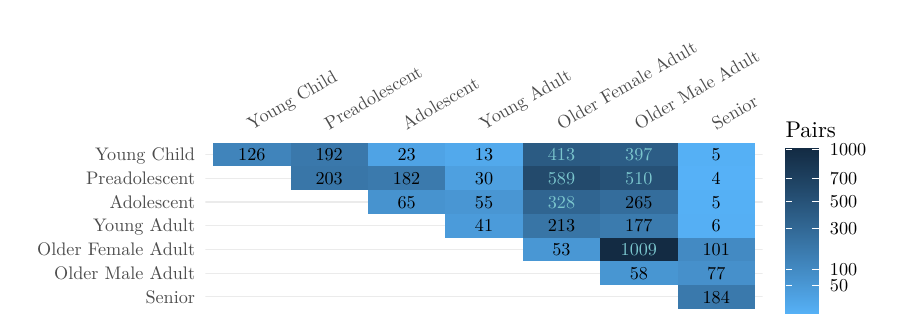} }\newline
\subfloat[Number of households with at least one pair observed: the conservative estimate\label{fig:mixmat_samp_size-2}]{\includegraphics[width=\maxwidth]{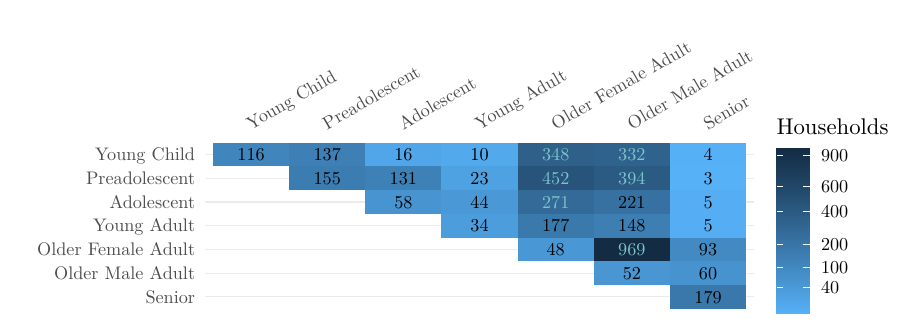} }

}

\caption[Each cell represents a measure of sample size for estimating the propensity of two individuals from those categories to interact]{Each cell represents a measure of sample size for estimating the propensity of two individuals from those categories to interact. Unobserved relations (e.g., those not incident on a respondent in the $E$ dataset) are not counted.}\label{fig:mixmat_samp_size}
\end{figure}

The number of relations whose state was observed was $2007$ from the $H$ and $3317$ from the $E$ datasets, for a total of $5324$. This number would be our sample size under a model that assumed independence among the relations within each household, and is therefore an ``optimistic'' measure of the information available. In contrast, the number of households informative about at least one instance of a particular feature of the model is the ``conservatve'' measure, since we do assume independence between households. Figure~\ref{fig:mixmat_samp_size} shows these counts for the different combinations of household roles that we use in the model developed in Section~\ref{sec:mixing-eff}.

\begin{figure}[p]

{\centering \subfloat[$H$: Household dataset\label{fig:mixmat_samp_size_by_subset-1}]{\includegraphics[width=\maxwidth]{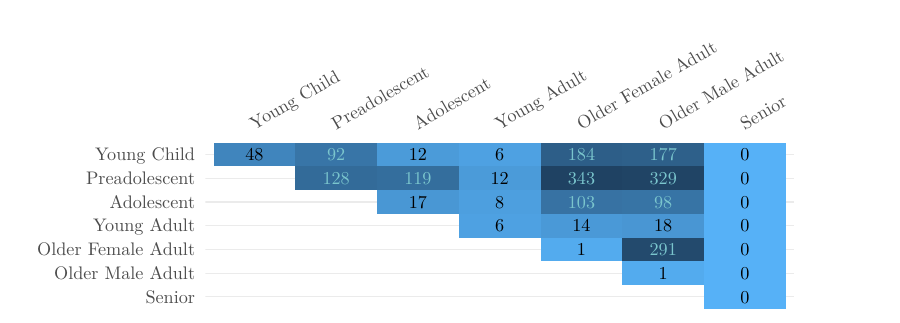} }\newline
\subfloat[\Ewc: Egocentric dataset, child at most 12 present\label{fig:mixmat_samp_size_by_subset-2}]{\includegraphics[width=\maxwidth]{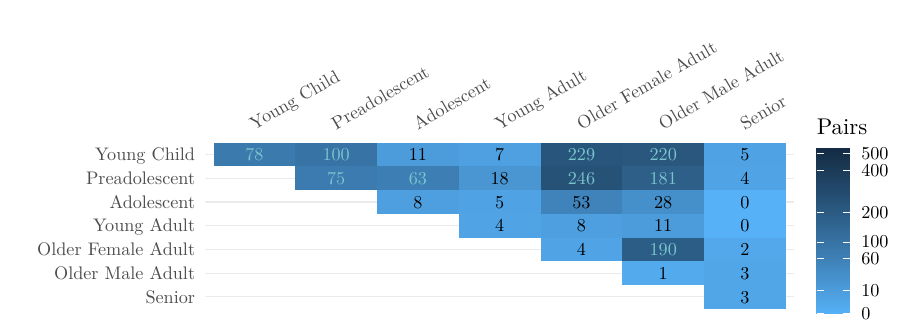} }\newline
\subfloat[\Enc: Egocentric dataset, child at most 12 absent\label{fig:mixmat_samp_size_by_subset-3}]{\includegraphics[width=\maxwidth]{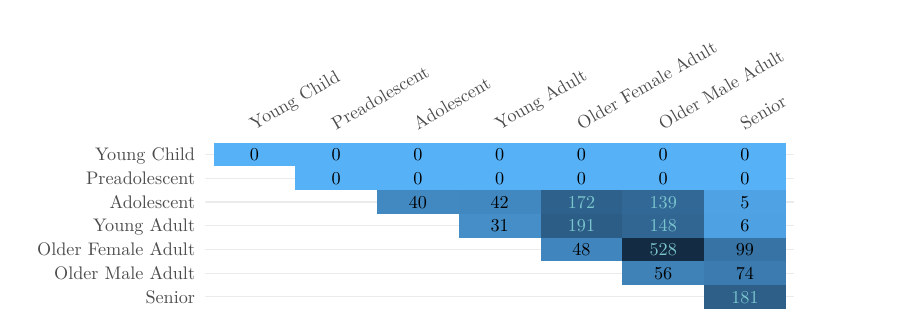} }

}

\caption[Each cell gives the total number of pairs of individuals from those categories whose relationship state was observed (the ``optimistic'' sample size)]{Each cell gives the total number of pairs of individuals from those categories whose relationship state was observed (the ``optimistic'' sample size). Unobserved relations (e.g., those not incident on a respondent in the $E$ dataset) are not counted.}\label{fig:mixmat_samp_size_by_subset}
\end{figure}

We are modelling the impact of a child in a household on mixing patterns, both as a matter of substantive interest and as a design effect. However, we do not have sufficient information to model this effect for each category individually. The sample sizes for these effects are the total number of potential relations for a given category pair observed in subsets $H$ and \Ewc{} (i.e., in the presence of a child) and this number for \Enc{} (i.e., in the absence of a child). These are given in Figure~\ref{fig:mixmat_samp_size_by_subset}. For example, we see that sample size is more than adequate for estimating a separate child effect for older female adult and older male adult contacts ($H+\Ewc$: 481, $\Enc$: 528) but not between two older male adults ($H+\Ewc$: 2, $\Enc$: 56) or two older female adults ($H+\Ewc$: 5, $\Enc$: 48). This informs our modelling choices.

\clearpage
\section[Curved ERGMs]{\label{app:curved}Results for Curved Exponential-Family Models}

\renewcommand{\ergmspec}{{\netspace, \edgecovv, \statv, \cnmapv}}

In Section~\ref{sec:model}, we focused on minimal exponential families, in which the free parameter vector $\paramv$ was also the canonical parameter vector. A more general specification, introduced by \citet{HuHa06i}, also includes a mapping $\cnmapv:\reals^{p'}\mapsto \reals^{p}$ with $p'\le p$ so that the free parameter vector $\paramv\in \reals^{p'}$ is first mapped to the canonical parameter vector $\cnmapv(\paramv)\in \reals^p$. It has since been used to represent less degeneracy-prone degree heterogeneity and triadic effects \citep[and others]{HuHa06i,ScKr20e,StSc19m}.

Here, we provide the definitions and the results for nontrivial $\cnmapv(\cdot)$ paralleling those given in Section~\ref{sec:model} for the $\cnmapv(\paramv)=\paramv$ special case.

\subsection*{Section~\ref{sec:ergm}}
For $\rnet\sim \ERGM_\ergmspec(\paramv)$, now
\begin{equation}\M_\ergmspec ({\Yy};\paramv)=\exp\enbc{\innerprod{\cnmapv\en(){\paramv}}{ \statv(\net\andedgecov)}}/{\nc_\ergmspec(\paramv)},\ {\net\in\netspace},\end{equation} where
$\nc_\ergmspec(\paramv)=\sum_{\ypnets} \exp\enbc{\innerprod{\cnmapv\en(){\paramv}}{\statv(\net'\andedgecov)}}.$

If the network is completely observed, the score function becomes
\[\score(\paramv)=\cnmapv'(\paramv)\t\enbc{\statv(\net) - \meanstatv(\paramv)},\]
where $\cnmapv': \reals^{p'}\mapsto\reals^{p\times p'}$ is the Jacobian matrix $\cnmapv'(\paramv)\defeq \frac{\partial \cnmapv(\paramv)}{\partial\paramv}$, and Fisher information becomes
\[{\infom(\paramv)} = \Var\en[]{\cnmapv'(\paramv)\t\enbc{\statv(\rnet) - \meanstatv(\paramv)}} = \cnmapv'(\paramv)\t\varstatm(\paramv)\cnmapv'(\paramv).\]

In the partially observed case \citep{OrWo72m,Su74m,HaGi10m}, the corresponding expressions are
\begin{equation}\score(\paramv)=\cnmapv'(\paramv)\t{\enbc{\meanstatv(\paramv\mid\net\obs) - \meanstatv(\paramv)}}\label{eq:curved-score}\end{equation}
and
\begin{align}
  \infom(\paramv) &= \Var_{\rnet}\en[]{\cnmapv'(\paramv)\t\meanstatv\enbc{\paramv\mid\observe(\rnet)}}\\
                  &= \cnmapv'(\paramv)\t\varstatm(\paramv)\cnmapv'(\paramv)-\cnmapv'(\paramv)\t\E_{\rnet}\en[]{\varstatm\enbc{\paramv\mid\observe(\rnet)}}\cnmapv'(\paramv).
\end{align}

\subsection*{Section~\ref{sec:mlergm}}
For a sample of networks, the mapping $\cnmapv_\sampidx(\cdot)$ is applied to $\paramv_\sampidx$, not $\lparamv$: $\cnmapv_\sampidx(\paramv_\sampidx) \equiv \cnmapv_\sampidx\enbc{(\netcovnet\lparamv)\t}$, so that
\[(\rnet_1,\rnet_2,\dotsc,\rnet_{\samp})\sim\ERGM_{\netcovv,\vec{\netspace},\vec{\edgecovv},\vec{\statv}, \vec{\cnmapv}}(\lparamv)\]
if $\rnet_{{\sampidx}}\simind\ERGM_{\netspace_{\sampidx},\edgecovv_{\sampidx},\statv_{\sampidx}, \cnmapv_{\sampidx}}(\paramv_{{\sampidx}})$ still.

\subsection*{Section~\ref{sec:inference}}
The complete-data likelihood becomes
\begin{align}
      \lk(\lparamv)&={\exp\EN[]{\sum_{\sampidx=1}^\samp \innerprod{\cnmapv_\sampidx\enbc{(\netcovnet\lparamv)\t}}{\statv_{\sampidx}(\net_{\sampidx}\andnetedgecov)}}}\big/\prod_{\sampidx=1}^\samp \nc_{\netspace_{\sampidx},\edgecovv_{\sampidx},\statv_{{\sampidx}}}\enbc{(\netcovnet\lparamv)\t}.
\end{align}
Recalling that $(\netcovnet\lparamv)\t = \netcovm \vecf(\lparamv)$ for $\netcovm\defeq\Ident_{p'}\kron\netcovnet\in \reals^{p'\times p'q}$ and applying the Chain Rule, the partially observed score function becomes
\[\score(\vecf\lparamv)=\sum_{\sampidx=1}^\samp\netcovm\t\cnmapv_\sampidx'\enbc{(\netcovnet\lparamv)\t}\t{\enbc{\meanstatv_\sampidx(\lparamv\mid\net\obs) - \meanstatv_\sampidx(\lparamv)}},\]
and the Fisher information becomes
\[\infom(\vecf\lparamv)=\sum_{\sampidx=1}^\samp\netcovm\t\cnmapv_\sampidx'\enbc{(\netcovnet\lparamv)\t}\t{\Var_{\rnet_{\sampidx}}\en[]{\meanstatv_\sampidx\enbc{\lparamv\mid\observe(\rnet_{\sampidx})}}}\cnmapv_\sampidx'\enbc{(\netcovnet\lparamv)\t}\netcovm.\]

\clearpage
\section{Examples of Nonidentifiability due to Network Sampling}
\subsection{Example 1: Exogenous Subgroups\label{app:nonident-1}}

Recall, an undirected network whose actor set $\actors$ is partitioned into disjoint sets $A$ and $B$. For actor sets $X\subseteq\actors$ and $Y\subseteq\actors$, let $\abs{\net_{X,Y}}$ be the number of edges between actors in sets $X$ and $Y$, and let $\abs{\net}$ be the total edge count, and consider an ERGM with 3 parameters, whose sufficient statistic is
\[\statv(\net)=\big[\abs{\net}, \abs{\net_{A,A}},\abs{\net_{A,B}}\big]\t.\]

This statistic does not induce dependence among the dyads, and so under this model,
\begin{align*}
  \abs{\rnet_{A,A}}&\sim\Binomial\enbc{\tsbinom{\abs{A}}{2}, \ilogit(\param_1+\param_2)}\\
  \abs{\rnet_{A,B}}&\sim\Binomial\enbc{\abs{A}\abs{B}, \ilogit(\param_1+\param_3)}\\
  \abs{\rnet_{B,B}}&\sim\Binomial\enbc{\tsbinom{\abs{B}}{2}, \ilogit(\param_1)},
\end{align*}
all mutually independent. Call their variances $\sigma_{A,A}$, $\sigma_{A,B}$, and $\sigma_{B,B}$, respectively.
Since $\abs{\net}\equiv \abs{\net_{A,A}}+\abs{\net_{A,B}}+\abs{\net_{B,B}}$,
\[\varstatm(\paramv)=\begin{bmatrix}
    \sigma_{A,A} + \sigma_{A,B} + \sigma_{B,B} & \sigma_{A,A} & \sigma_{A,B} \\
    \sigma_{A,A} & \sigma_{A,A} & 0 \\
    \sigma_{A,B} & 0 & \sigma_{A,B}
  \end{bmatrix},\]
and $\abs{\varstatm(\paramv)} = \sigma_{A,A} \sigma_{A,B} \sigma_{B,B}$, nonsingular as long as all elements of $\paramv$ are finite.

Now, consider an observation process in which only relationships incident on members of group $A$ ($A$--$A$ and $A$--$B$) are observed and $B$--$B$ relations are missing by design. $B$--$B$ relationships are not explicitly part of $\statv(\net)$, and, in fact, $\net\obs$ contains all of the information needed to compute $\stat_2(\cdot)$ and $\stat_3(\cdot)$. Then,
\begin{align*}
  \meanstatv\en(){\paramv\mid\net\obs} & =\big[ \abs{\net_{A,A}} + \abs{\net_{A,B}} + \E\enbc{\abs{\rnet_{B,B}}\mid\rnet\in\netspace(\net\obs)}, \abs{\net_{A,A}}, \abs{\net_{A,B}}\big]\t\\
                                    & =\big[ \abs{\net_{A,A}} + \abs{\net_{A,B}} + \tsbinom{\abs{B}}{2}p_{B,B}, \abs{\net_{A,A}}, \abs{\net_{A,B}}\big]\t,
\end{align*}
because $\abs{\rnet_{B,B}}$ does not depend on $\net\obs$. Thus, $\meanstat_1\en(){\paramv\mid\net\obs}$ is an affine combination of $\meanstat_2\en(){\paramv\mid\net\obs}$ and $\meanstat_3\en(){\paramv\mid\net\obs}$ for all $\net\obs$, making $\infom(\paramv)$ singular. More explicitly, from \eqref{eq:ergm-info-varcond},
\begin{align*}
  \infom(\paramv) & = \Var_{\rnet}\en[]{\meanstatv\enbc{\paramv\mid\observe(\rnet)}} \\
                  & = \Var \big[ \abs{\rnet_{A,A}} + \abs{\rnet_{A,B}} + \tsbinom{\abs{B}}{2}p_{B,B}, \abs{\rnet_{A,A}}, \abs{\rnet_{A,B}}\big]\t\\
                  & =
                    \begin{bmatrix}
                      \sigma_{A,A} + \sigma_{A,B} + 0 & \sigma_{A,A} & \sigma_{A,B} \\
                      \sigma_{A,A} & \sigma_{A,A} & 0 \\
                      \sigma_{A,B} & 0 & \sigma_{A,B}
                    \end{bmatrix}.
\end{align*}
Here, the first column of $\infom(\paramv)$ is the sum of its second and third.

\subsection{Example 2: Triadic Effects\label{app:nonident-2}}

Now, consider an independent, identically distributed sample of $\samp$ undirected networks of size $\nactors=3$ and an ERGM with the following sufficient statistic, counting edges, 2-stars, and triangles in network $\net$:
\[\statv(\net)=\big[\abs{\net}, \twostct, \trict\big],\]
where $\abs{\net_i}$ is the number of edges incident on actor $i$. Each network in the sample has a sample space of size $\abs{\netspace}=2^{\binom{3}{2}}=8$, enumerated in Table~\ref{tab:threenode-enum}.

For simplicity, we will set $\paramv=\0$, making all networks equiprobable. Then, we can evaluate the population covariance matrix and obtain
\[\infom(\0) = \frac{\samp}{64}\begin{bmatrix}{}
  48 & 48 & 12 \\ 
  48 & 60 & 18 \\ 
  12 & 18 & 7 \\ 
  \end{bmatrix}
\]
and $\abs{\infom(\0)}=(9/4096)\samp^3$. For $\samp=1$, the MLE for $\param_3$ will be infinite due to the sufficient statistic being at its highest or lowest possible value, but this is unrelated to nonidentifiability.

Now, consider an egocentric missing data regime in which relations between Actor 1 and the others are observed but the relationship between Actor 2 and Actor 3 ($\redge{2,3}$) is missing. The conditional expectations $\meanstatv\en(){\paramv\mid\net\obs}$ and probabilities $\Pr(\rnet\obs=\net\obs)$ for every possible $\net\obs$ are given in Table~\ref{tab:threenode-means}. Given those, the covariance matrix $\Var_{\rnet}\en[]{\meanstatv\enbc{\paramv\mid\observe(\rnet)}}$ gives
\[\infom(\0) = \frac{\samp}{64}\begin{bmatrix}{}
  32 & 32 & 8 \\ 
  32 & 36 & 10 \\ 
  8 & 10 & 3 \\ 
  \end{bmatrix}
\]
and $\abs{\infom(\0)}=0$.

\begin{table}[p]
  \caption{Possible undirected networks of size 3 and their edge, 2-star, and triangle statistics.\label{tab:threenode-enum}}
  \footnotesize
  \centering

\begin{tabular}{ccc@{\hskip 2em}cccccc@{\hskip 2em}cccccc@{\hskip 2em}cccccc@{\hskip 2em}cccccc@{\hskip 2em}cccccc@{\hskip 2em}ccc}
\toprule
$\edge{1,2}$ & $\edge{1,3}$ & $\edge{2,3}$ & edges & 2-stars & triangles\\
\midrule
0 & 0 & 0 & 0 & 0 & 0\\
0 & 0 & 1 & 1 & 0 & 0\\
0 & 1 & 0 & 1 & 0 & 0\\
1 & 0 & 0 & 1 & 0 & 0\\
0 & 1 & 1 & 2 & 1 & 0\\
1 & 0 & 1 & 2 & 1 & 0\\
1 & 1 & 0 & 2 & 1 & 0\\
1 & 1 & 1 & 3 & 3 & 1\\
\bottomrule
\end{tabular}

\end{table}
\begin{table}[p]
  \caption{Conditional expectations of edge, 2-star, and triangle statistics for undirected networks of size 3 with $\redge{2,3}$ unobserved and their probabilities under the null model.\label{tab:threenode-means}}
  \footnotesize
  \centering

\begin{tabular}{cc@{\hskip 2em}ccccc@{\hskip 2em}ccccc@{\hskip 2em}ccccc@{\hskip 2em}ccccc@{\hskip 2em}ccccc@{\hskip 2em}ccc}
\toprule
$\Pr(\redge{1,2}=\edge{1,2},\redge{1,3}=\edge{1,3})$ & $\edge{1,2}$ & $\edge{1,3}$ & edges & 2-stars & triangles\\
\midrule
0.25 & 0 & 0 & 0.5 & 0.0 & 0.0\\
0.25 & 0 & 1 & 1.5 & 0.5 & 0.0\\
0.25 & 1 & 0 & 1.5 & 0.5 & 0.0\\
0.25 & 1 & 1 & 2.5 & 2.0 & 0.5\\
\bottomrule
\end{tabular}

\end{table}

\clearpage
\section[Details of Diagnostic Methods]{Mathematical and Computational Details of Diagnostic Methods}

\subsection[Partially Observed Residual Variance]{Estimating the Variance of the Residuals for Partially Observed Networks\label{app:pearson-var}}
Here, we discuss estimation of $\Var_{\rnets}\en[]{\meantarget\enbc{\lmle\mid\observe(\rnets)}}$ in \eqref{eq:pearson-resid-miss} in Section~\ref{sec:pearson-calc}. This is a nested expectation, which calls for nested simulation of network statistics:
first, we simulate a sample of complete networks $\seq{\rnets^{(1)}}{\rnets^{(R_1)}}\sim \ERGM_{\netcovv,\vec{\netspace},\vec{\edgecovv},\vec{\statv}}(\lparamv)$, and then, for each $r_1\inseq{1}{R_1}$, simulate a sample $\seq{\rnets^{(r_1,1)}}{\rnets^{(r_1,R_2)}}\sim \ERGM_{\netcovv,\vec{\netspace}\enbc{\observe(\rnets^{(r_1)})},\vec{\edgecovv},\vec{\statv}}(\lparamv)$. We consider two ways to use these realisations to obtain the estimator.

\paragraph{Direct method\label{app:pearson-var-direct}}

Substituting simulated values into the expectation and the variance yields a consistent (as $R_1\to\infty$ and $R_2\to\infty$) estimator. This direct estimator for the variance of the conditional expectation in the divisor in \eqref{eq:pearson-resid-miss} has the following form:
\[\frac{1}{R_1-1}\sum_{r_1=1}^{R_1}\{\overline{\target(\rnets^{(r_1,\cdot)})}-\overline{\target(\rnets^{(\cdot,\cdot)})}\}^2,\] where $\overline{\target(\rnets^{(r_1,\cdot)})}\defeq R_2\inv\sum_{r_2=1}^{R_2}\target(\rnets^{(r_1,r_2)})$, an estimator of the inner expectation for a given $\observe(\rnets^{(r_1)})$, and $\overline{\target(\rnets^{(\cdot,\cdot)})}\defeq R_1\inv\sum_{r_1=1}^{R_1}\overline{\target(\rnets^{(r_1,\cdot)})}$, the grand mean. Here, we derive the approximate expression for its bias due to the variance capturing the additional variability (decreasing in $R_2$) in estimating the inner expectation by simulation.

Assuming that $R_1$ is sufficiently large that the variance of $\overline{\target(\rnets^{(\cdot,\cdot)})}$
is negligible, and since $\E\{\overline{\target(\rnets^{(r_1,\cdot)})}-\overline{\target(\rnets^{(\cdot,\cdot)})}\}=0$,
\begin{align*}
\E\{\overline{\target(\rnets^{(r_1,\cdot)})}-\overline{\target(\rnets^{(\cdot,\cdot)})}\}^2 & =\Var\{\overline{\target(\rnets^{(r_1,\cdot)})}-\overline{\target(\rnets^{(\cdot,\cdot)})}\} \approx\Var\{\overline{\target(\rnets^{(r_1,\cdot)})}\}\\
 & \approx\Var[\E\{\overline{\statv(\rnets^{(r_1,\cdot)})}\mid\seq{\rnets^{(r_1,1)}}{\rnets^{(r_1,R_2)}}\in\netspace((\rnets^{(r_1)})\obs)\}]\\
 & \hphantom{\approx}+\E[\Var\{\overline{\statv(\rnets^{(r_1,\cdot)})}\mid\seq{\rnets^{(r_1,1)}}{\rnets^{(r_1,R_2)}}\in\netspace((\rnets^{(r_1)})\obs)\}]\\
 & \approx\Var[\E\{\statv(\rnets')\mid\rnets'\in\netspace((\rnets)\obs)\}]\\
 & \hphantom{\approx}+\frac{1}{R_2}\E[\Var\{\statv(\rnets')\mid\rnets'\in\netspace((\rnets)\obs)\}]
\end{align*}
Therefore, this estimator is biased, with bias decreasing
as a function of $R_2$. An adjusted estimator could then be 
\[
\frac{1}{R_1-1}\sum_{r_1=1}^{R_1}\{\overline{\target(\rnets^{(r_1,\cdot)})}-\overline{\target(\rnets^{(\cdot,\cdot)})}\}^2
-\frac{1}{R_2}\frac{1}{R_1}\sum_{r_1=1}^{R_1}\frac{1}{R_2-1}\sum_{r_2=1}^{R_2}\{\target(\rnets^{(r_1,r_2)})-\overline{\target(\rnets^{(r_1,\cdot)})}\}^2.
\]

\paragraph{Law of Total Variance method\label{app:pearson-var-total}}

This approach takes advantage of the Law of Total Variance to write: \begin{align*}\Var_{\rnets}\en[]{\meantarget\enbc{\lmle\mid\observe(\rnets)}}&=\vartarget(\lmle)-\E_{\rnets}\en[]{\vartarget\enbc{\lmle\mid\observe(\rnets)}},\label{eq:condvar}\\
&\approx\frac{1}{R_1R_2 -1}\sum_{r_1=1}^{R_1}\sum_{r_2=1}^{R_2}\{\target(\rnets^{(i,j)})-\overline{\target(\rnets^{(\cdot,\cdot)})}\}^2\\
&\hphantom{\approx} -\frac{1}{R_1}\sum_{r_1=1}^{R_1}\frac{1}{R_2-1}\sum_{r_2=1}^{R_2}\{\target(\rnets^{(i,j)})-\overline{\target(\rnets^{(i,\cdot)})}\}^2,\end{align*}
an unbiased (except perhaps from MCMC autocorrelation) estimator with an added benefit that some of the simulation error in the minuend and the subtrahend cancels. We derive its properties as follows.

Suppose that each $\rnets^{(r_1)}$ induces a distribution
of $\target(\rnets^{(r_1,r_2)})$ with conditional variance $\sigma_{r_1}^{2}$
(i.e., $\Var\{\target(\rnets^{(r_1,r_2)})\mid\rnets^{(r_1,r_2)}\in\netspace((\rnets^{(r_1)})\obs)\}=\sigma_{r_1}^{2}$). The unbiased estimator is
\[
\tilde{\sigma}^2_{r_1}=\frac{1}{R_2-1}\sum_{r_2=1}^{R_2}\{\target(\rnets^{(r_1,r_2)})-\overline{\target(\rnets^{(r_1,\cdot)})}\}^2
\]
We wish to estimate the expected variance and, under the unconstrained process, $\E(\sigma_{r_1}^{2})=\sigma^2$. Suppose that $\Var(\sigma_{r_1}^{2})=\alpha$. Then, the estimator 
\[
\tilde{\sigma}^2=\frac{1}{R_1}\sum_{r_1=1}^{R_1}\tilde{\sigma}^2_{r_1}
\]
 has expectation 
\begin{align*}
\E(\tilde{\sigma}^2) & =\frac{1}{R_1}\sum_{r_1=1}^{R_1}\E(\tilde{\sigma}^2_{r_1})=\frac{1}{R_1}\sum_{r_1=1}^{R_1}\E_{\sigma^2_{r_1}}\{\E_{\tilde{\sigma}^2_{r_1}}(\tilde{\sigma}^2_{r_1}\mid\sigma^2_{r_1})\}\\
 & =\frac{1}{R_1}\sum_{r_1=1}^{R_1}\E_{\sigma^2_{r_1}}\{\sigma^2_{r_1}\}=\sigma^2,
\end{align*}
so it is unbiased regardless of $R_2$, $R_1$, and the distribution of $\target(\rnets^{(r_1,r_2)})$.

To approximate its variance, we must assume a conditional distribution $\target(\rnets^{(r_1,r_2)})\mid\rnets^{(r_1)}$. If it is approximately normal,
\[
\tilde{\sigma}^2_{r_1}\mid\sigma_{r_1}^{2}\sim\frac{\sigma_{r_1}^{2}}{R_2-1}\chi_{R_2-1}^{2}.
\]

Its variance is
\begin{align*}
  \Var(\tilde{\sigma}^2) & =\frac{1}{R_1^2}\sum_{r_1=1}^{R_1}\Var(\tilde{\sigma}^2_{r_1})\\
 & =\frac{1}{R_1^2}\sum_{r_1=1}^{R_1}[\E_{\sigma^2_{r_1}}\{\Var_{\tilde{\sigma}^2_{r_1}}(\tilde{\sigma}^2_{r_1}\mid\sigma^2_{r_1})\}+\Var_{\sigma^2_{r_1}}\{\E_{\tilde{\sigma}^2_{r_1}}(\tilde{\sigma}^2_{r_1}\mid\sigma^2_{r_1})\}]\\
 & =\frac{1}{R_1^2}\sum_{r_1=1}^{R_1}[\E_{\sigma^2_{r_1}}\{\frac{2\sigma_{r_1}^{4}(R_2-1)}{(R_2-1)^2}\}+\Var_{\sigma^2_{r_1}}\{\sigma^2_{r_1}\}]\\
                         & =\frac{1}{R_1^2}\sum_{r_1=1}^{R_1}[\frac{2}{R_2-1}\E_{\sigma^2_{r_1}}\{\sigma_{r_1}^{4}\}+\alpha]\\
   & =\frac{1}{R_1^2}\sum_{r_1=1}^{R_1}[\frac{2}{R_2-1}\{\E_{\sigma^2_{r_1}}^2(\sigma_{r_1}^{2})+\Var(\sigma_{r_1}^{2})\}+\alpha]\\
                         & =\frac{1}{R_1^2}\sum_{r_1=1}^{R_1}[\frac{2}{R_2-1}\{\sigma^{4}+\alpha\}+\alpha] =\frac{1}{R_1}[\frac{2}{R_2-1}\sigma^{4}+(\frac{2}{R_2-1}+1)\alpha].
\end{align*}
A corollary of this is that increasing $R_2$ can only increase the precision up to a point, whereas provided $R_2\ge 2$, increasing $R_1$ increases precision without limit. Nonetheless, if $\sigma^{4}\gg\alpha$, it may be worthwhile to increase $R_2$, since conditional simulation is computationally cheaper.

\subsection{Simulation-based Score Test for Dataset Effects\label{app:score}}

Here, we describe two variants of the simulation score test for dataset effects. For our purposes, the null hypothesis is the candidate model, which does not include explicit dataset effects, and the alternative is the same model with an explicit dataset effect. Rejection of the null hypothesis therefore implies that the dataset effect is not ignorable.

In a minimal exponential family, the score function is the difference between the observed value of sufficient statistic present in the full model but not in the reduced model and its expected value under the MLE of the reduced model. The score test in question can therefore be conducted by the following simulation \citep{Kr12e}:
\begin{enumerate}
\item Specify a statistic $\target[\text{dataset}](\nets)$ that is not a part of the candidate model and that contains an explicit dataset effect.
\item Simulate complete network datasets $\rnets^{(1)},\dotsc,\rnets^{(R)}\simiid\ERGM(\lmle)$ from the candidate model.
\item Evaluate the sample quantile $$q\defeq \frac{1}{R}\sum_{r=1}^R \I{\target[\text{dataset}](\nets) \le \target[\text{dataset}](\rnets^{(r)})}$$.
\item Then, $2\min(q,1-q)$ is a score test $P$-value for the null hypothesis $\param[\text{dataset}]=0$.
\end{enumerate}

In our application, a straightforward $\target[\text{dataset}](\nets)=\sum_{s\in H} \abs{\net_s}$: separate density for networks in the $H$ dataset, which only depends on the completely observed networks. (A test statistic that uses partially observed networks would require further adjustment along the lines of Section~\ref{sec:pearson-calc}.)

A joint test for several network features is also possible: if $\targetv_{\text{dataset}}(\nets)$ is vector-valued, then, with $\bm{m}$ the estimate of $\E\enbc{\targetv_{\text{dataset}}(\rnets)}$ and $V$ the estimate of $\Var\enbc{\targetv_{\text{dataset}}(\rnets)}$, both obtained by simulation, $\chi^2 = \enbc{\bm{m}-\targetv_{\text{dataset}}(\nets)}\t V\inv \enbc{\bm{m}-\targetv_{\text{dataset}}(\nets)}$ provides an omnibus score test. Unlike the quantile-based test, this test relies on approximate normality of $\targetv_{\text{dataset}}(\rnets)$, which can be guaranteed by a sufficiently large sample of networks.

\clearpage
\section{\label{app:parametrisation}Parametrisation and Network-Level Design Matrices}

In Section~\ref{sec:mlergm}, when we described the framework for modelling effects of network-level properties on its relational structure, we expressed it under the assumption that dimension $\nstat$ of the sufficient statistic $\statv_{\sampidx}(\cdot)$ did not vary between networks, and nor did the dimension $\nparam$ of the network-level covariate vector $\netcovnet$, and $\paramv_{\sampidx}\defeq(\netcovnet\lparamv)\t$. An implication of this is that every network-level covariate has the potential to affect every ERGM parameter.

This is often not the desired behaviour. For example, in \Model{1}, a network's edge count statistic is modelled by network size, location in Brussels, and whether the data were gathered on the weekend (but no intercept); whereas the 2-star and triangle counts are modelled with network size and intercept both. The family role mixing effects are assumed to be constant (net of others), with the exception of contacts between older female adults and older male adults, which also have the presence of a child as a predictor.

We claimed that provided that we could fix some elements of $\lparamv$ at 0, no generality was lost, and we illustrate this here for \Model{1} as follows.

Firstly, we construct a design matrix based on the superset of network properties used in \Model{1}. This is illustrated for a selection of households in Table~\ref{tab:model-matrix}.
\begin{table}[p]
  \caption{\label{tab:model-matrix}Network properties and resulting network-level covariates for a selection of households.}
  \footnotesize
  \centering

\begin{tabular}{lllccccccrrrrrr}
\toprule
\multicolumn{1}{c}{ } & \multicolumn{1}{c}{ } & \multicolumn{1}{c}{ } & \multicolumn{4}{c}{Network-level properties} & \multicolumn{1}{c}{ } & \multicolumn{1}{c}{ } & \multicolumn{6}{c}{Design matrix ($\netcovnet$)} \\
\cmidrule(l{3pt}r{3pt}){4-7} \cmidrule(l{3pt}r{3pt}){10-15}
ID &     &      & \rotatebox{90}{household size ($\nactors_\sampidx$)} & \rotatebox{90}{post code in Brussels} & \rotatebox{90}{day of the week} & \rotatebox{90}{a child aged $\le 12$} &   &    & \rotatebox{90}{Intercept} & \rotatebox{90}{$\log(\nactors_{\sampidx})$} & \rotatebox{90}{$\log^2(\nactors_{\sampidx})$} & \rotatebox{90}{Brussels post code} & \rotatebox{90}{on weekend} & \rotatebox{90}{child absent}\\
\midrule
H1 &  &  & 5 & No & Fri & Present &  &  & 1 & 1.61 & 2.59 & 0 & 0 & 0\\
H106 &  &  & 3 & No & Wed & Present &  &  & 1 & 1.10 & 1.21 & 0 & 0 & 0\\
H211 &  &  & 5 & No & Tue & Present &  &  & 1 & 1.61 & 2.59 & 0 & 0 & 0\\
H316 &  &  & 4 & Yes & Sat & Present &  &  & 1 & 1.39 & 1.92 & 1 & 1 & 0\\
E1 &  &  & 2 & No & Fri & Absent &  &  & 1 & 0.69 & 0.48 & 0 & 0 & 1\\
E488 &  &  & 5 & No & Mon & Present &  &  & 1 & 1.61 & 2.59 & 0 & 0 & 0\\
E975 &  &  & 2 & No & Sun & Absent &  &  & 1 & 0.69 & 0.48 & 0 & 1 & 1\\
E1462 &  &  & 6 & No & Mon & Present &  &  & 1 & 1.79 & 3.21 & 0 & 0 & 0\\
\bottomrule
\end{tabular}

\end{table}

Secondly, for all networks we construct a statistic vector that is a superset of all sufficient statistics used in the model. Where a statistic is inapplicable (e.g., number of triangles in a network of size 2), it will simply be 0. Recalling the definition of $\abs{\net_{X,Y}}$ from Section~\ref{app:nonident-1}, the following vector results:
\begin{multline*}\statv_{\sampidx}(\net)=\textstyle\Big[\abs{\net},\ \twostct[\nactors_{\sampidx}],\ \trict[\nactors_{\sampidx}],\\
  \abs{\net_{\text{OFA},\text{OMA}}},\ \text{\{other role mixing\}}\Big],\end{multline*}
containing, respectively, the number of edges, the number of 2-stars, the number of triangles, the number of relations between older female adults and older male adults, and the rest of the mixing counts for family roles enumerated in Section~\ref{sec:mixing-eff} and elsewhere.

Lastly, constrain $\lparamv$ as shown in Table~\ref{tab:M1-beta-constraints}.
\begin{table}
  \caption{\label{tab:M1-beta-constraints}Constraints on $\lparamv$ to implement \Model{1}. Zeros denote parameters fixed at 0, dots free parameters.}
  \centering
  \footnotesize
\begin{tabular}{lccccc}
  \toprule
  & \rotatebox{90}{$\abs{\net}$} & \rotatebox{90}{$\twostct[\nactors_{\sampidx}]$} & \rotatebox{90}{$\trict[\nactors_{\sampidx}]$} & \rotatebox{90}{$\abs{\net_{\text{OFA},\text{OMA}}}$} & \rotatebox{90}{other role mixing} \\
  \midrule
  Intercept & 0 & $\cdot$ & $\cdot$ & $\cdot$ & $\cdot$ \\
  $\log(\nactors_{\sampidx})$ & $\cdot$ & $\cdot$ & $\cdot$ & 0 & 0\\
  $\log^2(\nactors_{\sampidx})$ & $\cdot$ & $\cdot$ & $\cdot$ & 0 & 0\\
  Brussels post code & $\cdot$ & 0 & 0 & 0 & 0\\
  on weekend & $\cdot$ & 0 & 0 & 0 & 0\\
  child absent & 0 & 0 & 0 & $\cdot$ & 0\\
  \bottomrule
\end{tabular}
\end{table}

In practice, this approach, while technically supported by \pkg{ergm.multi}, is unnecessarily cumbersome and computationally inefficient. We refer the reader to the code to reproduce the analysis and \pkg{ergm.multi} documentation and tutorials for the recommended approach.

\clearpage
\section{Capacity to Detect Misspecification\label{app:misspec}}

To illustrate misspecification detection, we fit three reduced models based on \Model{1}:
\begin{description}
\item[\Model{\Mmin}: \underline{m}inimal model] a model with only edge count, 2-star, and triangle statistics, with their linear and quadratic network size effects, and weekend and Brussels edge effects;
\item[\Model{\Mind}: exclude dyad-\underline{d}ependent effects] a submodel of \Model{1} excluding the dyad-dependent 2-star and triangle effects;
\item[\Model{\Mnns}: exclude network size ($\nactors$) effects] a submodel of \Model{1} excluding linear and quadratic network size effects.
\end{description}
A comparison of their specification to that of the other models is given in Table~\ref{tab:coef-fit}.

\paragraph{Hypothesis testing for network size effect misspecification} Full ANOVA tables are given in Tables~\ref{tab:nns-anova-demo}. In \Model{\Mnns} (Table~\subref{tab:nns-anova-demo-mn}), which lacks network size effects, tests on residuals reject the hypothesis of fit for edges, 2-stars, and triangles (\pvals{} \ensuremath{<0.001}, \ensuremath{<0.001}, and \ensuremath{<0.001}, respectively). In \Model{\Mmin} (Table~\subref{tab:nns-anova-demo-mm}), which does incorporate them, none do (\pvals{} \ensuremath{0.85}, \ensuremath{0.77}, and \ensuremath{0.74}, respectively). \Model{\Mind} (Table~\subref{tab:nns-anova-demo-md}), which has network size effects for density (edge count) but not other features rejects hypothesis of fit for triangles ($\pv\ensuremath{<0.001}$) and nearly so for 2-stars ($\pv\ensuremath{=0.065}$) but not for edges ($\pv\ensuremath{=0.40}$). Thus, the method appears to be capable of detecting and identifying network size effect misspecification with some detail.
\begin{table}
  \caption{\label{tab:nns-anova-demo}Analyses of variance for fitting residuals against household size (represented as a categorical predictor, with a dummy variable for each size).}
  \centering
\subfloat[\Model{\Mnns} (no network size effects)\label{tab:nns-anova-demo-mn}]{
  \footnotesize

\begin{tabular}{lrrrrr}
\toprule
\quad{}Source & df & Sum Sq. & Mean Sq. & $F$ & \pval\\
\midrule
edges &  &  &  &  & \\
\quad{}$\nactors_{\sampidx}$ (categorical) & 7 & 33.6 & 4.8 & 4.6 & \ensuremath{<0.001}\\
\quad{}Residuals & 1773 & 1855.4 & 1.0 &  & \\
\addlinespace
2-stars &  &  &  &  & \\
\quad{}$\nactors_{\sampidx}$ (categorical) & 6 & 28.0 & 4.7 & 4.4 & \ensuremath{<0.001}\\
\quad{}Residuals & 1263 & 1338.2 & 1.1 &  & \\
\addlinespace
triangles &  &  &  &  & \\
\quad{}$\nactors_{\sampidx}$ (categorical) & 6 & 26.9 & 4.5 & 4.4 & \ensuremath{<0.001}\\
\quad{}Residuals & 1263 & 1288.9 & 1.0 &  & \\
\bottomrule
\end{tabular}

}

\subfloat[\Model{\Mmin} (minimal model)\label{tab:nns-anova-demo-mm}]{
  \footnotesize

\begin{tabular}{lrrrrr}
\toprule
\quad{}Source & df & Sum Sq. & Mean Sq. & $F$ & \pval\\
\midrule
edges &  &  &  &  & \\
\quad{}$\nactors_{\sampidx}$ (categorical) & 7 & 3.4 & 0.48 & 0.48 & \ensuremath{0.849}\\
\quad{}Residuals & 1773 & 1771.3 & 1.00 &  & \\
\addlinespace
2-stars &  &  &  &  & \\
\quad{}$\nactors_{\sampidx}$ (categorical) & 6 & 3.3 & 0.56 & 0.55 & \ensuremath{0.768}\\
\quad{}Residuals & 1263 & 1272.7 & 1.01 &  & \\
\addlinespace
triangles &  &  &  &  & \\
\quad{}$\nactors_{\sampidx}$ (categorical) & 6 & 3.6 & 0.59 & 0.58 & \ensuremath{0.743}\\
\quad{}Residuals & 1263 & 1283.8 & 1.02 &  & \\
\bottomrule
\end{tabular}

}

\subfloat[\Model{\Mind} (dyad-independent submodel)\label{tab:nns-anova-demo-md}]{
  \footnotesize

\begin{tabular}{lrrrrr}
\toprule
\quad{}Source & df & Sum Sq. & Mean Sq. & $F$ & \pval\\
\midrule
edges &  &  &  &  & \\
\quad{}$\nactors_{\sampidx}$ (categorical) & 7 & 8.2 & 1.17 & 1.0 & \ensuremath{0.402}\\
\quad{}Residuals & 1773 & 1996.5 & 1.13 &  & \\
\addlinespace
2-stars &  &  &  &  & \\
\quad{}$\nactors_{\sampidx}$ (categorical) & 6 & 11.2 & 1.86 & 2.0 & \ensuremath{0.065}\\
\quad{}Residuals & 1263 & 1183.4 & 0.94 &  & \\
\addlinespace
triangles &  &  &  &  & \\
\quad{}$\nactors_{\sampidx}$ (categorical) & 6 & 20.4 & 3.41 & 4.6 & \ensuremath{<0.001}\\
\quad{}Residuals & 1263 & 944.3 & 0.75 &  & \\
\bottomrule
\end{tabular}

}

\end{table}

\begin{figure}

{\centering \subfloat[\Model{1}: Edges (vs.\ fitted)\label{fig:resid-demo-1}]{\includegraphics[width=\maxwidth]{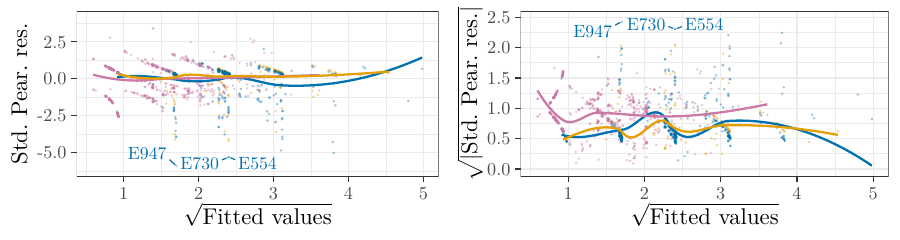} }\newline
\subfloat[\Model{\Mnns}: Edges (vs.\ fitted)\label{fig:resid-demo-2}]{\includegraphics[width=\maxwidth]{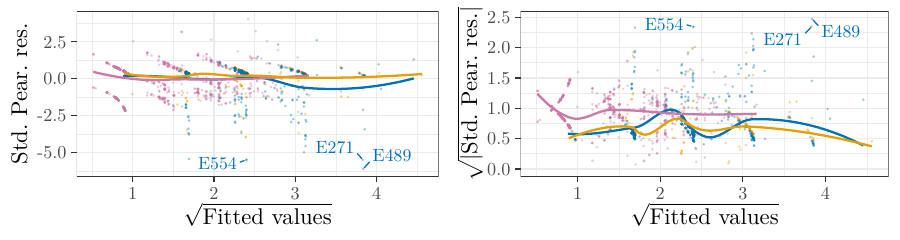} }\newline
\subfloat[\Model{\Mnns}: Edges (vs.\ network size)\label{fig:resid-demo-3}]{\includegraphics[width=\maxwidth]{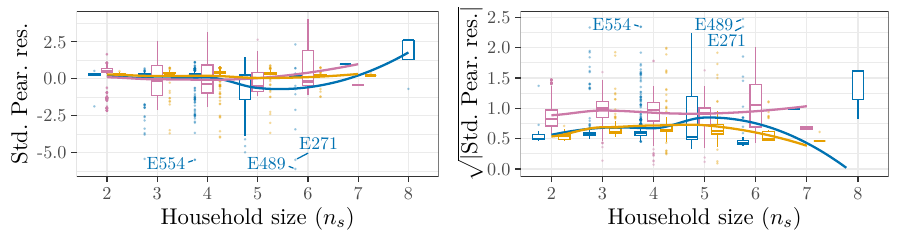} }\newline
\subfloat[\Model{\Mmin}: Edges (vs.\ fitted)\label{fig:resid-demo-4}]{\includegraphics[width=\maxwidth]{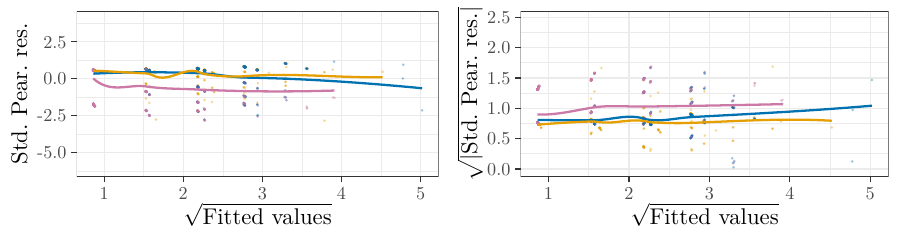} }

}

\caption[Pearson residual plots of network statistics for \Model{1} and its reduced models, using common vertical scales where appropriate]{Pearson residual plots of network statistics for \Model{1} and its reduced models, using common vertical scales where appropriate. \subdataleg}\label{fig:resid-demo}
\end{figure}

\paragraph{Residual plots for dataset and network size effects} Turning to residual plots, Figure~\ref{fig:resid-demo} shows the edge residual plots for \Model{1} (Panel~\subref{fig:resid-demo-1}), \Model{\Mnns} (Panel~\subref{fig:resid-demo-2} and~\subref{fig:resid-demo-3}), and \Model{\Mmin} (Panel~\subref{fig:resid-demo-4}). Observe that unlike the \Model{1}'s, \Model{\Mmin}'s residuals for households without a child are systematically different from those with a child (from either dataset). Thus, the dataset effect is not ignorable. On the other hand, the lack of network size effects in \Model{\Mnns} does not appear to be noticeable in either Panel~\subref{fig:resid-demo-2} or Panel~\subref{fig:resid-demo-3}.

\begin{figure}

{\centering \subfloat[ \Model{1}: Full model\label{fig:age-resid-demo-1}]{\includegraphics[width=\maxwidth]{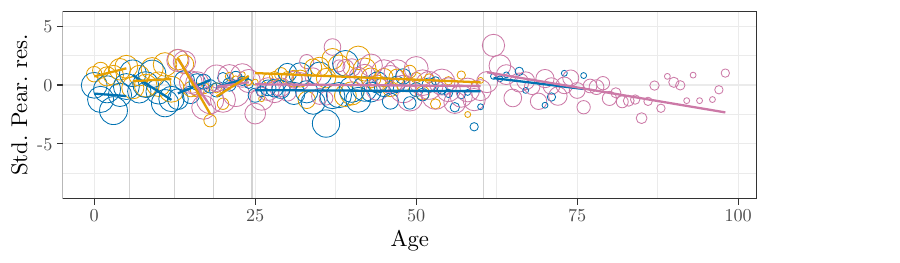} }\newline
\subfloat[ \Model{\Mind}: \Model{1} without dyad-dependent effects\label{fig:age-resid-demo-2}]{\includegraphics[width=\maxwidth]{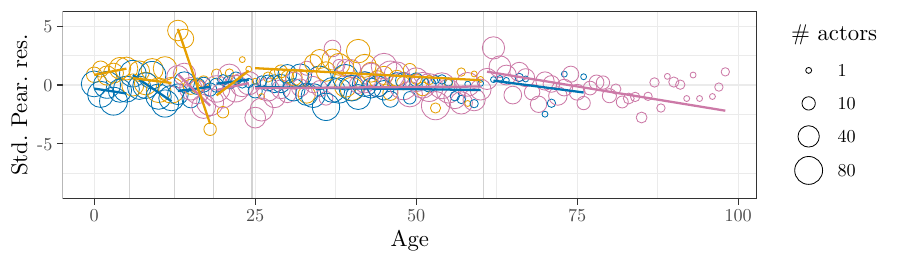} }\newline
\subfloat[\Model{\Mmin}: Minimal model with network size effects only\label{fig:age-resid-demo-3}]{\includegraphics[width=\maxwidth]{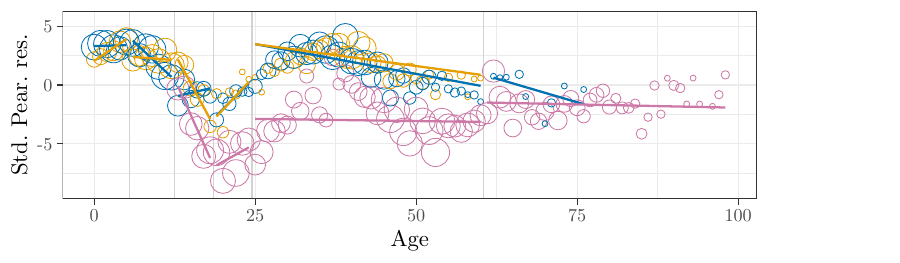} }

}

\caption[Pearson residual plots of edge counts broken down by actor's age and data subset]{Pearson residual plots of edge counts broken down by actor's age and data subset. Vertical lines denote age categories. Symbol size corresponds to the number of actors it represents. \subdataleg}\label{fig:age-resid-demo}
\end{figure}

\paragraph{Residual plots for age effects} We can also construct residual plots for actor-level statistics and plot them against actor attributes. This allows us to diagnose whether age effects in particular have been adequately modelled by our categories. Figure~\ref{fig:age-resid-demo} shows the edge residual plots for \Model{1} (Panel~\subref{fig:age-resid-demo-1}), \Model{\Mind} (Panel~\subref{fig:age-resid-demo-2}), and \Model{\Mmin} (Panel~\subref{fig:resid-demo-3}). \Model{1} and \Model{\Mind}, which model age effects in similar ways, display similar overall patterns (discussed in Section~\ref{sec:example-diagnostics} in the body of the paper), but \Model{\Mind} shows greater overall dispersion, consistent with the residual standard deviations discussed below. In contrast, \Model{\Mmin}, which does not represent age effects at all, shows much clearer patterns, with residuals forming continuous curves with respect to the age, and the data subsets diverging substantially.

\paragraph{Pearson residual standard deviation for unaccounted-for heterogeneity} In Section~\ref{sec:non-syst-diag}, we proposed a simple diagnostic for unexplained heterogeneity: that Pearson residual variances exceed 1. Residual standard deviations are given in Table~\ref{tab:overdispersion-demo} for \Model{1} and the three reduced models. The two models for which some standard deviations exceed 1 are \Model{\Mind} and \Model{\Mnns}; it is likely that the former fails to account for heterogeneity due to dependence among the relations in the network, and the latter fails to account for heterogeneity due to network size. Despite having about a third as many free parameters, \Model{\Mmin}, which incorporates 2-star, triangle, network size effects, and little else, shows residual standard deviations close to those of \Model{1}. This is likely because both the 2-star and the triangle statistics create positive dependence among relations in the network, absorbing the variation along the lines of \citet{Bu17b}. For small networks in particular, it might not be possible to distinguish anything but the most severe network-level heterogeneity---such as that due to network size---from more local heterogeneity.
\begin{table}
  \caption{\label{tab:overdispersion-demo}Standard deviations of Pearson residuals for the specified network statistic under each model.}
  \footnotesize
  \begin{center}

\begin{tabular}{lrrrr}
\toprule
Model & $p$ & edges & 2-stars & triangles\\
\midrule
\Model{\Mmin} & 11 & 1.00 & 1.00 & 1.01\\
\Model{\Mind} & 29 & 1.06 & 0.97 & 0.86\\
\Model{\Mnns} & 29 & 1.03 & 1.04 & 1.02\\
\Model{1} & 35 & 1.00 & 0.99 & 0.98\\
\bottomrule
\end{tabular}

\end{center}
\end{table}

\begin{figure}

{\centering \subfloat[ \Model{1}: Full model\label{fig:dataset-density-demo-1}]{\includegraphics[width=\maxwidth,height=1in]{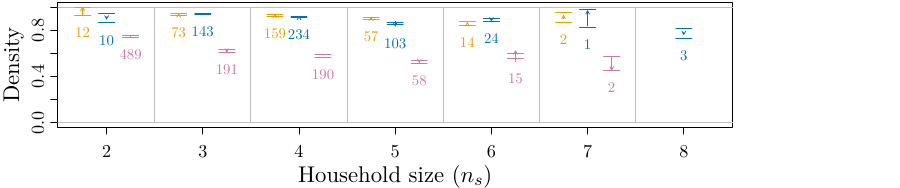} }\newline
\subfloat[ \Model{\Mnns}: \Model{1} without network size effects\label{fig:dataset-density-demo-2}]{\includegraphics[width=\maxwidth,height=1in]{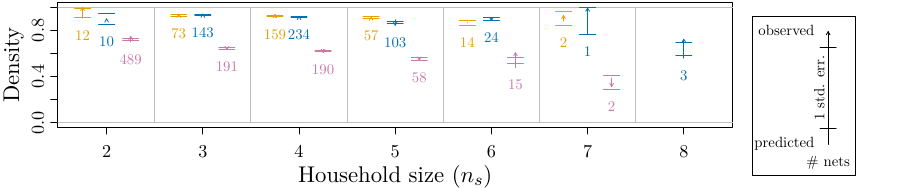} }\newline
\subfloat[\Model{\Mmin}: Minimal model with network size effects only\label{fig:dataset-density-demo-3}]{\includegraphics[width=\maxwidth,height=1in]{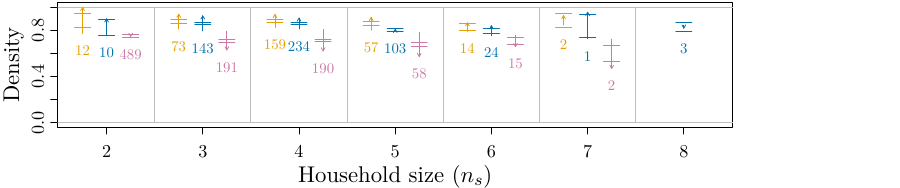} }

}

\caption[Average prediction errors of density]{Average prediction errors of density. Values are averaged over the networks grouped by size and subset. \subdataleg}\label{fig:dataset-density-demo}
\end{figure}

\paragraph{Density error plots for network size and dataset effects} Lastly, we describe an approach for assessing practical significance of lack of fit in predicting density of contacts by contrasting the predicted and the observed/imputed densities for each combination of household size and data subset. Figure~\ref{fig:dataset-density-demo} provides these plots for \Model{1}, \Model{\Mnns}, and \Model{\Mmin}. Generally, households in $H$ and in $\Ewc$ tend to have similar densities for a given size, and both \Model{1} and \Model{\Mnns} (Panels~\subref{fig:dataset-density-demo-1} and~\subref{fig:dataset-density-demo-2}) capture this. We see in Panel~\subref{fig:dataset-density-demo-2} (\Model{\Mnns}) that removing the network size effects does not significantly affect density predictions, as the other 29 parameters adjust to compensate, but we can observe a fairly consistent trend in density errors as a function of network size: for very small and large networks, the densities are underpredicted, but for mid-sized networks, they are overpredicted. (This pattern is not found in \Model{1}'s Panel~\subref{fig:dataset-density-demo-1}.) \Model{\Mmin} (Panel~\subref{fig:dataset-density-demo-3}), as expected, shows large density prediction errors, since it does not attempt to account for household composition.

\clearpage
\section{\label{app:extra-results}Additional Results and Diagnostics}

\setcounter{tocdepth}{2}
\localtableofcontents

A total of eight models were fit: one (\emph{0}) initial, two (\emph{1} and \emph{2}) substantive, two (\emph{\MIp} and \emph{\MIIp}) illustrative, and three (\emph{\Mmin}, \emph{\Mind}, and \emph{\Mnns}) reduced to study misspecification detection. Their motivation is described in Section~\ref{sec:additional-eff} in the body of the article and in Appendix~\ref{app:misspec}, respectively.

Table~\ref{tab:coef-fit} shows a side-by-side comparison of which effects were included in each model, and Table~\ref{tab:models-aod} gives their fit summaries (number of parameters, log-likelihood, AIC, and BIC).

\begin{table}
  \caption{\label{tab:coef-fit}Effects included in each model.}
  \begin{center}
    \footnotesize

\begin{tabular}{lrrrrrrrr}
\toprule
Relationship Effect\\\quad{}$\times$ Network-Level Effect & \emph{\Mmin} & \emph{\Mind} & \emph{\Mnns} & \emph{0} & \emph{1} & \emph{\MIp} & \emph{2} & \emph{\MIIp}\\
\midrule
edges & \checkmark &  &  &  &  &  &  & \\
\quad $\times$ $\log(\nactors_{\sampidx})$ & \checkmark & \checkmark &  & \checkmark & \checkmark & \checkmark & \checkmark & \checkmark\\
\quad $\times$ $\log^2(\nactors_{\sampidx})$ & \checkmark & \checkmark &  & \checkmark & \checkmark & \checkmark & \checkmark & \checkmark\\
\quad   if Brussels post code & \checkmark & \checkmark & \checkmark & \checkmark & \checkmark & \checkmark & \checkmark & \checkmark\\
\quad   if on weekend & \checkmark & \checkmark & \checkmark & \checkmark & \checkmark & \checkmark & \checkmark & \checkmark\\
\quad   if child absent &  &  &  &  &  & \checkmark &  & \\
\quad $\times$ $\log(\text{pop.\ dens.\ in post code})$ &  &  &  &  &  &  & \checkmark & \\
\quad   if city post code &  &  &  &  &  &  &  & \checkmark\\
2-stars & \checkmark &  & \checkmark & \checkmark & \checkmark & \checkmark & \checkmark & \checkmark\\
\quad $\times$ $\log(\nactors_{\sampidx})$ & \checkmark &  &  & \checkmark & \checkmark & \checkmark & \checkmark & \checkmark\\
\quad $\times$ $\log^2(\nactors_{\sampidx})$ & \checkmark &  &  & \checkmark & \checkmark & \checkmark & \checkmark & \checkmark\\
triangles & \checkmark &  & \checkmark & \checkmark & \checkmark & \checkmark & \checkmark & \checkmark\\
\quad $\times$ $\log(\nactors_{\sampidx})$ & \checkmark &  &  & \checkmark & \checkmark & \checkmark & \checkmark & \checkmark\\
\quad $\times$ $\log^2(\nactors_{\sampidx})$ & \checkmark &  &  & \checkmark & \checkmark & \checkmark & \checkmark & \checkmark\\
Young Child with Young Child &  & \checkmark & \checkmark & \checkmark & \checkmark & \checkmark & \checkmark & \checkmark\\
Young Child with Preadolescent &  & \checkmark & \checkmark & \checkmark & \checkmark & \checkmark & \checkmark & \checkmark\\
Preadolescent with Preadolescent &  & \checkmark & \checkmark & \checkmark & \checkmark & \checkmark & \checkmark & \checkmark\\
Adolescent with Adolescent &  & \checkmark & \checkmark & \checkmark & \checkmark & \checkmark & \checkmark & \checkmark\\
Young Child with Young Adult &  & \checkmark & \checkmark & \checkmark & \checkmark & \checkmark & \checkmark & \checkmark\\
Preadolescent with Young Adult &  & \checkmark & \checkmark & \checkmark & \checkmark & \checkmark & \checkmark & \checkmark\\
Adolescent with Young Adult &  & \checkmark & \checkmark & \checkmark & \checkmark & \checkmark & \checkmark & \checkmark\\
Young Adult with Young Adult &  & \checkmark & \checkmark & \checkmark & \checkmark & \checkmark & \checkmark & \checkmark\\
Young Child with Older Female Adult &  & \checkmark & \checkmark & \checkmark & \checkmark & \checkmark & \checkmark & \checkmark\\
Preadolescent with Older Female Adult &  & \checkmark & \checkmark & \checkmark & \checkmark & \checkmark & \checkmark & \checkmark\\
Adolescent with Older Female Adult &  & \checkmark & \checkmark & \checkmark & \checkmark & \checkmark & \checkmark & \checkmark\\
Older Female Adult with Older Female Adult &  & \checkmark & \checkmark & \checkmark & \checkmark & \checkmark & \checkmark & \checkmark\\
Young Child with Older Male Adult &  & \checkmark & \checkmark & \checkmark & \checkmark & \checkmark & \checkmark & \checkmark\\
Preadolescent with Older Male Adult &  & \checkmark & \checkmark & \checkmark & \checkmark & \checkmark & \checkmark & \checkmark\\
Adolescent with Older Male Adult &  & \checkmark & \checkmark & \checkmark & \checkmark & \checkmark & \checkmark & \checkmark\\
Older Female Adult with Older Male Adult &  & \checkmark & \checkmark & \checkmark & \checkmark & \checkmark & \checkmark & \checkmark\\
\quad   if child absent &  & \checkmark & \checkmark &  & \checkmark &  & \checkmark & \checkmark\\
Older Male Adult with Older Male Adult &  & \checkmark & \checkmark & \checkmark & \checkmark & \checkmark & \checkmark & \checkmark\\
Older Female Adult with Senior &  & \checkmark & \checkmark & \checkmark & \checkmark & \checkmark & \checkmark & \checkmark\\
Older Male Adult with Senior &  & \checkmark & \checkmark & \checkmark & \checkmark & \checkmark & \checkmark & \checkmark\\
Senior with Senior &  & \checkmark & \checkmark & \checkmark & \checkmark & \checkmark & \checkmark & \checkmark\\
Adolescent with Young Child or Preadolescent &  & \checkmark & \checkmark & \checkmark & \checkmark & \checkmark & \checkmark & \checkmark\\
Young Adult with Older Adult &  & \checkmark & \checkmark & \checkmark & \checkmark & \checkmark & \checkmark & \checkmark\\
Young Child or Preadolescent with Senior &  & \checkmark & \checkmark & \checkmark & \checkmark & \checkmark & \checkmark & \checkmark\\
Adolescent or Young Adult with Senior &  & \checkmark & \checkmark & \checkmark & \checkmark & \checkmark & \checkmark & \checkmark\\
\bottomrule
\end{tabular}

\end{center}
\end{table}
\begin{table}
  \caption{\label{tab:models-aod}Model fit summaries. MCMC standard errors are given in parentheses.}
  \footnotesize
  \begin{center}

\begin{tabular}{lrrrr}
\toprule
Model & $p$ & log-likelihood & AIC & BIC\\
\midrule
\Model{\Mmin} & 11 & $ -2131.0  \; ( 0.1 )$ & $ 4283.9  \; ( 0.2 )$ & $ 4356.3  \; ( 0.2 )$\\
\Model{\Mind} & 29 & $ -2015.6  \; ( 0.0 )$ & $ 4089.2  \; ( 0.0 )$ & $ 4280.0  \; ( 0.0 )$\\
\Model{\Mnns} & 29 & $ -1844.5  \; ( 0.1 )$ & $ 3747.0  \; ( 0.2 )$ & $ 3937.9  \; ( 0.2 )$\\
\addlinespace
\Model{0} & 34 & $ -1823.3  \; ( 0.1 )$ & $ 3714.6  \; ( 0.2 )$ & $ 3938.3  \; ( 0.2 )$\\
\Model{1} & 35 & $ -1813.3  \; ( 0.1 )$ & $ 3696.6  \; ( 0.2 )$ & $ 3926.9  \; ( 0.2 )$\\
\Model{\MIp} & 35 & $ -1821.8  \; ( 0.1 )$ & $ 3713.6  \; ( 0.2 )$ & $ 3943.9  \; ( 0.2 )$\\
\Model{2} & 36 & $ -1812.4  \; ( 0.1 )$ & $ 3696.8  \; ( 0.2 )$ & $ 3933.7  \; ( 0.2 )$\\
\Model{\MIIp} & 36 & $ -1812.9  \; ( 0.1 )$ & $ 3697.8  \; ( 0.2 )$ & $ 3934.7  \; ( 0.2 )$\\
\bottomrule
\end{tabular}

\end{center}
\end{table}

In the following subsections, we provide the details for each of these models:
\begin{itemize}
\item full coefficient table;
\item full results for common omnibus tests;
\item plots of residuals vs.\ fitted values and network size for edge, 2-star, and triangle counts;
\item plot of residuals for edges incident on actors of each age against age;
\item plot of prediction errors in density by network size and subset;
\item ANOVA tables for testing for unaccounted-for network size effects;
\item results from regressing residuals for edges, 2-stars, and triangles on candidate network-level predictors,
\item tests for dataset effect net of the model;
\item sample standard deviations of Pearson residuals for some network statistics under the model; and
\item residuals for the edge, 2-star, and triangle counts, and counts of relations for each combination of age categories older than 12, all broken down by sub-dataset.
\end{itemize}

\clearpage

\subsection[\Model{\Mmin}]{\label{app:extra-results-mm}Full results and diagnostics for \Model{\Mmin} (the minimal model)}
\small

\begin{table}[H]
  \caption{\label{tab:coef-mm}Parameter estimates (and standard errors) for \Model{\Mmin}.}
  \footnotesize
  \begin{center}

\begin{tabular}{lr}
\toprule
Relationship Effect\\\quad{}$\times$ Network-Level Effect & Coefficient (SE)$^{\hphantom{\star\star\star}}$\\
\midrule
edges & $8.06 \; (1.37)^{\star\star\star}$\\
\quad $\times$ $\log(\nactors_{\sampidx})$ & $-13.69 \; (2.76)^{\star\star\star}$\\
\quad $\times$ $\log^2(\nactors_{\sampidx})$ & $5.26 \; (1.23)^{\star\star\star}$\\
\quad   if Brussels post code & $0.45 \; (0.19)^{\star\phantom{\star}\phantom{\star}}$\\
\quad   if on weekend & $0.14 \; (0.04)^{\star\star\star}$\\
2-stars & $6.84 \; (0.73)^{\star\star\star}$\\
\quad $\times$ $\log(\nactors_{\sampidx})$ & $-7.61 \; (0.31)^{\star\star\star}$\\
\quad $\times$ $\log^2(\nactors_{\sampidx})$ & $1.90 \; (0.14)^{\star\star\star}$\\
triangles & $-4.53 \; (0.98)^{\star\star\star}$\\
\quad $\times$ $\log(\nactors_{\sampidx})$ & $8.57 \; (1.37)^{\star\star\star}$\\
\quad $\times$ $\log^2(\nactors_{\sampidx})$ & $-2.67 \; (0.69)^{\star\star\star}$\\
\bottomrule
\end{tabular}

Significance: $^{\star\star\star}\le 0.001<^{\star\star}\le 0.01< ^\star \le 0.05$
\end{center}
\end{table}
\begin{table}[H]
  \caption{\label{tab:omnibus-mm} Omnibus tests for selected groups of effects in \Model{\Mmin}. Effects are net of the rest of the model.}
  \footnotesize
  \begin{center}

\begin{tabular}{lrr}
\toprule
Effects & Wald $\chi^2$ (df) & $\pval$\\
\midrule
any 2-star & 1300.5 (3) & \ensuremath{<0.001}\\
any triangle & 149.9 (3) & \ensuremath{<0.001}\\
any $\log(\nactors_\sampidx)$ or $\log^2(\nactors_\sampidx)$ & 67787.2 (6) & \ensuremath{<0.001}\\
any $\log^2(\nactors_\sampidx)$ & 428.2 (3) & \ensuremath{<0.001}\\
2-star or triangle $\log^2(\nactors_\sampidx)$ & 281.8 (2) & \ensuremath{<0.001}\\
\bottomrule
\end{tabular}

\end{center}
\end{table}

\begin{figure}[H]

{\centering \subfloat[Edges\label{fig:unnamed-chunk-42-1}]{\includegraphics[width=\maxwidth]{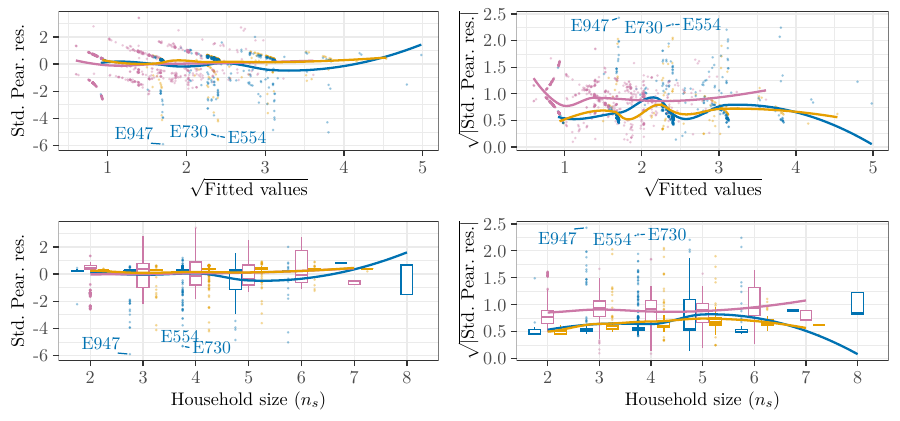} }\newline
\subfloat[2-stars\label{fig:unnamed-chunk-42-2}]{\includegraphics[width=\maxwidth]{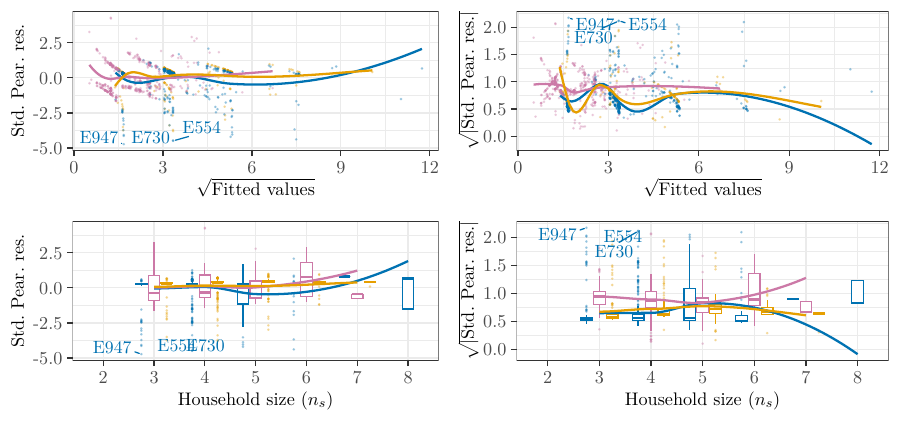} }\newline
\subfloat[Triangles\label{fig:unnamed-chunk-42-3}]{\includegraphics[width=\maxwidth]{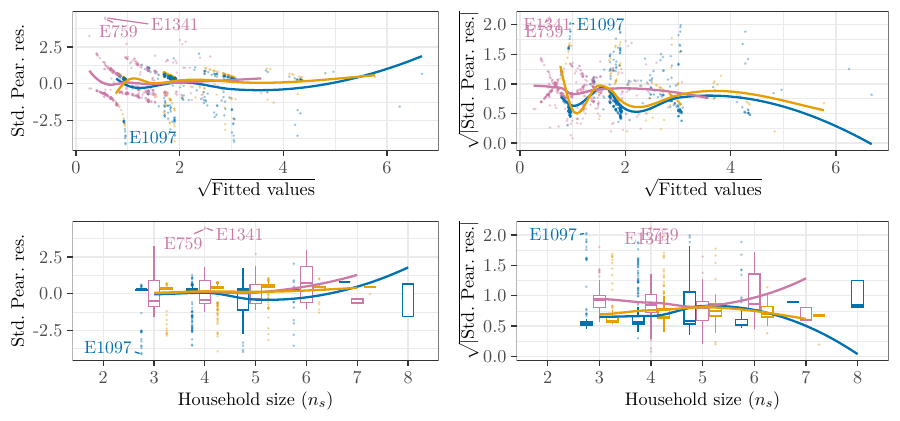} }

}

\caption{\label{fig:resid-plots-mm}Residual plots of network statistics against fitted values and network size for \Model{\Mmin}. \subdataleg}\label{fig:unnamed-chunk-42}
\end{figure}

\begin{figure}[H]

{\centering \includegraphics[width=\maxwidth]{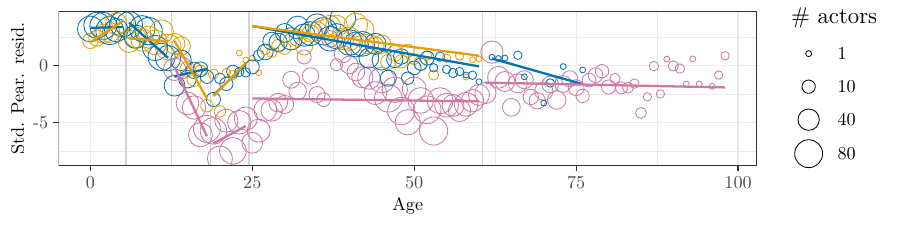} 

}

\caption{\label{fig:age-resid-mm}Residual plot for \Model{\Mmin} of edges incident on actors of a given age against age. \subdataleg}\label{fig:unnamed-chunk-43}
\end{figure}

\begin{figure}[H]

{\centering \includegraphics[width=\maxwidth,height=1in]{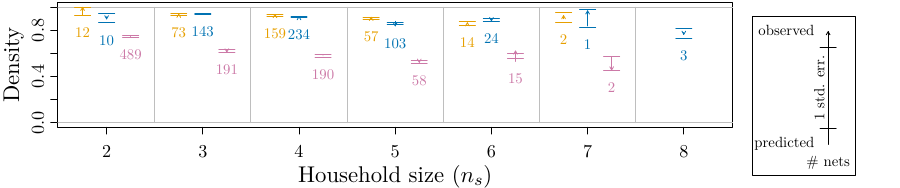} 

}

\caption{\label{fig:density-error-mm}Average prediction errors of density in \Model{\Mmin}. Values are averaged over the networks grouped by size and subset. \subdataleg}\label{fig:unnamed-chunk-44}
\end{figure}

\begin{table}[H]
  \caption{\label{tab:nns-anova-mm}Analyses of variance for fitting residuals for \Model{\Mmin} against household size (represented as a categorical predictor, with a dummy variable for each size).}
  \centering
  \footnotesize

\begin{tabular}{lrrrrr}
\toprule
\quad{}Source & df & Sum Sq. & Mean Sq. & $F$ & \pval\\
\midrule
edges &  &  &  &  & \\
\quad{}$\nactors_{\sampidx}$ (categorical) & 7 & 3.4 & 0.48 & 0.48 & \ensuremath{0.849}\\
\quad{}Residuals & 1773 & 1771.3 & 1.00 &  & \\
\addlinespace
2-stars &  &  &  &  & \\
\quad{}$\nactors_{\sampidx}$ (categorical) & 6 & 3.3 & 0.56 & 0.55 & \ensuremath{0.768}\\
\quad{}Residuals & 1263 & 1272.7 & 1.01 &  & \\
\addlinespace
triangles &  &  &  &  & \\
\quad{}$\nactors_{\sampidx}$ (categorical) & 6 & 3.6 & 0.59 & 0.58 & \ensuremath{0.743}\\
\quad{}Residuals & 1263 & 1283.8 & 1.02 &  & \\
\bottomrule
\end{tabular}

\end{table}

\begin{table}[H]
  \caption{\label{tab:reg-res-mm} Regression of residuals of \Model{\Mmin} for each network structural property on each candidate predictor.}
  \footnotesize
  \begin{center}

\begin{tabular}{lrr}
\toprule
Relationship Effect\\\quad{}$\times$ Network-Level Effect & Wald $\chi^2$ (df) & $\pval$\\
\midrule
edges &  & \\
$\quad\times\log(\text{pop.\ dens.\ in post code})$ (linear) & 6.1 (1) & \ensuremath{0.014}\\
\quad$\times\log(\text{pop.\ dens.\ in post code})$ (quadratic) & 7.0 (2) & \ensuremath{0.030}\\
\quad if city post code & 1.9 (1) & \ensuremath{0.169}\\
\quad if Brussels post code & 0.5 (1) & \ensuremath{0.464}\\
\quad if weekend & 0.4 (1) & \ensuremath{0.510}\\
2-stars &  & \\
$\quad\times\log(\text{pop.\ dens.\ in post code})$ (linear) & 5.0 (1) & \ensuremath{0.025}\\
\quad$\times\log(\text{pop.\ dens.\ in post code})$ (quadratic) & 5.0 (2) & \ensuremath{0.081}\\
\quad if city post code & 5.1 (1) & \ensuremath{0.024}\\
\quad if Brussels post code & 0.5 (1) & \ensuremath{0.480}\\
\quad if weekend & 0.0 (1) & \ensuremath{0.971}\\
triangles &  & \\
$\quad\times\log(\text{pop.\ dens.\ in post code})$ (linear) & 5.8 (1) & \ensuremath{0.016}\\
\quad$\times\log(\text{pop.\ dens.\ in post code})$ (quadratic) & 5.8 (2) & \ensuremath{0.055}\\
\quad if city post code & 5.3 (1) & \ensuremath{0.021}\\
\quad if Brussels post code & 0.6 (1) & \ensuremath{0.437}\\
\quad if weekend & 0.0 (1) & \ensuremath{0.877}\\
\bottomrule
\end{tabular}

\end{center}
\end{table}
\begin{table}[H]
  \caption{\label{tab:dataset-mm} Tests of the null hypothesis of no effect of dataset $H$ on the specified network statistic over and above \Model{\Mmin}.}
  \footnotesize
  \begin{center}

\begin{tabular}{lrr}
\toprule
Statistic & Score $\chi^2$ (df) & $\pval$\\
\midrule
Omnibus & 50.8 (2) & \ensuremath{<0.0001}\\
\quad edges & nonparam. & \ensuremath{<0.0001}\\
\quad 2-stars & nonparam. & \ensuremath{<0.0001}\\
\bottomrule
\end{tabular}

\end{center}
\end{table}

\begin{table}[H]
  \caption{\label{tab:pearson-sd-mm}Sample standard deviations of Pearson residuals for edge, two-star, and triangle counts in \Model{\Mmin} for the dataset and its subsets. Values substantially greater than 1 indicate overdispersion relative to the model.}
  \footnotesize
\begin{center}

\begin{tabular}{lrrrr}
\toprule
Statistic & Overall & $H$ & \Ewc & \Enc\\
\midrule
edges & 0.999 & 0.596 & 0.728 & 1.12\\
2-stars & 1.003 & 0.698 & 0.761 & 1.08\\
triangles & 1.008 & 0.763 & 0.779 & 1.06\\
\bottomrule
\end{tabular}

\end{center}
\end{table}

\begin{table}[H]
  \caption{\label{tab:pearson-mix-mm}Pearson residuals for edge, two-star, triangle, and mixing counts in \Model{\Mmin} for the dataset and its subsets. Extreme positive or negative residual values indicate that the model poorly accounts for effect of the presence of a child (the main selection criterion for the $H$ dataset) and/or that cells in the mixing model should not have been merged. Mixing effects for which a child effect is inherent (i.e., involving a young child and/or a preadolescent but no other on one side of the relation) or which are not found in the $H$ dataset (i.e., seniors) are also excluded.}
  \footnotesize
\begin{center}

\begin{tabular}{lrrrr}
\toprule
Effect & Overall & $H$ & \Ewc & \Enc\\
\midrule
edges & $  0.02$ & $ 6.93$ & $ 5.86$ & $-14.23$\\
2-stars & $  0.06$ & $ 5.64$ & $ 4.17$ & $-11.65$\\
triangles & $  0.05$ & $ 5.58$ & $ 4.13$ & $-11.69$\\
Adolescent with Adolescent & $ -4.77$ & $-2.79$ & $-1.10$ & $ -4.09$\\
Adolescent with Young Adult & $ -8.42$ & $-3.47$ & $-2.33$ & $ -7.14$\\
Young Adult with Young Adult & $ -8.02$ & $ 1.44$ & $-2.84$ & $ -8.41$\\
Adolescent with Older Female Adult & $ -1.93$ & $ 2.55$ & $-0.67$ & $ -3.73$\\
Young Adult with Older Female Adult & $-10.02$ & $-1.53$ & $-2.69$ & $ -9.69$\\
Older Female Adult with Older Female Adult & $ -5.21$ & $ 0.51$ & $ 0.47$ & $ -5.75$\\
Adolescent with Older Male Adult & $ -4.57$ & $ 1.08$ & $-1.16$ & $ -6.09$\\
Young Adult with Older Male Adult & $-10.73$ & $-2.01$ & $-1.56$ & $-10.76$\\
Older Female Adult with Older Male Adult & $  3.21$ & $ 6.75$ & $ 7.71$ & $ -4.80$\\
Older Male Adult with Older Male Adult & $ -9.96$ & $-1.47$ & $-1.72$ & $ -9.59$\\
\bottomrule
\end{tabular}

\end{center}
\end{table}

\clearpage

\subsection[\Model{\Mind}]{\label{app:extra-results-mi}Full results and diagnostics for \Model{\Mind} (dyad-independent model)}
\small

\begin{table}[H]
  \caption{\label{tab:coef-mi}Parameter estimates (and standard errors) for \Model{\Mind}.}
  \footnotesize
  \begin{center}

\begin{tabular}{lr}
\toprule
Relationship Effect\\\quad{}$\times$ Network-Level Effect & Coefficient (SE)$^{\hphantom{\star\star\star}}$\\
\midrule
edges $\times$ $\log(\nactors_{\sampidx})$ & $-4.50 \; (0.91)^{\star\star\star}$\\
\quad $\times$ $\log^2(\nactors_{\sampidx})$ & $1.54 \; (0.35)^{\star\star\star}$\\
\quad   if Brussels post code & $0.38 \; (0.30)^{\phantom{\star}\phantom{\star}\phantom{\star}}$\\
\quad   if on weekend & $0.34 \; (0.10)^{\star\star\star}$\\
Young Child with Young Child & $5.99 \; (0.71)^{\star\star\star}$\\
Young Child with Preadolescent & $6.26 \; (0.69)^{\star\star\star}$\\
Preadolescent with Preadolescent & $5.08 \; (0.63)^{\star\star\star}$\\
Adolescent with Adolescent & $3.28 \; (0.64)^{\star\star\star}$\\
Young Child with Young Adult & $5.58 \; (1.18)^{\star\star\star}$\\
Preadolescent with Young Adult & $2.80 \; (0.68)^{\star\star\star}$\\
Adolescent with Young Adult & $2.61 \; (0.64)^{\star\star\star}$\\
Young Adult with Young Adult & $2.76 \; (0.66)^{\star\star\star}$\\
Young Child with Older Female Adult & $7.02 \; (0.69)^{\star\star\star}$\\
Preadolescent with Older Female Adult & $6.01 \; (0.62)^{\star\star\star}$\\
Adolescent with Older Female Adult & $4.47 \; (0.61)^{\star\star\star}$\\
Older Female Adult with Older Female Adult & $3.02 \; (0.64)^{\star\star\star}$\\
Young Child with Older Male Adult & $6.10 \; (0.64)^{\star\star\star}$\\
Preadolescent with Older Male Adult & $5.45 \; (0.62)^{\star\star\star}$\\
Adolescent with Older Male Adult & $4.04 \; (0.61)^{\star\star\star}$\\
Older Female Adult with Older Male Adult & $6.54 \; (0.65)^{\star\star\star}$\\
\quad   if child absent & $-2.56 \; (0.29)^{\star\star\star}$\\
Older Male Adult with Older Male Adult & $2.19 \; (0.64)^{\star\star\star}$\\
Older Female Adult with Senior & $3.29 \; (0.56)^{\star\star\star}$\\
Older Male Adult with Senior & $2.70 \; (0.60)^{\star\star\star}$\\
Senior with Senior & $2.94 \; (0.51)^{\star\star\star}$\\
Adolescent with Young Child or Preadolescent & $4.46 \; (0.61)^{\star\star\star}$\\
Young Adult with Older Adult & $3.22 \; (0.59)^{\star\star\star}$\\
Young Child or Preadolescent with Senior & $3.89 \; (0.92)^{\star\star\star}$\\
Adolescent or Young Adult with Senior & $5.35 \; (1.20)^{\star\star\star}$\\
\bottomrule
\end{tabular}

Significance: $^{\star\star\star}\le 0.001<^{\star\star}\le 0.01< ^\star \le 0.05$
\end{center}
\end{table}
\begin{table}[H]
  \caption{\label{tab:omnibus-mi} Omnibus tests for selected groups of effects in \Model{\Mind}. Effects are net of the rest of the model.}
  \footnotesize
  \begin{center}

\begin{tabular}{lrr}
\toprule
Effects & Wald $\chi^2$ (df) & $\pval$\\
\midrule
any $\log(\nactors_\sampidx)$ or $\log^2(\nactors_\sampidx)$ & 32.3 (2) & \ensuremath{<0.001}\\
any $\log^2(\nactors_\sampidx)$ & 19.4 (1) & \ensuremath{<0.001}\\
\bottomrule
\end{tabular}

\end{center}
\end{table}

\begin{figure}[H]

{\centering \subfloat[Edges\label{fig:unnamed-chunk-57-1}]{\includegraphics[width=\maxwidth]{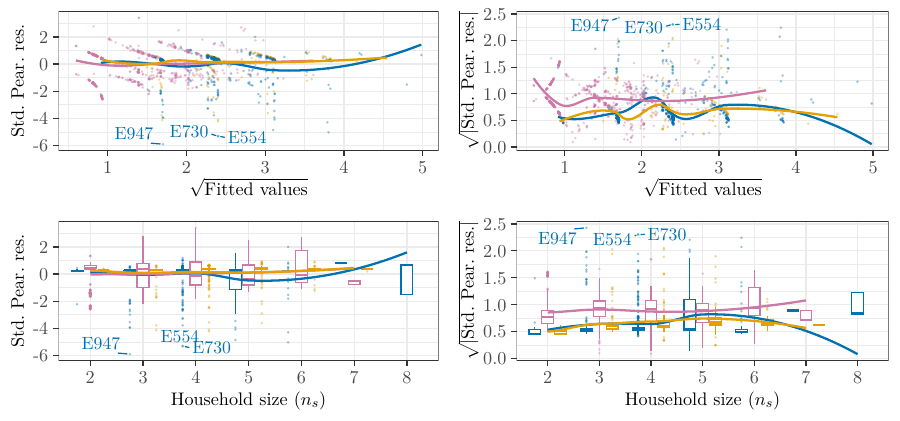} }\newline
\subfloat[2-stars\label{fig:unnamed-chunk-57-2}]{\includegraphics[width=\maxwidth]{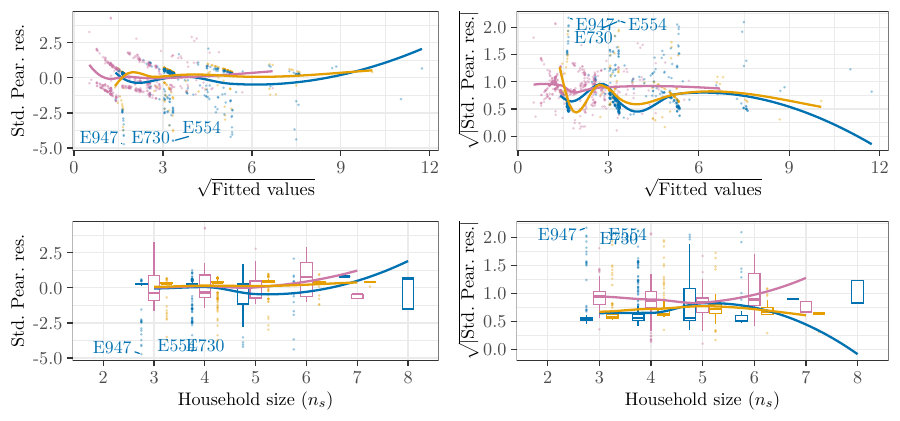} }\newline
\subfloat[Triangles\label{fig:unnamed-chunk-57-3}]{\includegraphics[width=\maxwidth]{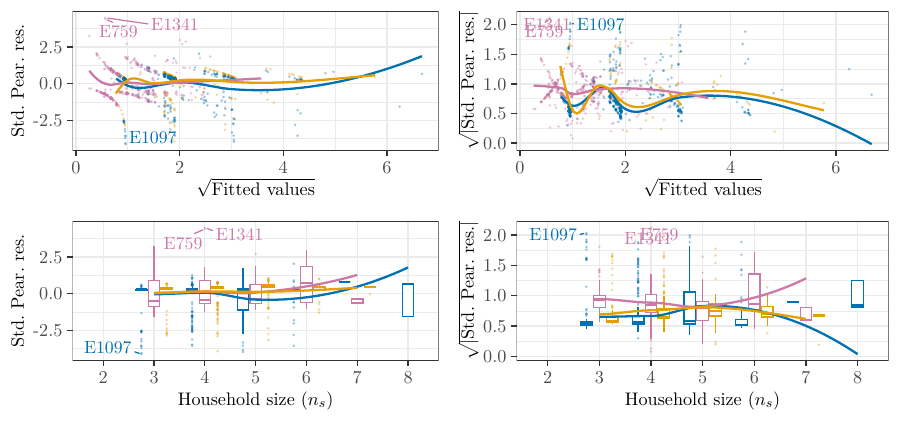} }

}

\caption{\label{fig:resid-plots-mi}Residual plots of network statistics against fitted values and network size for \Model{\Mind}. \subdataleg}\label{fig:unnamed-chunk-57}
\end{figure}

\begin{figure}[H]

{\centering \includegraphics[width=\maxwidth]{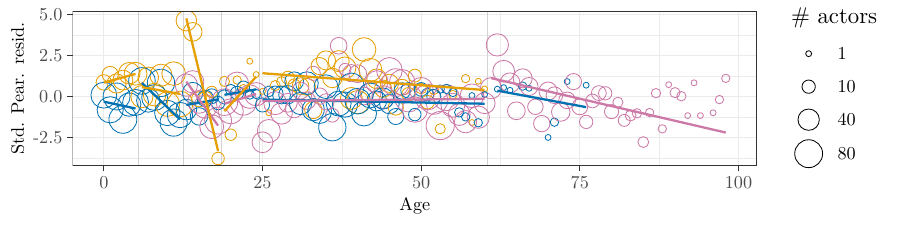} 

}

\caption{\label{fig:age-resid-mi}Residual plot for \Model{\Mind} of edges incident on actors of a given age against age. \subdataleg}\label{fig:unnamed-chunk-58}
\end{figure}

\begin{figure}[H]

{\centering \includegraphics[width=\maxwidth,height=1in]{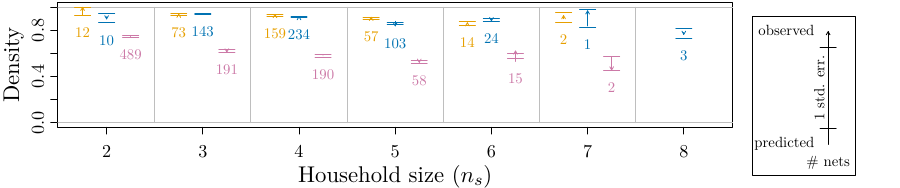} 

}

\caption{\label{fig:density-error-mi}Average prediction errors of density in \Model{\Mind}. Values are averaged over the networks grouped by size and subset. \subdataleg}\label{fig:unnamed-chunk-59}
\end{figure}

\begin{table}[H]
  \caption{\label{tab:nns-anova-mi}Analyses of variance for fitting residuals for \Model{\Mind} against household size (represented as a categorical predictor, with a dummy variable for each size).}
  \centering
  \footnotesize

\begin{tabular}{lrrrrr}
\toprule
\quad{}Source & df & Sum Sq. & Mean Sq. & $F$ & \pval\\
\midrule
edges &  &  &  &  & \\
\quad{}$\nactors_{\sampidx}$ (categorical) & 7 & 8.2 & 1.17 & 1.0 & \ensuremath{0.402}\\
\quad{}Residuals & 1773 & 1996.5 & 1.13 &  & \\
\addlinespace
2-stars &  &  &  &  & \\
\quad{}$\nactors_{\sampidx}$ (categorical) & 6 & 11.2 & 1.86 & 2.0 & \ensuremath{0.065}\\
\quad{}Residuals & 1263 & 1183.4 & 0.94 &  & \\
\addlinespace
triangles &  &  &  &  & \\
\quad{}$\nactors_{\sampidx}$ (categorical) & 6 & 20.4 & 3.41 & 4.6 & \ensuremath{<0.001}\\
\quad{}Residuals & 1263 & 944.3 & 0.75 &  & \\
\bottomrule
\end{tabular}

\end{table}

\begin{table}[H]
  \caption{\label{tab:reg-res-mi} Regression of residuals of \Model{\Mind} for each network structural property on each candidate predictor.}
  \footnotesize
  \begin{center}

\begin{tabular}{lrr}
\toprule
Relationship Effect\\\quad{}$\times$ Network-Level Effect & Wald $\chi^2$ (df) & $\pval$\\
\midrule
edges &  & \\
$\quad\times\log(\text{pop.\ dens.\ in post code})$ (linear) & 3.3 (1) & \ensuremath{0.069}\\
\quad$\times\log(\text{pop.\ dens.\ in post code})$ (quadratic) & 6.9 (2) & \ensuremath{0.031}\\
\quad if city post code & 0.6 (1) & \ensuremath{0.444}\\
\quad if Brussels post code & 0.0 (1) & \ensuremath{0.984}\\
\quad if weekend & 2.9 (1) & \ensuremath{0.087}\\
2-stars &  & \\
$\quad\times\log(\text{pop.\ dens.\ in post code})$ (linear) & 0.3 (1) & \ensuremath{0.591}\\
\quad$\times\log(\text{pop.\ dens.\ in post code})$ (quadratic) & 0.9 (2) & \ensuremath{0.642}\\
\quad if city post code & 0.4 (1) & \ensuremath{0.529}\\
\quad if Brussels post code & 0.1 (1) & \ensuremath{0.720}\\
\quad if weekend & 3.8 (1) & \ensuremath{0.052}\\
triangles &  & \\
$\quad\times\log(\text{pop.\ dens.\ in post code})$ (linear) & 0.1 (1) & \ensuremath{0.711}\\
\quad$\times\log(\text{pop.\ dens.\ in post code})$ (quadratic) & 0.5 (2) & \ensuremath{0.767}\\
\quad if city post code & 0.2 (1) & \ensuremath{0.675}\\
\quad if Brussels post code & 0.1 (1) & \ensuremath{0.808}\\
\quad if weekend & 3.2 (1) & \ensuremath{0.073}\\
\bottomrule
\end{tabular}

\end{center}
\end{table}
\begin{table}[H]
  \caption{\label{tab:dataset-mi} Tests of the null hypothesis of no effect of dataset $H$ on the specified network statistic over and above \Model{\Mind}.}
  \footnotesize
  \begin{center}

\begin{tabular}{lrr}
\toprule
Statistic & Score $\chi^2$ (df) & $\pval$\\
\midrule
Omnibus & 57.5 (2) & \ensuremath{<0.0001}\\
\quad edges & nonparam. & \ensuremath{<0.0001}\\
\quad 2-stars & nonparam. & \ensuremath{<0.0001}\\
\bottomrule
\end{tabular}

\end{center}
\end{table}

\begin{table}[H]
  \caption{\label{tab:pearson-sd-mi}Sample standard deviations of Pearson residuals for edge, two-star, and triangle counts in \Model{\Mind} for the dataset and its subsets. Values substantially greater than 1 indicate overdispersion relative to the model.}
  \footnotesize
\begin{center}

\begin{tabular}{lrrrr}
\toprule
Statistic & Overall & $H$ & \Ewc & \Enc\\
\midrule
edges & 1.062 & 1.220 & 1.053 & 1.000\\
2-stars & 0.968 & 1.098 & 0.917 & 0.908\\
triangles & 0.863 & 0.999 & 0.805 & 0.797\\
\bottomrule
\end{tabular}

\end{center}
\end{table}

\begin{table}[H]
  \caption{\label{tab:pearson-mix-mi}Pearson residuals for edge, two-star, triangle, and mixing counts in \Model{\Mind} for the dataset and its subsets. Extreme positive or negative residual values indicate that the model poorly accounts for effect of the presence of a child (the main selection criterion for the $H$ dataset) and/or that cells in the mixing model should not have been merged. Mixing effects for which a child effect is inherent (i.e., involving a young child and/or a preadolescent but no other on one side of the relation) or which are not found in the $H$ dataset (i.e., seniors) are also excluded.}
  \footnotesize
\begin{center}

\begin{tabular}{lrrrr}
\toprule
Effect & Overall & $H$ & \Ewc & \Enc\\
\midrule
edges & $ 0.16$ & $ 3.81$ & $-1.14$ & $-0.86$\\
2-stars & $ 2.43$ & $ 5.47$ & $-0.47$ & $ 0.49$\\
triangles & $ 4.49$ & $ 6.77$ & $ 0.23$ & $ 1.89$\\
Adolescent with Adolescent & $ 0.07$ & $-0.21$ & $-0.12$ & $ 0.24$\\
Adolescent with Young Adult & $-0.03$ & $-0.63$ & $-0.16$ & $ 0.22$\\
Young Adult with Young Adult & $ 0.14$ & $ 3.04$ & $-0.57$ & $-0.46$\\
Adolescent with Older Female Adult & $ 0.05$ & $ 3.00$ & $-0.44$ & $-1.40$\\
Young Adult with Older Female Adult & $ 0.55$ & $ 0.43$ & $-0.23$ & $ 0.59$\\
Older Female Adult with Older Female Adult & $ 0.06$ & $ 0.98$ & $ 1.97$ & $-0.67$\\
Adolescent with Older Male Adult & $-0.02$ & $ 3.00$ & $-0.60$ & $-1.35$\\
Young Adult with Older Male Adult & $-0.58$ & $-0.40$ & $ 1.06$ & $-0.91$\\
Older Female Adult with Older Male Adult & $ 0.14$ & $ 0.08$ & $ 0.17$ & $ 0.08$\\
Older Male Adult with Older Male Adult & $-0.18$ & $-0.76$ & $-0.35$ & $-0.02$\\
\bottomrule
\end{tabular}

\end{center}
\end{table}

\clearpage

\subsection[\Model{\Mnns}]{\label{app:extra-results-mn}Full results and diagnostics for \Model{\Mnns} (\Model{1} without network size effects)}
\small

\begin{table}[H]
  \caption{\label{tab:coef-mn}Parameter estimates (and standard errors) for \Model{\Mnns}.}
  \footnotesize
  \begin{center}

\begin{tabular}{lr}
\toprule
Relationship Effect\\\quad{}$\times$ Network-Level Effect & Coefficient (SE)$^{\hphantom{\star\star\star}}$\\
\midrule
edges   if Brussels post code & $0.08 \; (0.19)^{\phantom{\star}\phantom{\star}\phantom{\star}}$\\
\quad   if on weekend & $0.15 \; (0.06)^{\star\star\phantom{\star}}$\\
2-stars & $-1.04 \; (0.07)^{\star\star\star}$\\
triangles & $3.01 \; (0.18)^{\star\star\star}$\\
Young Child with Young Child & $1.06 \; (0.40)^{\star\star\phantom{\star}}$\\
Young Child with Preadolescent & $1.45 \; (0.37)^{\star\star\star}$\\
Preadolescent with Preadolescent & $0.58 \; (0.23)^{\star\phantom{\star}\phantom{\star}}$\\
Adolescent with Adolescent & $0.26 \; (0.25)^{\phantom{\star}\phantom{\star}\phantom{\star}}$\\
Young Child with Young Adult & $2.41 \; (1.03)^{\star\phantom{\star}\phantom{\star}}$\\
Preadolescent with Young Adult & $-0.06 \; (0.36)^{\phantom{\star}\phantom{\star}\phantom{\star}}$\\
Adolescent with Young Adult & $0.46 \; (0.24)^{\phantom{\star}\phantom{\star}\phantom{\star}}$\\
Young Adult with Young Adult & $0.14 \; (0.29)^{\phantom{\star}\phantom{\star}\phantom{\star}}$\\
Young Child with Older Female Adult & $2.87 \; (0.36)^{\star\star\star}$\\
Preadolescent with Older Female Adult & $2.18 \; (0.20)^{\star\star\star}$\\
Adolescent with Older Female Adult & $1.24 \; (0.16)^{\star\star\star}$\\
Older Female Adult with Older Female Adult & $-0.12 \; (0.28)^{\phantom{\star}\phantom{\star}\phantom{\star}}$\\
Young Child with Older Male Adult & $1.75 \; (0.24)^{\star\star\star}$\\
Preadolescent with Older Male Adult & $1.27 \; (0.18)^{\star\star\star}$\\
Adolescent with Older Male Adult & $0.62 \; (0.16)^{\star\star\star}$\\
Older Female Adult with Older Male Adult & $2.62 \; (0.27)^{\star\star\star}$\\
\quad   if child absent & $-1.25 \; (0.29)^{\star\star\star}$\\
Older Male Adult with Older Male Adult & $-1.01 \; (0.30)^{\star\star\star}$\\
Older Female Adult with Senior & $0.77 \; (0.20)^{\star\star\star}$\\
Older Male Adult with Senior & $-0.12 \; (0.21)^{\phantom{\star}\phantom{\star}\phantom{\star}}$\\
Senior with Senior & $0.53 \; (0.15)^{\star\star\star}$\\
Adolescent with Young Child or Preadolescent & $0.59 \; (0.19)^{\star\star\phantom{\star}}$\\
Young Adult with Older Adult & $0.39 \; (0.10)^{\star\star\star}$\\
Young Child or Preadolescent with Senior & $0.55 \; (0.46)^{\phantom{\star}\phantom{\star}\phantom{\star}}$\\
Adolescent or Young Adult with Senior & $2.25 \; (0.88)^{\star\phantom{\star}\phantom{\star}}$\\
\bottomrule
\end{tabular}

Significance: $^{\star\star\star}\le 0.001<^{\star\star}\le 0.01< ^\star \le 0.05$
\end{center}
\end{table}
\begin{table}[H]
  \caption{\label{tab:omnibus-mn} Omnibus tests for selected groups of effects in \Model{\Mnns}. Effects are net of the rest of the model.}
  \footnotesize
  \begin{center}

\begin{tabular}{lrr}
\toprule
Effects & Wald $\chi^2$ (df) & $\pval$\\
\midrule
any 2-star & 252.2 (1) & \ensuremath{<0.001}\\
any triangle & 290.2 (1) & \ensuremath{<0.001}\\
\bottomrule
\end{tabular}

\end{center}
\end{table}

\begin{figure}[H]

{\centering \subfloat[Edges\label{fig:unnamed-chunk-72-1}]{\includegraphics[width=\maxwidth]{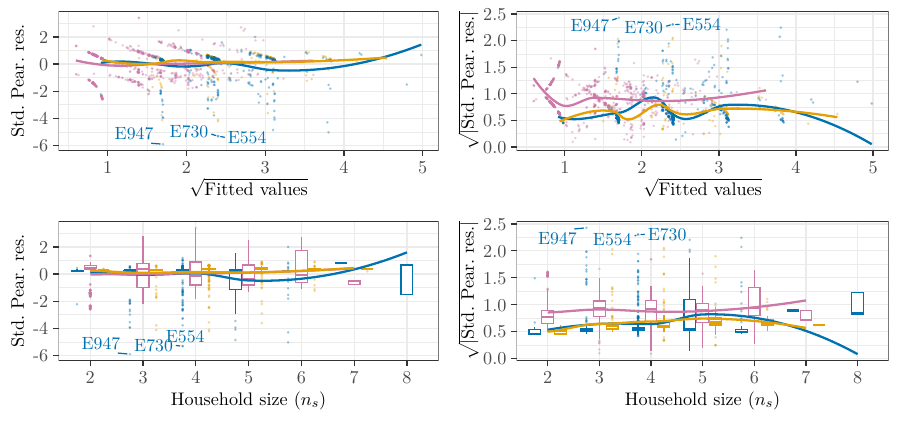} }\newline
\subfloat[2-stars\label{fig:unnamed-chunk-72-2}]{\includegraphics[width=\maxwidth]{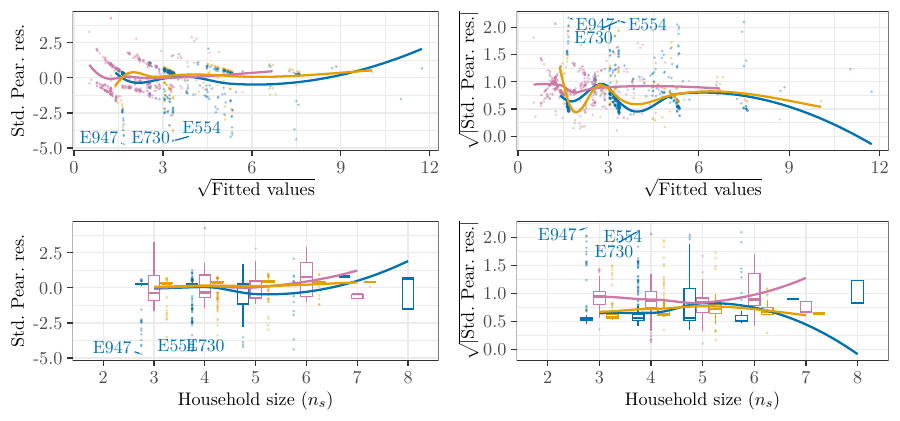} }\newline
\subfloat[Triangles\label{fig:unnamed-chunk-72-3}]{\includegraphics[width=\maxwidth]{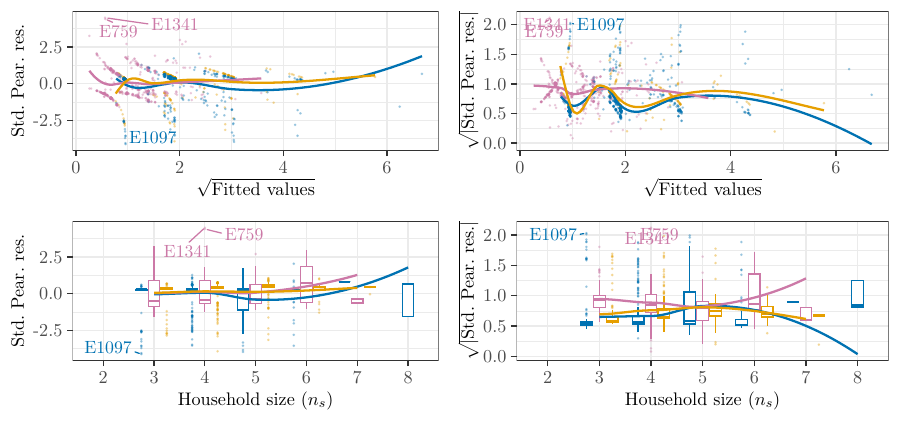} }

}

\caption{\label{fig:resid-plots-mn}Residual plots of network statistics against fitted values and network size for \Model{\Mnns}. \subdataleg}\label{fig:unnamed-chunk-72}
\end{figure}

\begin{figure}[H]

{\centering \includegraphics[width=\maxwidth]{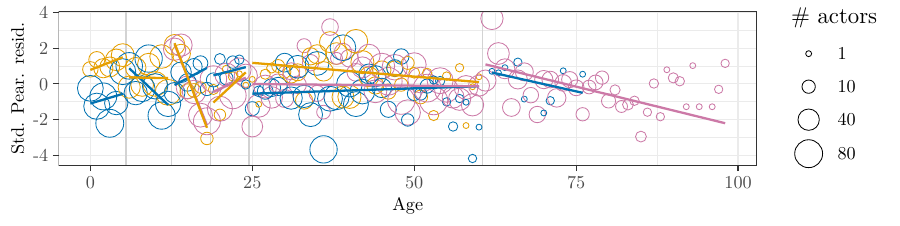} 

}

\caption{\label{fig:age-resid-mn}Residual plot for \Model{\Mnns} of edges incident on actors of a given age against age. \subdataleg}\label{fig:unnamed-chunk-73}
\end{figure}

\begin{figure}[H]

{\centering \includegraphics[width=\maxwidth,height=1in]{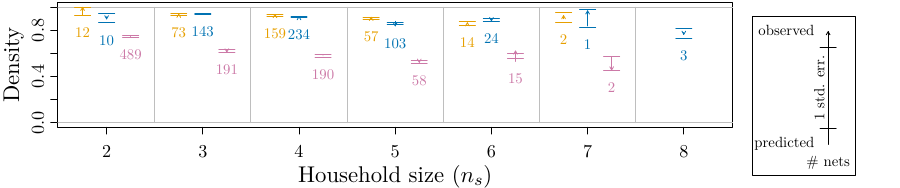} 

}

\caption{\label{fig:density-error-mn}Average prediction errors of density in \Model{\Mnns}. Values are averaged over the networks grouped by size and subset. \subdataleg}\label{fig:unnamed-chunk-74}
\end{figure}

\begin{table}[H]
  \caption{\label{tab:nns-anova-mn}Analyses of variance for fitting residuals for \Model{\Mnns} against household size (represented as a categorical predictor, with a dummy variable for each size).}
  \centering
  \footnotesize

\begin{tabular}{lrrrrr}
\toprule
\quad{}Source & df & Sum Sq. & Mean Sq. & $F$ & \pval\\
\midrule
edges &  &  &  &  & \\
\quad{}$\nactors_{\sampidx}$ (categorical) & 7 & 33.6 & 4.8 & 4.6 & \ensuremath{<0.001}\\
\quad{}Residuals & 1773 & 1855.4 & 1.0 &  & \\
\addlinespace
2-stars &  &  &  &  & \\
\quad{}$\nactors_{\sampidx}$ (categorical) & 6 & 28.0 & 4.7 & 4.4 & \ensuremath{<0.001}\\
\quad{}Residuals & 1263 & 1338.2 & 1.1 &  & \\
\addlinespace
triangles &  &  &  &  & \\
\quad{}$\nactors_{\sampidx}$ (categorical) & 6 & 26.9 & 4.5 & 4.4 & \ensuremath{<0.001}\\
\quad{}Residuals & 1263 & 1288.9 & 1.0 &  & \\
\bottomrule
\end{tabular}

\end{table}

\begin{table}[H]
  \caption{\label{tab:reg-res-mn} Regression of residuals of \Model{\Mnns} for each network structural property on each candidate predictor.}
  \footnotesize
  \begin{center}

\begin{tabular}{lrr}
\toprule
Relationship Effect\\\quad{}$\times$ Network-Level Effect & Wald $\chi^2$ (df) & $\pval$\\
\midrule
edges &  & \\
$\quad\times\log(\text{pop.\ dens.\ in post code})$ (linear) & 4.3 (1) & \ensuremath{0.037}\\
\quad$\times\log(\text{pop.\ dens.\ in post code})$ (quadratic) & 4.9 (2) & \ensuremath{0.085}\\
\quad if city post code & 1.5 (1) & \ensuremath{0.221}\\
\quad if Brussels post code & 0.4 (1) & \ensuremath{0.516}\\
\quad if weekend & 0.6 (1) & \ensuremath{0.449}\\
2-stars &  & \\
$\quad\times\log(\text{pop.\ dens.\ in post code})$ (linear) & 1.4 (1) & \ensuremath{0.239}\\
\quad$\times\log(\text{pop.\ dens.\ in post code})$ (quadratic) & 1.4 (2) & \ensuremath{0.500}\\
\quad if city post code & 1.6 (1) & \ensuremath{0.202}\\
\quad if Brussels post code & 0.2 (1) & \ensuremath{0.618}\\
\quad if weekend & 0.4 (1) & \ensuremath{0.503}\\
triangles &  & \\
$\quad\times\log(\text{pop.\ dens.\ in post code})$ (linear) & 1.3 (1) & \ensuremath{0.246}\\
\quad$\times\log(\text{pop.\ dens.\ in post code})$ (quadratic) & 1.4 (2) & \ensuremath{0.509}\\
\quad if city post code & 1.3 (1) & \ensuremath{0.250}\\
\quad if Brussels post code & 0.4 (1) & \ensuremath{0.540}\\
\quad if weekend & 0.2 (1) & \ensuremath{0.656}\\
\bottomrule
\end{tabular}

\end{center}
\end{table}
\begin{table}[H]
  \caption{\label{tab:dataset-mn} Tests of the null hypothesis of no effect of dataset $H$ on the specified network statistic over and above \Model{\Mnns}.}
  \footnotesize
  \begin{center}

\begin{tabular}{lrr}
\toprule
Statistic & Score $\chi^2$ (df) & $\pval$\\
\midrule
Omnibus & 4.8 (2) & \ensuremath{0.0923}\\
\quad edges & nonparam. & \ensuremath{0.0874}\\
\quad 2-stars & nonparam. & \ensuremath{0.2505}\\
\bottomrule
\end{tabular}

\end{center}
\end{table}

\begin{table}[H]
  \caption{\label{tab:pearson-sd-mn}Sample standard deviations of Pearson residuals for edge, two-star, and triangle counts in \Model{\Mnns} for the dataset and its subsets. Values substantially greater than 1 indicate overdispersion relative to the model.}
  \footnotesize
\begin{center}

\begin{tabular}{lrrrr}
\toprule
Statistic & Overall & $H$ & \Ewc & \Enc\\
\midrule
edges & 1.03 & 0.832 & 1.15 & 1.02\\
2-stars & 1.04 & 0.834 & 1.12 & 1.06\\
triangles & 1.02 & 0.855 & 1.09 & 1.03\\
\bottomrule
\end{tabular}

\end{center}
\end{table}

\begin{table}[H]
  \caption{\label{tab:pearson-mix-mn}Pearson residuals for edge, two-star, triangle, and mixing counts in \Model{\Mnns} for the dataset and its subsets. Extreme positive or negative residual values indicate that the model poorly accounts for effect of the presence of a child (the main selection criterion for the $H$ dataset) and/or that cells in the mixing model should not have been merged. Mixing effects for which a child effect is inherent (i.e., involving a young child and/or a preadolescent but no other on one side of the relation) or which are not found in the $H$ dataset (i.e., seniors) are also excluded.}
  \footnotesize
\begin{center}

\begin{tabular}{lrrrr}
\toprule
Effect & Overall & $H$ & \Ewc & \Enc\\
\midrule
edges & $ 0.34$ & $ 1.59$ & $-0.53$ & $-0.21$\\
2-stars & $ 0.82$ & $ 1.05$ & $ 0.16$ & $ 0.16$\\
triangles & $ 0.76$ & $ 0.90$ & $ 0.03$ & $ 0.36$\\
Adolescent with Adolescent & $ 0.48$ & $-1.39$ & $ 1.43$ & $ 0.36$\\
Adolescent with Young Adult & $ 0.36$ & $-1.81$ & $ 2.46$ & $-0.05$\\
Young Adult with Young Adult & $ 0.04$ & $ 1.75$ & $-0.46$ & $-0.42$\\
Adolescent with Older Female Adult & $ 0.22$ & $ 1.97$ & $-0.29$ & $-0.76$\\
Young Adult with Older Female Adult & $ 0.62$ & $-0.29$ & $ 0.63$ & $ 0.56$\\
Older Female Adult with Older Female Adult & $ 0.01$ & $ 0.78$ & $-0.10$ & $-0.07$\\
Adolescent with Older Male Adult & $ 0.21$ & $ 1.56$ & $-0.39$ & $-0.44$\\
Young Adult with Older Male Adult & $-0.41$ & $-1.00$ & $ 1.28$ & $-0.61$\\
Older Female Adult with Older Male Adult & $-0.08$ & $ 0.07$ & $ 0.04$ & $-0.11$\\
Older Male Adult with Older Male Adult & $-0.16$ & $-1.98$ & $-2.75$ & $ 0.52$\\
\bottomrule
\end{tabular}

\end{center}
\end{table}

\clearpage

\subsection[\Model{0}]{\label{app:extra-results-m0}Full results and diagnostics for \Model{0} (initial model)}
\small

\begin{table}[H]
  \caption{\label{tab:coef-m0}Parameter estimates (and standard errors) for \Model{0}.}
  \footnotesize
  \begin{center}

\begin{tabular}{lr}
\toprule
Relationship Effect\\\quad{}$\times$ Network-Level Effect & Coefficient (SE)$^{\hphantom{\star\star\star}}$\\
\midrule
edges $\times$ $\log(\nactors_{\sampidx})$ & $-14.30 \; (2.99)^{\star\star\star}$\\
\quad $\times$ $\log^2(\nactors_{\sampidx})$ & $5.72 \; (1.34)^{\star\star\star}$\\
\quad   if Brussels post code & $0.12 \; (0.19)^{\phantom{\star}\phantom{\star}\phantom{\star}}$\\
\quad   if on weekend & $0.14 \; (0.05)^{\star\phantom{\star}\phantom{\star}}$\\
2-stars & $0.77 \; (0.86)^{\phantom{\star}\phantom{\star}\phantom{\star}}$\\
\quad $\times$ $\log(\nactors_{\sampidx})$ & $-0.66 \; (0.47)^{\phantom{\star}\phantom{\star}\phantom{\star}}$\\
\quad $\times$ $\log^2(\nactors_{\sampidx})$ & $-0.14 \; (0.11)^{\phantom{\star}\phantom{\star}\phantom{\star}}$\\
triangles & $9.45 \; (0.93)^{\star\star\star}$\\
\quad $\times$ $\log(\nactors_{\sampidx})$ & $-8.30 \; (1.45)^{\star\star\star}$\\
\quad $\times$ $\log^2(\nactors_{\sampidx})$ & $2.43 \; (0.75)^{\star\star\phantom{\star}}$\\
Young Child with Young Child & $8.41 \; (1.55)^{\star\star\star}$\\
Young Child with Preadolescent & $8.91 \; (1.54)^{\star\star\star}$\\
Preadolescent with Preadolescent & $7.99 \; (1.51)^{\star\star\star}$\\
Adolescent with Adolescent & $7.58 \; (1.49)^{\star\star\star}$\\
Young Child with Young Adult & $9.60 \; (1.82)^{\star\star\star}$\\
Preadolescent with Young Adult & $7.20 \; (1.51)^{\star\star\star}$\\
Adolescent with Young Adult & $7.65 \; (1.51)^{\star\star\star}$\\
Young Adult with Young Adult & $7.57 \; (1.50)^{\star\star\star}$\\
Young Child with Older Female Adult & $10.28 \; (1.51)^{\star\star\star}$\\
Preadolescent with Older Female Adult & $9.66 \; (1.49)^{\star\star\star}$\\
Adolescent with Older Female Adult & $8.80 \; (1.48)^{\star\star\star}$\\
Older Female Adult with Older Female Adult & $7.21 \; (1.52)^{\star\star\star}$\\
Young Child with Older Male Adult & $9.14 \; (1.49)^{\star\star\star}$\\
Preadolescent with Older Male Adult & $8.84 \; (1.48)^{\star\star\star}$\\
Adolescent with Older Male Adult & $8.08 \; (1.48)^{\star\star\star}$\\
Older Female Adult with Older Male Adult & $9.06 \; (1.48)^{\star\star\star}$\\
Older Male Adult with Older Male Adult & $6.30 \; (1.50)^{\star\star\star}$\\
Older Female Adult with Senior & $8.06 \; (1.47)^{\star\star\star}$\\
Older Male Adult with Senior & $7.43 \; (1.50)^{\star\star\star}$\\
Senior with Senior & $7.81 \; (1.46)^{\star\star\star}$\\
Adolescent with Young Child or Preadolescent & $7.99 \; (1.49)^{\star\star\star}$\\
Young Adult with Older Adult & $7.90 \; (1.48)^{\star\star\star}$\\
Young Child or Preadolescent with Senior & $8.14 \; (1.58)^{\star\star\star}$\\
Adolescent or Young Adult with Senior & $9.75 \; (1.74)^{\star\star\star}$\\
\bottomrule
\end{tabular}

Significance: $^{\star\star\star}\le 0.001<^{\star\star}\le 0.01< ^\star \le 0.05$
\end{center}
\end{table}
\begin{table}[H]
  \caption{\label{tab:omnibus-m0} Omnibus tests for selected groups of effects in \Model{0}. Effects are net of the rest of the model.}
  \footnotesize
  \begin{center}

\begin{tabular}{lrr}
\toprule
Effects & Wald $\chi^2$ (df) & $\pval$\\
\midrule
any 2-star & 24.4 (3) & \ensuremath{<0.001}\\
any triangle & 142.6 (3) & \ensuremath{<0.001}\\
any $\log(\nactors_\sampidx)$ or $\log^2(\nactors_\sampidx)$ & 3383.1 (6) & \ensuremath{<0.001}\\
any $\log^2(\nactors_\sampidx)$ & 19.5 (3) & \ensuremath{<0.001}\\
2-star or triangle $\log^2(\nactors_\sampidx)$ & 10.9 (2) & \ensuremath{0.004}\\
\bottomrule
\end{tabular}

\end{center}
\end{table}

\begin{figure}[H]

{\centering \subfloat[Edges\label{fig:unnamed-chunk-87-1}]{\includegraphics[width=\maxwidth]{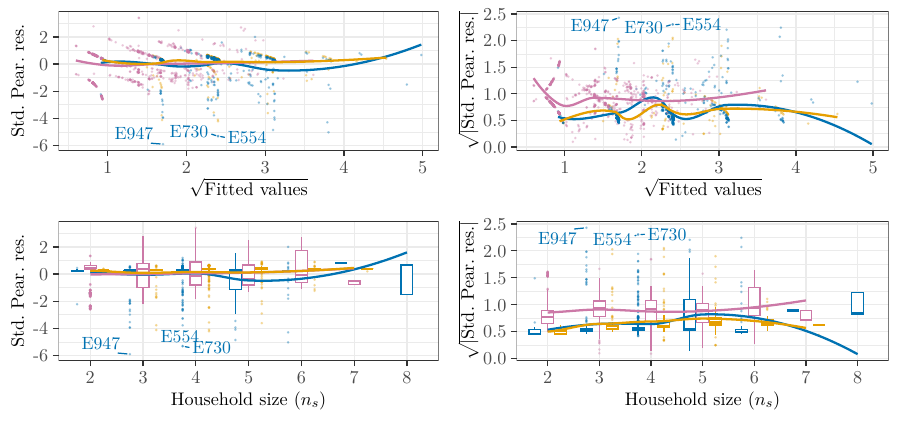} }\newline
\subfloat[2-stars\label{fig:unnamed-chunk-87-2}]{\includegraphics[width=\maxwidth]{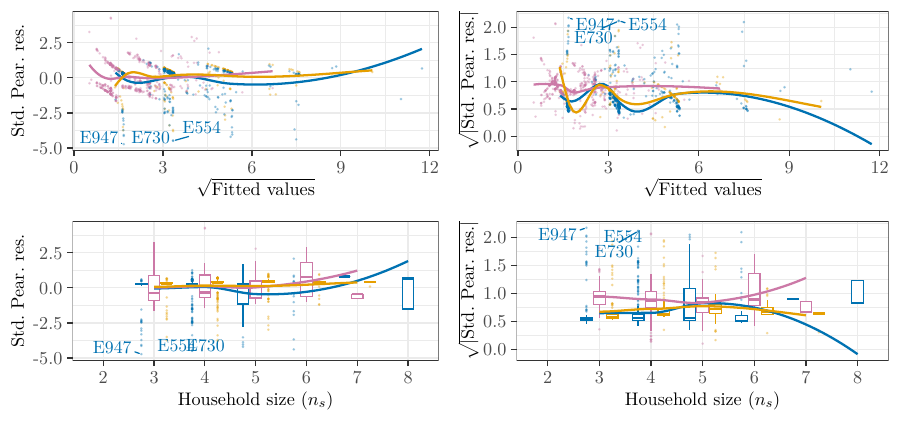} }\newline
\subfloat[Triangles\label{fig:unnamed-chunk-87-3}]{\includegraphics[width=\maxwidth]{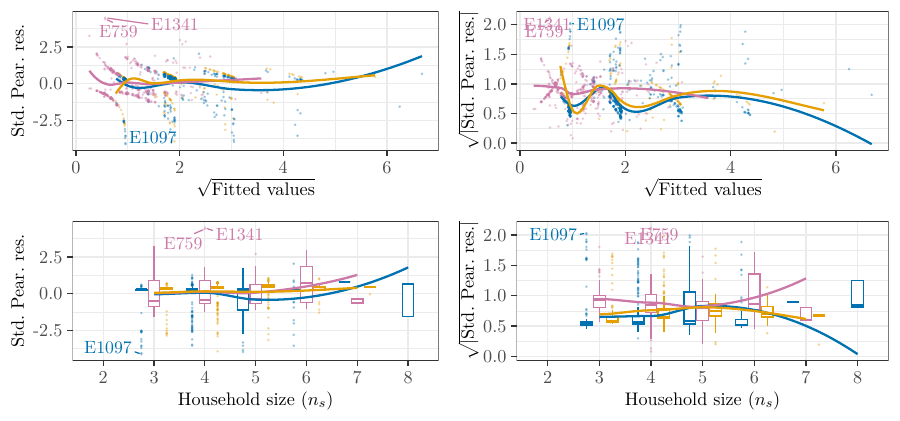} }

}

\caption{\label{fig:resid-plots-m0}Residual plots of network statistics against fitted values and network size for \Model{0}. \subdataleg}\label{fig:unnamed-chunk-87}
\end{figure}

\begin{figure}[H]

{\centering \includegraphics[width=\maxwidth]{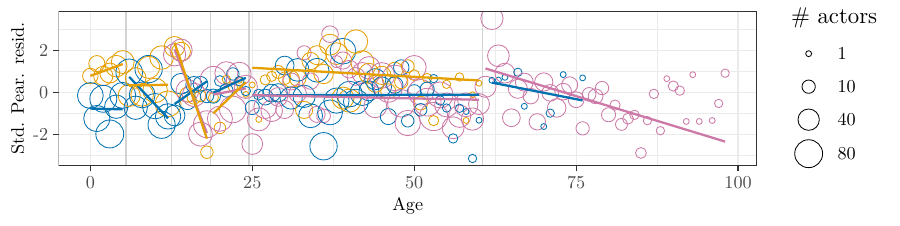} 

}

\caption{\label{fig:age-resid-m0}Residual plot for \Model{0} of edges incident on actors of a given age against age. \subdataleg}\label{fig:unnamed-chunk-88}
\end{figure}

\begin{figure}[H]

{\centering \includegraphics[width=\maxwidth,height=1in]{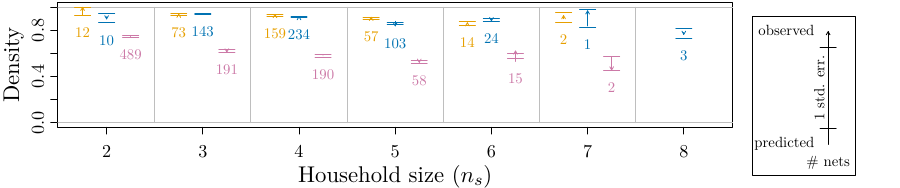} 

}

\caption{\label{fig:density-error-m0}Average prediction errors of density in \Model{0}. Values are averaged over the networks grouped by size and subset. \subdataleg}\label{fig:unnamed-chunk-89}
\end{figure}

\begin{table}[H]
  \caption{\label{tab:nns-anova-m0}Analyses of variance for fitting residuals for \Model{0} against household size (represented as a categorical predictor, with a dummy variable for each size).}
  \centering
  \footnotesize

\begin{tabular}{lrrrrr}
\toprule
\quad{}Source & df & Sum Sq. & Mean Sq. & $F$ & \pval\\
\midrule
edges &  &  &  &  & \\
\quad{}$\nactors_{\sampidx}$ (categorical) & 7 & 8.7 & 1.25 & 1.3 & \ensuremath{0.258}\\
\quad{}Residuals & 1773 & 1731.4 & 0.98 &  & \\
\addlinespace
2-stars &  &  &  &  & \\
\quad{}$\nactors_{\sampidx}$ (categorical) & 6 & 8.0 & 1.33 & 1.4 & \ensuremath{0.205}\\
\quad{}Residuals & 1263 & 1183.0 & 0.94 &  & \\
\addlinespace
triangles &  &  &  &  & \\
\quad{}$\nactors_{\sampidx}$ (categorical) & 6 & 7.2 & 1.19 & 1.3 & \ensuremath{0.258}\\
\quad{}Residuals & 1263 & 1165.7 & 0.92 &  & \\
\bottomrule
\end{tabular}

\end{table}

\begin{table}[H]
  \caption{\label{tab:reg-res-m0} Regression of residuals of \Model{0} for each network structural property on each candidate predictor.}
  \footnotesize
  \begin{center}

\begin{tabular}{lrr}
\toprule
Relationship Effect\\\quad{}$\times$ Network-Level Effect & Wald $\chi^2$ (df) & $\pval$\\
\midrule
edges &  & \\
$\quad\times\log(\text{pop.\ dens.\ in post code})$ (linear) & 4.3 (1) & \ensuremath{0.037}\\
\quad$\times\log(\text{pop.\ dens.\ in post code})$ (quadratic) & 5.5 (2) & \ensuremath{0.065}\\
\quad if city post code & 1.0 (1) & \ensuremath{0.307}\\
\quad if Brussels post code & 0.7 (1) & \ensuremath{0.416}\\
\quad if weekend & 0.9 (1) & \ensuremath{0.332}\\
2-stars &  & \\
$\quad\times\log(\text{pop.\ dens.\ in post code})$ (linear) & 0.9 (1) & \ensuremath{0.352}\\
\quad$\times\log(\text{pop.\ dens.\ in post code})$ (quadratic) & 0.9 (2) & \ensuremath{0.641}\\
\quad if city post code & 1.5 (1) & \ensuremath{0.215}\\
\quad if Brussels post code & 0.1 (1) & \ensuremath{0.729}\\
\quad if weekend & 0.4 (1) & \ensuremath{0.517}\\
triangles &  & \\
$\quad\times\log(\text{pop.\ dens.\ in post code})$ (linear) & 0.8 (1) & \ensuremath{0.357}\\
\quad$\times\log(\text{pop.\ dens.\ in post code})$ (quadratic) & 0.9 (2) & \ensuremath{0.623}\\
\quad if city post code & 1.4 (1) & \ensuremath{0.233}\\
\quad if Brussels post code & 0.2 (1) & \ensuremath{0.628}\\
\quad if weekend & 0.3 (1) & \ensuremath{0.599}\\
\bottomrule
\end{tabular}

\end{center}
\end{table}
\begin{table}[H]
  \caption{\label{tab:dataset-m0} Tests of the null hypothesis of no effect of dataset $H$ on the specified network statistic over and above \Model{0}.}
  \footnotesize
  \begin{center}

\begin{tabular}{lrr}
\toprule
Statistic & Score $\chi^2$ (df) & $\pval$\\
\midrule
Omnibus & 5.6 (2) & \ensuremath{0.0617}\\
\quad edges & nonparam. & \ensuremath{0.0187}\\
\quad 2-stars & nonparam. & \ensuremath{0.0604}\\
\bottomrule
\end{tabular}

\end{center}
\end{table}

\begin{table}[H]
  \caption{\label{tab:pearson-sd-m0}Sample standard deviations of Pearson residuals for edge, two-star, and triangle counts in \Model{0} for the dataset and its subsets. Values substantially greater than 1 indicate overdispersion relative to the model.}
  \footnotesize
\begin{center}

\begin{tabular}{lrrrr}
\toprule
Statistic & Overall & $H$ & \Ewc & \Enc\\
\midrule
edges & 0.989 & 0.764 & 1.021 & 1.03\\
2-stars & 0.969 & 0.811 & 1.001 & 1.02\\
triangles & 0.962 & 0.855 & 0.972 & 1.01\\
\bottomrule
\end{tabular}

\end{center}
\end{table}

\begin{table}[H]
  \caption{\label{tab:pearson-mix-m0}Pearson residuals for edge, two-star, triangle, and mixing counts in \Model{0} for the dataset and its subsets. Extreme positive or negative residual values indicate that the model poorly accounts for effect of the presence of a child (the main selection criterion for the $H$ dataset) and/or that cells in the mixing model should not have been merged. Mixing effects for which a child effect is inherent (i.e., involving a young child and/or a preadolescent but no other on one side of the relation) or which are not found in the $H$ dataset (i.e., seniors) are also excluded.}
  \footnotesize
\begin{center}

\begin{tabular}{lrrrr}
\toprule
Effect & Overall & $H$ & \Ewc & \Enc\\
\midrule
edges & $ 0.31$ & $ 2.26$ & $-0.64$ & $-0.46$\\
2-stars & $ 0.53$ & $ 1.66$ & $-0.53$ & $ 0.05$\\
triangles & $ 0.51$ & $ 1.56$ & $-0.53$ & $ 0.06$\\
Adolescent with Adolescent & $ 0.27$ & $-1.15$ & $ 0.48$ & $ 0.49$\\
Adolescent with Young Adult & $ 0.24$ & $-1.35$ & $ 1.03$ & $ 0.19$\\
Young Adult with Young Adult & $ 0.07$ & $ 1.72$ & $-0.75$ & $-0.14$\\
Adolescent with Older Female Adult & $ 0.21$ & $ 1.96$ & $-0.84$ & $-0.28$\\
Young Adult with Older Female Adult & $ 0.61$ & $-0.42$ & $ 0.29$ & $ 0.65$\\
Older Female Adult with Older Female Adult & $-0.03$ & $ 0.58$ & $ 0.02$ & $-0.10$\\
Adolescent with Older Male Adult & $ 0.22$ & $ 1.81$ & $-0.53$ & $-0.45$\\
Young Adult with Older Male Adult & $-0.55$ & $-1.17$ & $ 1.07$ & $-0.64$\\
Older Female Adult with Older Male Adult & $ 0.08$ & $ 2.62$ & $ 1.76$ & $-1.77$\\
Older Male Adult with Older Male Adult & $ 0.04$ & $-1.58$ & $-1.81$ & $ 0.64$\\
\bottomrule
\end{tabular}

\end{center}
\end{table}

\clearpage

\subsection[\Model{1}]{\label{app:extra-results-m1}Full results and diagnostics for \Model{1} }
\small

\begin{table}[H]
  \caption{\label{tab:coef-m1}Parameter estimates (and standard errors) for \Model{1}.}
  \footnotesize
  \begin{center}

\begin{tabular}{lr}
\toprule
Relationship Effect\\\quad{}$\times$ Network-Level Effect & Coefficient (SE)$^{\hphantom{\star\star\star}}$\\
\midrule
edges $\times$ $\log(\nactors_{\sampidx})$ & $-14.28 \; (2.87)^{\star\star\star}$\\
\quad $\times$ $\log^2(\nactors_{\sampidx})$ & $5.69 \; (1.29)^{\star\star\star}$\\
\quad   if Brussels post code & $0.08 \; (0.19)^{\phantom{\star}\phantom{\star}\phantom{\star}}$\\
\quad   if on weekend & $0.14 \; (0.06)^{\star\phantom{\star}\phantom{\star}}$\\
2-stars & $1.91 \; (0.78)^{\star\phantom{\star}\phantom{\star}}$\\
\quad $\times$ $\log(\nactors_{\sampidx})$ & $-2.15 \; (0.41)^{\star\star\star}$\\
\quad $\times$ $\log^2(\nactors_{\sampidx})$ & $0.34 \; (0.11)^{\star\star\phantom{\star}}$\\
triangles & $5.55 \; (0.97)^{\star\star\star}$\\
\quad $\times$ $\log(\nactors_{\sampidx})$ & $-3.46 \; (1.39)^{\star\phantom{\star}\phantom{\star}}$\\
\quad $\times$ $\log^2(\nactors_{\sampidx})$ & $0.93 \; (0.70)^{\phantom{\star}\phantom{\star}\phantom{\star}}$\\
Young Child with Young Child & $8.60 \; (1.49)^{\star\star\star}$\\
Young Child with Preadolescent & $9.10 \; (1.48)^{\star\star\star}$\\
Preadolescent with Preadolescent & $8.17 \; (1.45)^{\star\star\star}$\\
Adolescent with Adolescent & $7.70 \; (1.43)^{\star\star\star}$\\
Young Child with Young Adult & $9.64 \; (1.76)^{\star\star\star}$\\
Preadolescent with Young Adult & $7.25 \; (1.46)^{\star\star\star}$\\
Adolescent with Young Adult & $7.73 \; (1.45)^{\star\star\star}$\\
Young Adult with Young Adult & $7.66 \; (1.44)^{\star\star\star}$\\
Young Child with Older Female Adult & $10.26 \; (1.45)^{\star\star\star}$\\
Preadolescent with Older Female Adult & $9.67 \; (1.43)^{\star\star\star}$\\
Adolescent with Older Female Adult & $8.90 \; (1.43)^{\star\star\star}$\\
Older Female Adult with Older Female Adult & $7.45 \; (1.46)^{\star\star\star}$\\
Young Child with Older Male Adult & $9.09 \; (1.43)^{\star\star\star}$\\
Preadolescent with Older Male Adult & $8.76 \; (1.42)^{\star\star\star}$\\
Adolescent with Older Male Adult & $8.20 \; (1.42)^{\star\star\star}$\\
Older Female Adult with Older Male Adult & $10.11 \; (1.44)^{\star\star\star}$\\
\quad   if child absent & $-1.22 \; (0.30)^{\star\star\star}$\\
Older Male Adult with Older Male Adult & $6.59 \; (1.45)^{\star\star\star}$\\
Older Female Adult with Senior & $8.12 \; (1.42)^{\star\star\star}$\\
Older Male Adult with Senior & $7.51 \; (1.45)^{\star\star\star}$\\
Senior with Senior & $7.82 \; (1.40)^{\star\star\star}$\\
Adolescent with Young Child or Preadolescent & $8.07 \; (1.43)^{\star\star\star}$\\
Young Adult with Older Adult & $8.02 \; (1.43)^{\star\star\star}$\\
Young Child or Preadolescent with Senior & $8.29 \; (1.52)^{\star\star\star}$\\
Adolescent or Young Adult with Senior & $9.93 \; (1.70)^{\star\star\star}$\\
\bottomrule
\end{tabular}

Significance: $^{\star\star\star}\le 0.001<^{\star\star}\le 0.01< ^\star \le 0.05$
\end{center}
\end{table}
\begin{table}[H]
  \caption{\label{tab:omnibus-m1} Omnibus tests for selected groups of effects in \Model{1}. Effects are net of the rest of the model.}
  \footnotesize
  \begin{center}

\begin{tabular}{lrr}
\toprule
Effects & Wald $\chi^2$ (df) & $\pval$\\
\midrule
any 2-star & 48.1 (3) & \ensuremath{<0.001}\\
any triangle & 101.5 (3) & \ensuremath{<0.001}\\
any $\log(\nactors_\sampidx)$ or $\log^2(\nactors_\sampidx)$ & 2979.3 (6) & \ensuremath{<0.001}\\
any $\log^2(\nactors_\sampidx)$ & 57.1 (3) & \ensuremath{<0.001}\\
2-star or triangle $\log^2(\nactors_\sampidx)$ & 22.2 (2) & \ensuremath{<0.001}\\
\bottomrule
\end{tabular}

\end{center}
\end{table}

\begin{figure}[H]

{\centering \subfloat[Edges\label{fig:unnamed-chunk-102-1}]{\includegraphics[width=\maxwidth]{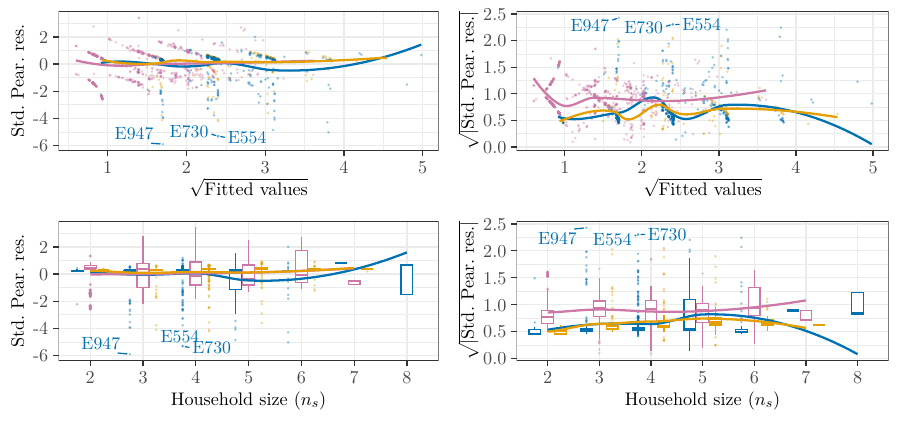} }\newline
\subfloat[2-stars\label{fig:unnamed-chunk-102-2}]{\includegraphics[width=\maxwidth]{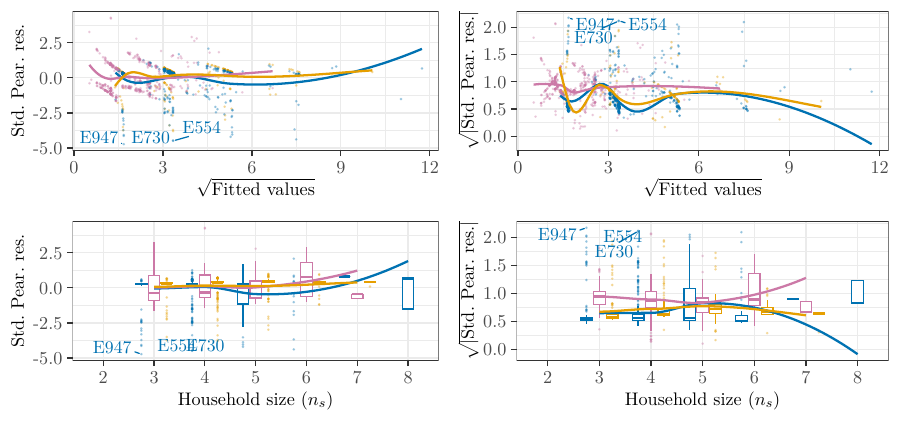} }\newline
\subfloat[Triangles\label{fig:unnamed-chunk-102-3}]{\includegraphics[width=\maxwidth]{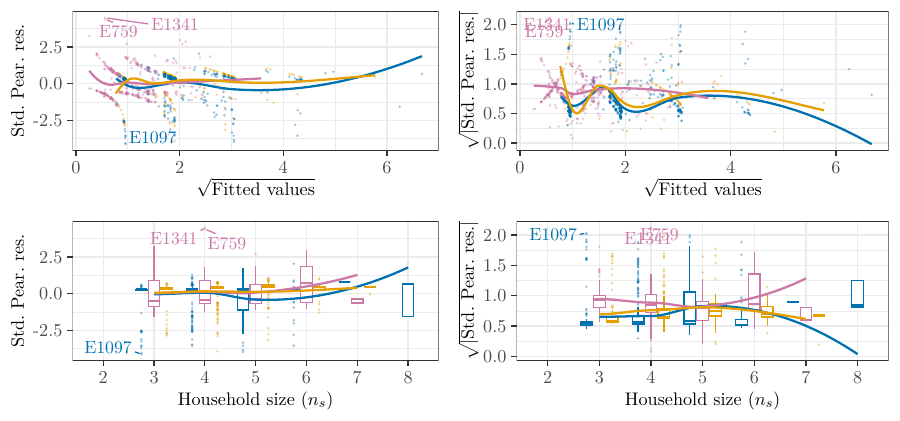} }

}

\caption{\label{fig:resid-plots-m1}Residual plots of network statistics against fitted values and network size for \Model{1}. \subdataleg}\label{fig:unnamed-chunk-102}
\end{figure}

\begin{figure}[H]

{\centering \includegraphics[width=\maxwidth]{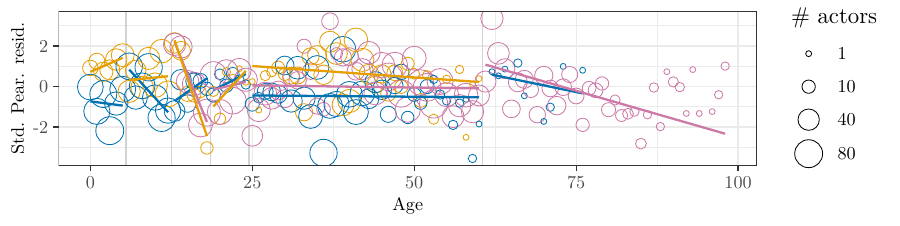} 

}

\caption{\label{fig:age-resid-m1}Residual plot for \Model{1} of edges incident on actors of a given age against age. \subdataleg}\label{fig:unnamed-chunk-103}
\end{figure}

\begin{figure}[H]

{\centering \includegraphics[width=\maxwidth,height=1in]{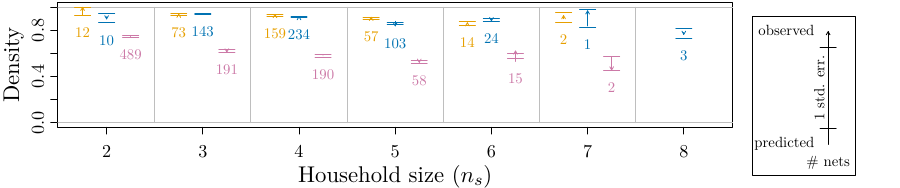} 

}

\caption{\label{fig:density-error-m1}Average prediction errors of density in \Model{1}. Values are averaged over the networks grouped by size and subset. \subdataleg}\label{fig:unnamed-chunk-104}
\end{figure}

\begin{table}[H]
  \caption{\label{tab:nns-anova-m1}Analyses of variance for fitting residuals for \Model{1} against household size (represented as a categorical predictor, with a dummy variable for each size).}
  \centering
  \footnotesize

\begin{tabular}{lrrrrr}
\toprule
\quad{}Source & df & Sum Sq. & Mean Sq. & $F$ & \pval\\
\midrule
edges &  &  &  &  & \\
\quad{}$\nactors_{\sampidx}$ (categorical) & 7 & 9.9 & 1.41 & 1.4 & \ensuremath{0.198}\\
\quad{}Residuals & 1773 & 1778.1 & 1.00 &  & \\
\addlinespace
2-stars &  &  &  &  & \\
\quad{}$\nactors_{\sampidx}$ (categorical) & 6 & 8.9 & 1.48 & 1.5 & \ensuremath{0.175}\\
\quad{}Residuals & 1263 & 1244.7 & 0.99 &  & \\
\addlinespace
triangles &  &  &  &  & \\
\quad{}$\nactors_{\sampidx}$ (categorical) & 6 & 7.9 & 1.32 & 1.4 & \ensuremath{0.223}\\
\quad{}Residuals & 1263 & 1212.8 & 0.96 &  & \\
\bottomrule
\end{tabular}

\end{table}

\begin{table}[H]
  \caption{\label{tab:reg-res-m1} Regression of residuals of \Model{1} for each network structural property on each candidate predictor.}
  \footnotesize
  \begin{center}

\begin{tabular}{lrr}
\toprule
Relationship Effect\\\quad{}$\times$ Network-Level Effect & Wald $\chi^2$ (df) & $\pval$\\
\midrule
edges &  & \\
$\quad\times\log(\text{pop.\ dens.\ in post code})$ (linear) & 4.0 (1) & \ensuremath{0.046}\\
\quad$\times\log(\text{pop.\ dens.\ in post code})$ (quadratic) & 5.2 (2) & \ensuremath{0.073}\\
\quad if city post code & 1.1 (1) & \ensuremath{0.305}\\
\quad if Brussels post code & 0.4 (1) & \ensuremath{0.553}\\
\quad if weekend & 1.2 (1) & \ensuremath{0.278}\\
2-stars &  & \\
$\quad\times\log(\text{pop.\ dens.\ in post code})$ (linear) & 1.1 (1) & \ensuremath{0.303}\\
\quad$\times\log(\text{pop.\ dens.\ in post code})$ (quadratic) & 1.1 (2) & \ensuremath{0.581}\\
\quad if city post code & 1.3 (1) & \ensuremath{0.251}\\
\quad if Brussels post code & 0.1 (1) & \ensuremath{0.800}\\
\quad if weekend & 1.1 (1) & \ensuremath{0.305}\\
triangles &  & \\
$\quad\times\log(\text{pop.\ dens.\ in post code})$ (linear) & 1.1 (1) & \ensuremath{0.294}\\
\quad$\times\log(\text{pop.\ dens.\ in post code})$ (quadratic) & 1.1 (2) & \ensuremath{0.576}\\
\quad if city post code & 1.3 (1) & \ensuremath{0.255}\\
\quad if Brussels post code & 0.2 (1) & \ensuremath{0.682}\\
\quad if weekend & 0.7 (1) & \ensuremath{0.409}\\
\bottomrule
\end{tabular}

\end{center}
\end{table}
\begin{table}[H]
  \caption{\label{tab:dataset-m1} Tests of the null hypothesis of no effect of dataset $H$ on the specified network statistic over and above \Model{1}.}
  \footnotesize
  \begin{center}

\begin{tabular}{lrr}
\toprule
Statistic & Score $\chi^2$ (df) & $\pval$\\
\midrule
Omnibus & 3.4 (2) & \ensuremath{0.1829}\\
\quad edges & nonparam. & \ensuremath{0.0675}\\
\quad 2-stars & nonparam. & \ensuremath{0.1108}\\
\bottomrule
\end{tabular}

\end{center}
\end{table}

\begin{table}[H]
  \caption{\label{tab:pearson-sd-m1}Sample standard deviations of Pearson residuals for edge, two-star, and triangle counts in \Model{1} for the dataset and its subsets. Values substantially greater than 1 indicate overdispersion relative to the model.}
  \footnotesize
\begin{center}

\begin{tabular}{lrrrr}
\toprule
Statistic & Overall & $H$ & \Ewc & \Enc\\
\midrule
edges & 1.003 & 0.822 & 1.07 & 1.01\\
2-stars & 0.994 & 0.852 & 1.05 & 1.02\\
triangles & 0.981 & 0.879 & 1.01 & 1.01\\
\bottomrule
\end{tabular}

\end{center}
\end{table}

\begin{table}[H]
  \caption{\label{tab:pearson-mix-m1}Pearson residuals for edge, two-star, triangle, and mixing counts in \Model{1} for the dataset and its subsets. Extreme positive or negative residual values indicate that the model poorly accounts for effect of the presence of a child (the main selection criterion for the $H$ dataset) and/or that cells in the mixing model should not have been merged. Mixing effects for which a child effect is inherent (i.e., involving a young child and/or a preadolescent but no other on one side of the relation) or which are not found in the $H$ dataset (i.e., seniors) are also excluded.}
  \footnotesize
\begin{center}

\begin{tabular}{lrrrr}
\toprule
Effect & Overall & $H$ & \Ewc & \Enc\\
\midrule
edges & $ 0.21$ & $ 1.92$ & $-1.36$ & $ 0.13$\\
2-stars & $ 0.43$ & $ 1.72$ & $-1.08$ & $ 0.47$\\
triangles & $ 0.44$ & $ 1.66$ & $-1.06$ & $ 0.47$\\
Adolescent with Adolescent & $ 0.26$ & $-1.29$ & $ 0.33$ & $ 0.62$\\
Adolescent with Young Adult & $ 0.27$ & $-1.47$ & $ 0.56$ & $ 0.40$\\
Young Adult with Young Adult & $ 0.07$ & $ 2.05$ & $-0.99$ & $-0.22$\\
Adolescent with Older Female Adult & $ 0.10$ & $ 1.79$ & $-1.26$ & $-0.17$\\
Young Adult with Older Female Adult & $ 0.62$ & $-0.29$ & $-0.03$ & $ 0.70$\\
Older Female Adult with Older Female Adult & $-0.09$ & $ 0.60$ & $ 0.03$ & $-0.16$\\
Adolescent with Older Male Adult & $ 0.05$ & $ 1.53$ & $-1.21$ & $-0.12$\\
Young Adult with Older Male Adult & $-0.61$ & $-0.97$ & $ 0.86$ & $-0.66$\\
Older Female Adult with Older Male Adult & $ 0.01$ & $ 0.21$ & $-0.03$ & $-0.04$\\
Older Male Adult with Older Male Adult & $ 0.02$ & $-1.38$ & $-1.67$ & $ 0.50$\\
\bottomrule
\end{tabular}

\end{center}
\end{table}

\clearpage

\subsection[\Model{\MIp}]{\label{app:extra-results-m1p}Full results and diagnostics for \Model{\MIp} }
\small

\begin{table}[H]
  \caption{\label{tab:coef-m1p}Parameter estimates (and standard errors) for \Model{\MIp}.}
  \footnotesize
  \begin{center}

\begin{tabular}{lr}
\toprule
Relationship Effect\\\quad{}$\times$ Network-Level Effect & Coefficient (SE)$^{\hphantom{\star\star\star}}$\\
\midrule
edges $\times$ $\log(\nactors_{\sampidx})$ & $-14.38 \; (3.55)^{\star\star\star}$\\
\quad $\times$ $\log^2(\nactors_{\sampidx})$ & $5.78 \; (1.60)^{\star\star\star}$\\
\quad   if Brussels post code & $0.10 \; (0.19)^{\phantom{\star}\phantom{\star}\phantom{\star}}$\\
\quad   if on weekend & $0.13 \; (0.05)^{\star\phantom{\star}\phantom{\star}}$\\
\quad   if child absent & $-0.17 \; (0.09)^{\phantom{\star}\phantom{\star}\phantom{\star}}$\\
2-stars & $0.57 \; (4.27)^{\phantom{\star}\phantom{\star}\phantom{\star}}$\\
\quad $\times$ $\log(\nactors_{\sampidx})$ & $-0.42 \; (5.04)^{\phantom{\star}\phantom{\star}\phantom{\star}}$\\
\quad $\times$ $\log^2(\nactors_{\sampidx})$ & $-0.22 \; (1.47)^{\phantom{\star}\phantom{\star}\phantom{\star}}$\\
triangles & $9.94 \; (10.24)^{\phantom{\star}\phantom{\star}\phantom{\star}}$\\
\quad $\times$ $\log(\nactors_{\sampidx})$ & $-8.96 \; (12.67)^{\phantom{\star}\phantom{\star}\phantom{\star}}$\\
\quad $\times$ $\log^2(\nactors_{\sampidx})$ & $2.67 \; (3.90)^{\phantom{\star}\phantom{\star}\phantom{\star}}$\\
Young Child with Young Child & $8.49 \; (1.78)^{\star\star\star}$\\
Young Child with Preadolescent & $9.00 \; (1.78)^{\star\star\star}$\\
Preadolescent with Preadolescent & $8.08 \; (1.75)^{\star\star\star}$\\
Adolescent with Adolescent & $7.80 \; (1.74)^{\star\star\star}$\\
Young Child with Young Adult & $9.42 \; (1.99)^{\star\star\star}$\\
Preadolescent with Young Adult & $7.05 \; (1.76)^{\star\star\star}$\\
Adolescent with Young Adult & $7.87 \; (1.73)^{\star\star\star}$\\
Young Adult with Young Adult & $7.76 \; (1.76)^{\star\star\star}$\\
Young Child with Older Female Adult & $10.30 \; (1.75)^{\star\star\star}$\\
Preadolescent with Older Female Adult & $9.70 \; (1.75)^{\star\star\star}$\\
Adolescent with Older Female Adult & $8.97 \; (1.74)^{\star\star\star}$\\
Older Female Adult with Older Female Adult & $7.46 \; (1.77)^{\star\star\star}$\\
Young Child with Older Male Adult & $9.15 \; (1.75)^{\star\star\star}$\\
Preadolescent with Older Male Adult & $8.85 \; (1.73)^{\star\star\star}$\\
Adolescent with Older Male Adult & $8.26 \; (1.75)^{\star\star\star}$\\
Older Female Adult with Older Male Adult & $9.24 \; (1.73)^{\star\star\star}$\\
Older Male Adult with Older Male Adult & $6.54 \; (1.76)^{\star\star\star}$\\
Older Female Adult with Senior & $8.28 \; (1.72)^{\star\star\star}$\\
Older Male Adult with Senior & $7.64 \; (1.76)^{\star\star\star}$\\
Senior with Senior & $8.00 \; (1.72)^{\star\star\star}$\\
Adolescent with Young Child or Preadolescent & $7.94 \; (1.74)^{\star\star\star}$\\
Young Adult with Older Adult & $8.11 \; (1.74)^{\star\star\star}$\\
Young Child or Preadolescent with Senior & $8.09 \; (1.82)^{\star\star\star}$\\
Adolescent or Young Adult with Senior & $10.04 \; (1.99)^{\star\star\star}$\\
\bottomrule
\end{tabular}

Significance: $^{\star\star\star}\le 0.001<^{\star\star}\le 0.01< ^\star \le 0.05$
\end{center}
\end{table}
\begin{table}[H]
  \caption{\label{tab:omnibus-m1p} Omnibus tests for selected groups of effects in \Model{\MIp}. Effects are net of the rest of the model.}
  \footnotesize
  \begin{center}

\begin{tabular}{lrr}
\toprule
Effects & Wald $\chi^2$ (df) & $\pval$\\
\midrule
any 2-star & 21.7 (3) & \ensuremath{<0.001}\\
any triangle & 100.1 (3) & \ensuremath{<0.001}\\
any $\log(\nactors_\sampidx)$ or $\log^2(\nactors_\sampidx)$ & 79.1 (6) & \ensuremath{<0.001}\\
any $\log^2(\nactors_\sampidx)$ & 19.3 (3) & \ensuremath{<0.001}\\
2-star or triangle $\log^2(\nactors_\sampidx)$ & 10.5 (2) & \ensuremath{0.005}\\
\bottomrule
\end{tabular}

\end{center}
\end{table}

\begin{figure}[H]

{\centering \subfloat[Edges\label{fig:unnamed-chunk-117-1}]{\includegraphics[width=\maxwidth]{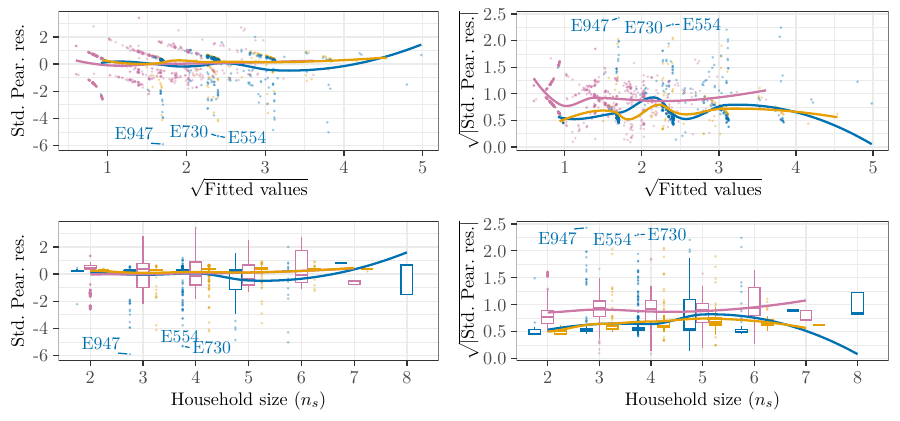} }\newline
\subfloat[2-stars\label{fig:unnamed-chunk-117-2}]{\includegraphics[width=\maxwidth]{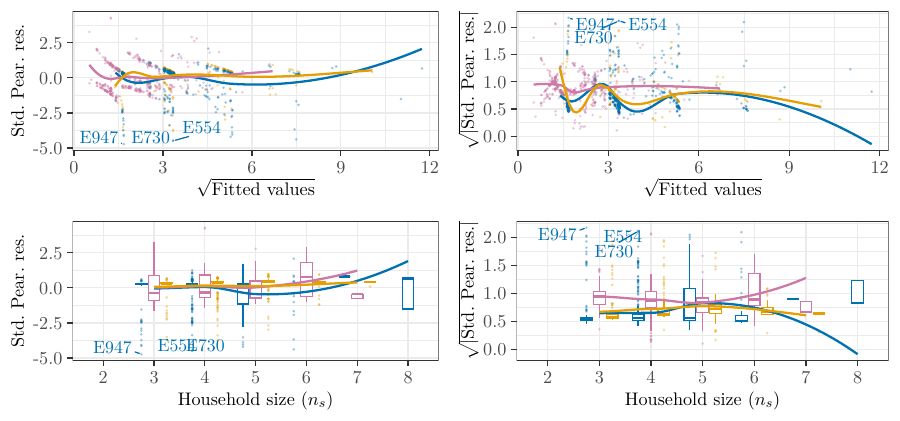} }\newline
\subfloat[Triangles\label{fig:unnamed-chunk-117-3}]{\includegraphics[width=\maxwidth]{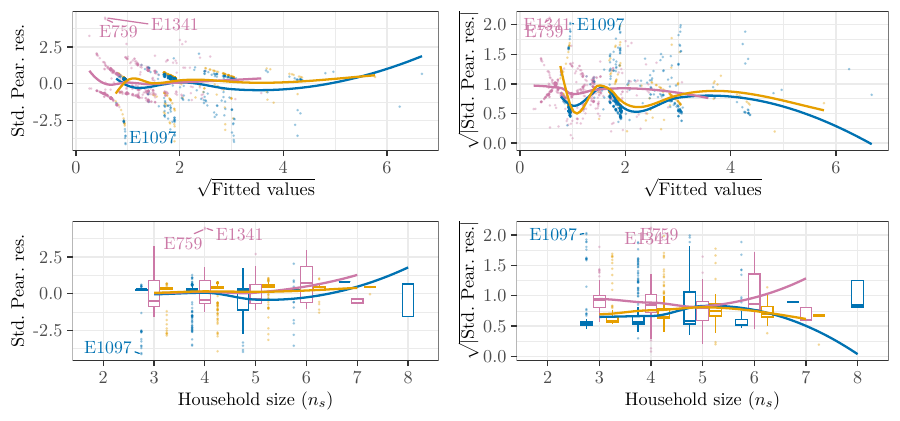} }

}

\caption{\label{fig:resid-plots-m1p}Residual plots of network statistics against fitted values and network size for \Model{\MIp}. \subdataleg}\label{fig:unnamed-chunk-117}
\end{figure}

\begin{figure}[H]

{\centering \includegraphics[width=\maxwidth]{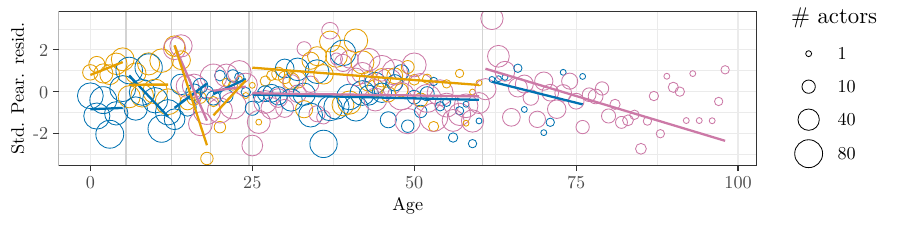} 

}

\caption{\label{fig:age-resid-m1p}Residual plot for \Model{\MIp} of edges incident on actors of a given age against age. \subdataleg}\label{fig:unnamed-chunk-118}
\end{figure}

\begin{figure}[H]

{\centering \includegraphics[width=\maxwidth,height=1in]{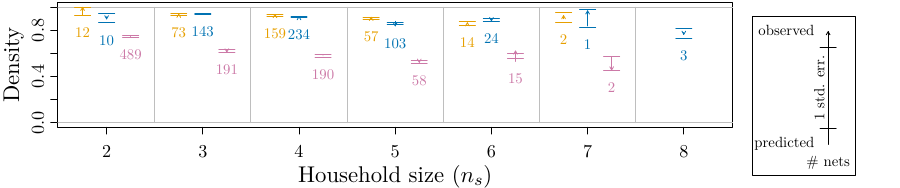} 

}

\caption{\label{fig:density-error-m1p}Average prediction errors of density in \Model{\MIp}. Values are averaged over the networks grouped by size and subset. \subdataleg}\label{fig:unnamed-chunk-119}
\end{figure}

\begin{table}[H]
  \caption{\label{tab:nns-anova-m1p}Analyses of variance for fitting residuals for \Model{\MIp} against household size (represented as a categorical predictor, with a dummy variable for each size).}
  \centering
  \footnotesize

\begin{tabular}{lrrrrr}
\toprule
\quad{}Source & df & Sum Sq. & Mean Sq. & $F$ & \pval\\
\midrule
edges &  &  &  &  & \\
\quad{}$\nactors_{\sampidx}$ (categorical) & 7 & 8.7 & 1.24 & 1.3 & \ensuremath{0.267}\\
\quad{}Residuals & 1773 & 1751.5 & 0.99 &  & \\
\addlinespace
2-stars &  &  &  &  & \\
\quad{}$\nactors_{\sampidx}$ (categorical) & 6 & 8.0 & 1.33 & 1.4 & \ensuremath{0.212}\\
\quad{}Residuals & 1263 & 1206.6 & 0.96 &  & \\
\addlinespace
triangles &  &  &  &  & \\
\quad{}$\nactors_{\sampidx}$ (categorical) & 6 & 7.3 & 1.22 & 1.3 & \ensuremath{0.253}\\
\quad{}Residuals & 1263 & 1187.3 & 0.94 &  & \\
\bottomrule
\end{tabular}

\end{table}

\begin{table}[H]
  \caption{\label{tab:reg-res-m1p} Regression of residuals of \Model{\MIp} for each network structural property on each candidate predictor.}
  \footnotesize
  \begin{center}

\begin{tabular}{lrr}
\toprule
Relationship Effect\\\quad{}$\times$ Network-Level Effect & Wald $\chi^2$ (df) & $\pval$\\
\midrule
edges &  & \\
$\quad\times\log(\text{pop.\ dens.\ in post code})$ (linear) & 3.8 (1) & \ensuremath{0.050}\\
\quad$\times\log(\text{pop.\ dens.\ in post code})$ (quadratic) & 4.9 (2) & \ensuremath{0.086}\\
\quad if city post code & 0.9 (1) & \ensuremath{0.331}\\
\quad if Brussels post code & 0.6 (1) & \ensuremath{0.458}\\
\quad if weekend & 0.9 (1) & \ensuremath{0.336}\\
2-stars &  & \\
$\quad\times\log(\text{pop.\ dens.\ in post code})$ (linear) & 0.7 (1) & \ensuremath{0.401}\\
\quad$\times\log(\text{pop.\ dens.\ in post code})$ (quadratic) & 0.7 (2) & \ensuremath{0.700}\\
\quad if city post code & 1.3 (1) & \ensuremath{0.247}\\
\quad if Brussels post code & 0.1 (1) & \ensuremath{0.791}\\
\quad if weekend & 0.5 (1) & \ensuremath{0.467}\\
triangles &  & \\
$\quad\times\log(\text{pop.\ dens.\ in post code})$ (linear) & 0.7 (1) & \ensuremath{0.406}\\
\quad$\times\log(\text{pop.\ dens.\ in post code})$ (quadratic) & 0.8 (2) & \ensuremath{0.684}\\
\quad if city post code & 1.2 (1) & \ensuremath{0.266}\\
\quad if Brussels post code & 0.2 (1) & \ensuremath{0.691}\\
\quad if weekend & 0.4 (1) & \ensuremath{0.552}\\
\bottomrule
\end{tabular}

\end{center}
\end{table}
\begin{table}[H]
  \caption{\label{tab:dataset-m1p} Tests of the null hypothesis of no effect of dataset $H$ on the specified network statistic over and above \Model{\MIp}.}
  \footnotesize
  \begin{center}

\begin{tabular}{lrr}
\toprule
Statistic & Score $\chi^2$ (df) & $\pval$\\
\midrule
Omnibus & 3.9 (2) & \ensuremath{0.1410}\\
\quad edges & nonparam. & \ensuremath{0.0586}\\
\quad 2-stars & nonparam. & \ensuremath{0.1255}\\
\bottomrule
\end{tabular}

\end{center}
\end{table}

\begin{table}[H]
  \caption{\label{tab:pearson-sd-m1p}Sample standard deviations of Pearson residuals for edge, two-star, and triangle counts in \Model{\MIp} for the dataset and its subsets. Values substantially greater than 1 indicate overdispersion relative to the model.}
  \footnotesize
\begin{center}

\begin{tabular}{lrrrr}
\toprule
Statistic & Overall & $H$ & \Ewc & \Enc\\
\midrule
edges & 0.995 & 0.781 & 1.034 & 1.03\\
2-stars & 0.979 & 0.826 & 1.012 & 1.03\\
triangles & 0.971 & 0.868 & 0.983 & 1.02\\
\bottomrule
\end{tabular}

\end{center}
\end{table}

\begin{table}[H]
  \caption{\label{tab:pearson-mix-m1p}Pearson residuals for edge, two-star, triangle, and mixing counts in \Model{\MIp} for the dataset and its subsets. Extreme positive or negative residual values indicate that the model poorly accounts for effect of the presence of a child (the main selection criterion for the $H$ dataset) and/or that cells in the mixing model should not have been merged. Mixing effects for which a child effect is inherent (i.e., involving a young child and/or a preadolescent but no other on one side of the relation) or which are not found in the $H$ dataset (i.e., seniors) are also excluded.}
  \footnotesize
\begin{center}

\begin{tabular}{lrrrr}
\toprule
Effect & Overall & $H$ & \Ewc & \Enc\\
\midrule
edges & $ 0.26$ & $ 1.76$ & $-1.07$ & $ 0.05$\\
2-stars & $ 0.57$ & $ 1.44$ & $-0.73$ & $ 0.61$\\
triangles & $ 0.55$ & $ 1.35$ & $-0.72$ & $ 0.60$\\
Adolescent with Adolescent & $ 0.19$ & $-1.62$ & $ 0.05$ & $ 0.89$\\
Adolescent with Young Adult & $ 0.65$ & $-1.69$ & $ 1.14$ & $ 0.65$\\
Young Adult with Young Adult & $-0.08$ & $ 1.57$ & $-1.16$ & $-0.14$\\
Adolescent with Older Female Adult & $ 0.18$ & $ 1.53$ & $-1.39$ & $ 0.13$\\
Young Adult with Older Female Adult & $ 0.64$ & $-0.59$ & $-0.01$ & $ 0.79$\\
Older Female Adult with Older Female Adult & $-0.05$ & $ 0.57$ & $-0.03$ & $-0.10$\\
Adolescent with Older Male Adult & $ 0.17$ & $ 1.20$ & $-1.24$ & $ 0.15$\\
Young Adult with Older Male Adult & $-0.43$ & $-1.43$ & $ 0.82$ & $-0.35$\\
Older Female Adult with Older Male Adult & $ 0.04$ & $ 1.94$ & $ 1.29$ & $-1.29$\\
Older Male Adult with Older Male Adult & $ 0.04$ & $-1.51$ & $-1.68$ & $ 0.59$\\
\bottomrule
\end{tabular}

\end{center}
\end{table}

\clearpage

\subsection[\Model{2}]{\label{app:extra-results-m2}Full results and diagnostics for \Model{2} }
\small

\begin{table}[H]
  \caption{\label{tab:coef-m2}Parameter estimates (and standard errors) for \Model{2}.}
  \footnotesize
  \begin{center}

\begin{tabular}{lr}
\toprule
Relationship Effect\\\quad{}$\times$ Network-Level Effect & Coefficient (SE)$^{\hphantom{\star\star\star}}$\\
\midrule
edges $\times$ $\log(\nactors_{\sampidx})$ & $-13.78 \; (2.98)^{\star\star\star}$\\
\quad $\times$ $\log^2(\nactors_{\sampidx})$ & $5.47 \; (1.34)^{\star\star\star}$\\
\quad   if Brussels post code & $-0.02 \; (0.20)^{\phantom{\star}\phantom{\star}\phantom{\star}}$\\
\quad $\times$ $\log(\text{pop.\ dens.\ in post code})$ & $0.04 \; (0.03)^{\phantom{\star}\phantom{\star}\phantom{\star}}$\\
\quad   if on weekend & $0.13 \; (0.06)^{\star\phantom{\star}\phantom{\star}}$\\
2-stars & $1.14 \; (0.82)^{\phantom{\star}\phantom{\star}\phantom{\star}}$\\
\quad $\times$ $\log(\nactors_{\sampidx})$ & $-1.22 \; (0.44)^{\star\star\phantom{\star}}$\\
\quad $\times$ $\log^2(\nactors_{\sampidx})$ & $0.07 \; (0.11)^{\phantom{\star}\phantom{\star}\phantom{\star}}$\\
triangles & $7.30 \; (0.96)^{\star\star\star}$\\
\quad $\times$ $\log(\nactors_{\sampidx})$ & $-5.65 \; (1.44)^{\star\star\star}$\\
\quad $\times$ $\log^2(\nactors_{\sampidx})$ & $1.60 \; (0.74)^{\star\phantom{\star}\phantom{\star}}$\\
Young Child with Young Child & $8.66 \; (1.54)^{\star\star\star}$\\
Young Child with Preadolescent & $9.15 \; (1.54)^{\star\star\star}$\\
Preadolescent with Preadolescent & $8.24 \; (1.51)^{\star\star\star}$\\
Adolescent with Adolescent & $7.75 \; (1.49)^{\star\star\star}$\\
Young Child with Young Adult & $9.67 \; (1.81)^{\star\star\star}$\\
Preadolescent with Young Adult & $7.28 \; (1.51)^{\star\star\star}$\\
Adolescent with Young Adult & $7.82 \; (1.51)^{\star\star\star}$\\
Young Adult with Young Adult & $7.70 \; (1.49)^{\star\star\star}$\\
Young Child with Older Female Adult & $10.32 \; (1.51)^{\star\star\star}$\\
Preadolescent with Older Female Adult & $9.73 \; (1.49)^{\star\star\star}$\\
Adolescent with Older Female Adult & $8.96 \; (1.48)^{\star\star\star}$\\
Older Female Adult with Older Female Adult & $7.50 \; (1.52)^{\star\star\star}$\\
Young Child with Older Male Adult & $9.14 \; (1.49)^{\star\star\star}$\\
Preadolescent with Older Male Adult & $8.83 \; (1.48)^{\star\star\star}$\\
Adolescent with Older Male Adult & $8.26 \; (1.48)^{\star\star\star}$\\
Older Female Adult with Older Male Adult & $10.17 \; (1.49)^{\star\star\star}$\\
\quad   if child absent & $-1.20 \; (0.30)^{\star\star\star}$\\
Older Male Adult with Older Male Adult & $6.66 \; (1.50)^{\star\star\star}$\\
Older Female Adult with Senior & $8.20 \; (1.47)^{\star\star\star}$\\
Older Male Adult with Senior & $7.58 \; (1.50)^{\star\star\star}$\\
Senior with Senior & $7.89 \; (1.46)^{\star\star\star}$\\
Adolescent with Young Child or Preadolescent & $8.13 \; (1.48)^{\star\star\star}$\\
Young Adult with Older Adult & $8.07 \; (1.48)^{\star\star\star}$\\
Young Child or Preadolescent with Senior & $8.34 \; (1.57)^{\star\star\star}$\\
Adolescent or Young Adult with Senior & $10.01 \; (1.76)^{\star\star\star}$\\
\bottomrule
\end{tabular}

Significance: $^{\star\star\star}\le 0.001<^{\star\star}\le 0.01< ^\star \le 0.05$
\end{center}
\end{table}
\begin{table}[H]
  \caption{\label{tab:omnibus-m2} Omnibus tests for selected groups of effects in \Model{2}. Effects are net of the rest of the model.}
  \footnotesize
  \begin{center}

\begin{tabular}{lrr}
\toprule
Effects & Wald $\chi^2$ (df) & $\pval$\\
\midrule
any 2-star & 20.9 (3) & \ensuremath{<0.001}\\
any triangle & 109.0 (3) & \ensuremath{<0.001}\\
any $\log(\nactors_\sampidx)$ or $\log^2(\nactors_\sampidx)$ & 142.5 (6) & \ensuremath{<0.001}\\
any $\log^2(\nactors_\sampidx)$ & 28.0 (3) & \ensuremath{<0.001}\\
2-star or triangle $\log^2(\nactors_\sampidx)$ & 10.0 (2) & \ensuremath{0.007}\\
\bottomrule
\end{tabular}

\end{center}
\end{table}

\begin{figure}[H]

{\centering \subfloat[Edges\label{fig:unnamed-chunk-132-1}]{\includegraphics[width=\maxwidth]{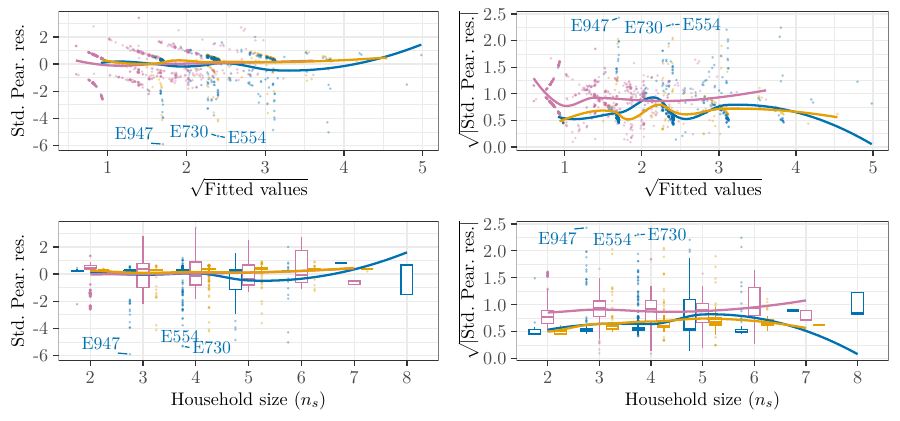} }\newline
\subfloat[2-stars\label{fig:unnamed-chunk-132-2}]{\includegraphics[width=\maxwidth]{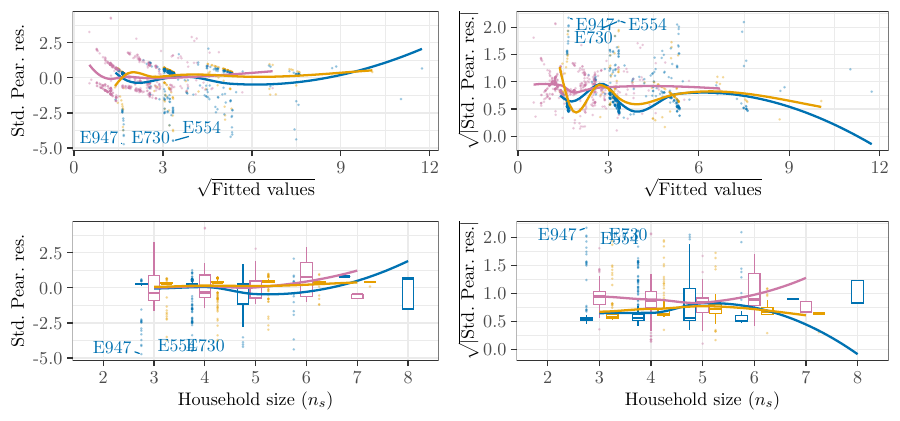} }\newline
\subfloat[Triangles\label{fig:unnamed-chunk-132-3}]{\includegraphics[width=\maxwidth]{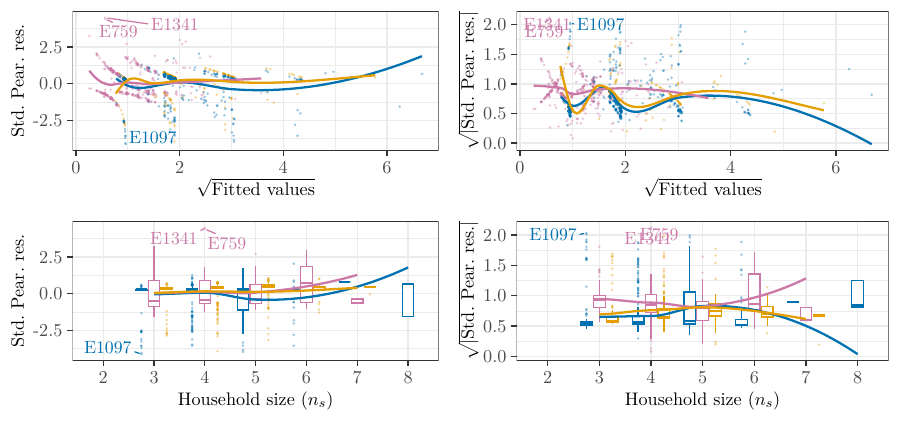} }

}

\caption{\label{fig:resid-plots-m2}Residual plots of network statistics against fitted values and network size for \Model{2}. \subdataleg}\label{fig:unnamed-chunk-132}
\end{figure}

\begin{figure}[H]

{\centering \includegraphics[width=\maxwidth]{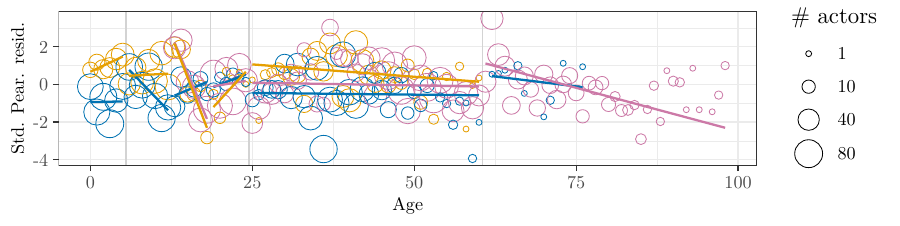} 

}

\caption{\label{fig:age-resid-m2}Residual plot for \Model{2} of edges incident on actors of a given age against age. \subdataleg}\label{fig:unnamed-chunk-133}
\end{figure}

\begin{figure}[H]

{\centering \includegraphics[width=\maxwidth,height=1in]{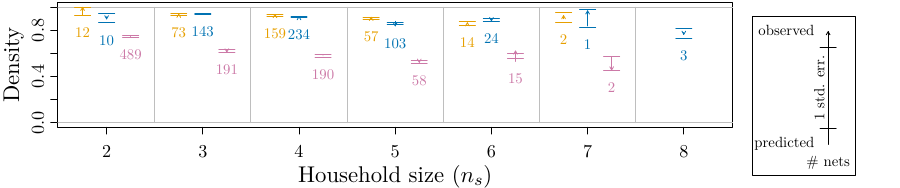} 

}

\caption{\label{fig:density-error-m2}Average prediction errors of density in \Model{2}. Values are averaged over the networks grouped by size and subset. \subdataleg}\label{fig:unnamed-chunk-134}
\end{figure}

\begin{table}[H]
  \caption{\label{tab:nns-anova-m2}Analyses of variance for fitting residuals for \Model{2} against household size (represented as a categorical predictor, with a dummy variable for each size).}
  \centering
  \footnotesize

\begin{tabular}{lrrrrr}
\toprule
\quad{}Source & df & Sum Sq. & Mean Sq. & $F$ & \pval\\
\midrule
edges &  &  &  &  & \\
\quad{}$\nactors_{\sampidx}$ (categorical) & 7 & 9.5 & 1.36 & 1.4 & \ensuremath{0.218}\\
\quad{}Residuals & 1773 & 1771.8 & 1.00 &  & \\
\addlinespace
2-stars &  &  &  &  & \\
\quad{}$\nactors_{\sampidx}$ (categorical) & 6 & 8.6 & 1.44 & 1.5 & \ensuremath{0.187}\\
\quad{}Residuals & 1263 & 1241.2 & 0.98 &  & \\
\addlinespace
triangles &  &  &  &  & \\
\quad{}$\nactors_{\sampidx}$ (categorical) & 6 & 7.8 & 1.29 & 1.3 & \ensuremath{0.232}\\
\quad{}Residuals & 1263 & 1209.5 & 0.96 &  & \\
\bottomrule
\end{tabular}

\end{table}

\begin{table}[H]
  \caption{\label{tab:reg-res-m2} Regression of residuals of \Model{2} for each network structural property on each candidate predictor.}
  \footnotesize
  \begin{center}

\begin{tabular}{lrr}
\toprule
Relationship Effect\\\quad{}$\times$ Network-Level Effect & Wald $\chi^2$ (df) & $\pval$\\
\midrule
edges &  & \\
$\quad\times\log(\text{pop.\ dens.\ in post code})$ (linear) & 1.9 (1) & \ensuremath{0.171}\\
\quad$\times\log(\text{pop.\ dens.\ in post code})$ (quadratic) & 2.6 (2) & \ensuremath{0.277}\\
\quad if city post code & 0.6 (1) & \ensuremath{0.458}\\
\quad if Brussels post code & 0.3 (1) & \ensuremath{0.594}\\
\quad if weekend & 1.0 (1) & \ensuremath{0.328}\\
2-stars &  & \\
$\quad\times\log(\text{pop.\ dens.\ in post code})$ (linear) & 0.3 (1) & \ensuremath{0.587}\\
\quad$\times\log(\text{pop.\ dens.\ in post code})$ (quadratic) & 0.3 (2) & \ensuremath{0.855}\\
\quad if city post code & 0.9 (1) & \ensuremath{0.330}\\
\quad if Brussels post code & 0.0 (1) & \ensuremath{0.876}\\
\quad if weekend & 1.0 (1) & \ensuremath{0.321}\\
triangles &  & \\
$\quad\times\log(\text{pop.\ dens.\ in post code})$ (linear) & 0.3 (1) & \ensuremath{0.594}\\
\quad$\times\log(\text{pop.\ dens.\ in post code})$ (quadratic) & 0.4 (2) & \ensuremath{0.828}\\
\quad if city post code & 0.9 (1) & \ensuremath{0.330}\\
\quad if Brussels post code & 0.1 (1) & \ensuremath{0.743}\\
\quad if weekend & 0.7 (1) & \ensuremath{0.420}\\
\bottomrule
\end{tabular}

\end{center}
\end{table}
\begin{table}[H]
  \caption{\label{tab:dataset-m2} Tests of the null hypothesis of no effect of dataset $H$ on the specified network statistic over and above \Model{2}.}
  \footnotesize
  \begin{center}

\begin{tabular}{lrr}
\toprule
Statistic & Score $\chi^2$ (df) & $\pval$\\
\midrule
Omnibus & 3.2 (2) & \ensuremath{0.2060}\\
\quad edges & nonparam. & \ensuremath{0.0829}\\
\quad 2-stars & nonparam. & \ensuremath{0.1443}\\
\bottomrule
\end{tabular}

\end{center}
\end{table}

\begin{table}[H]
  \caption{\label{tab:pearson-sd-m2}Sample standard deviations of Pearson residuals for edge, two-star, and triangle counts in \Model{2} for the dataset and its subsets. Values substantially greater than 1 indicate overdispersion relative to the model.}
  \footnotesize
\begin{center}

\begin{tabular}{lrrrr}
\toprule
Statistic & Overall & $H$ & \Ewc & \Enc\\
\midrule
edges & 1.001 & 0.822 & 1.07 & 1.01\\
2-stars & 0.993 & 0.852 & 1.04 & 1.02\\
triangles & 0.980 & 0.879 & 1.01 & 1.01\\
\bottomrule
\end{tabular}

\end{center}
\end{table}

\begin{table}[H]
  \caption{\label{tab:pearson-mix-m2}Pearson residuals for edge, two-star, triangle, and mixing counts in \Model{2} for the dataset and its subsets. Extreme positive or negative residual values indicate that the model poorly accounts for effect of the presence of a child (the main selection criterion for the $H$ dataset) and/or that cells in the mixing model should not have been merged. Mixing effects for which a child effect is inherent (i.e., involving a young child and/or a preadolescent but no other on one side of the relation) or which are not found in the $H$ dataset (i.e., seniors) are also excluded.}
  \footnotesize
\begin{center}

\begin{tabular}{lrrrr}
\toprule
Effect & Overall & $H$ & \Ewc & \Enc\\
\midrule
edges & $ 0.12$ & $ 1.84$ & $-1.44$ & $ 0.15$\\
2-stars & $ 0.33$ & $ 1.62$ & $-1.01$ & $ 0.56$\\
triangles & $ 0.32$ & $ 1.60$ & $-1.01$ & $ 0.54$\\
Adolescent with Adolescent & $ 0.24$ & $-1.25$ & $ 0.38$ & $ 0.60$\\
Adolescent with Young Adult & $ 0.10$ & $-1.40$ & $ 0.44$ & $ 0.24$\\
Young Adult with Young Adult & $ 0.05$ & $ 1.70$ & $-0.76$ & $-0.26$\\
Adolescent with Older Female Adult & $-0.02$ & $ 1.79$ & $-1.32$ & $-0.29$\\
Young Adult with Older Female Adult & $ 0.55$ & $-0.59$ & $-0.13$ & $ 0.74$\\
Older Female Adult with Older Female Adult & $ 0.10$ & $ 0.48$ & $ 0.11$ & $ 0.03$\\
Adolescent with Older Male Adult & $-0.03$ & $ 1.52$ & $-1.36$ & $-0.20$\\
Young Adult with Older Male Adult & $-0.58$ & $-1.47$ & $ 0.76$ & $-0.52$\\
Older Female Adult with Older Male Adult & $-0.04$ & $ 0.01$ & $-0.07$ & $-0.02$\\
Older Male Adult with Older Male Adult & $ 0.02$ & $-1.87$ & $-1.69$ & $ 0.58$\\
\bottomrule
\end{tabular}

\end{center}
\end{table}

\clearpage

\subsection[\Model{\MIIp}]{\label{app:extra-results-m2p}Full results and diagnostics for \Model{\MIIp} }
\small

\begin{table}[H]
  \caption{\label{tab:coef-m2p}Parameter estimates (and standard errors) for \Model{\MIIp}.}
  \footnotesize
  \begin{center}

\begin{tabular}{lr}
\toprule
Relationship Effect\\\quad{}$\times$ Network-Level Effect & Coefficient (SE)$^{\hphantom{\star\star\star}}$\\
\midrule
edges $\times$ $\log(\nactors_{\sampidx})$ & $-14.29 \; (2.97)^{\star\star\star}$\\
\quad $\times$ $\log^2(\nactors_{\sampidx})$ & $5.69 \; (1.33)^{\star\star\star}$\\
\quad   if Brussels post code & $0.02 \; (0.21)^{\phantom{\star}\phantom{\star}\phantom{\star}}$\\
\quad   if city post code & $0.07 \; (0.09)^{\phantom{\star}\phantom{\star}\phantom{\star}}$\\
\quad   if on weekend & $0.13 \; (0.06)^{\star\phantom{\star}\phantom{\star}}$\\
2-stars & $1.92 \; (0.83)^{\star\phantom{\star}\phantom{\star}}$\\
\quad $\times$ $\log(\nactors_{\sampidx})$ & $-2.13 \; (0.43)^{\star\star\star}$\\
\quad $\times$ $\log^2(\nactors_{\sampidx})$ & $0.33 \; (0.12)^{\star\star\phantom{\star}}$\\
triangles & $5.65 \; (0.97)^{\star\star\star}$\\
\quad $\times$ $\log(\nactors_{\sampidx})$ & $-3.63 \; (1.42)^{\star\phantom{\star}\phantom{\star}}$\\
\quad $\times$ $\log^2(\nactors_{\sampidx})$ & $1.00 \; (0.73)^{\phantom{\star}\phantom{\star}\phantom{\star}}$\\
Young Child with Young Child & $8.59 \; (1.53)^{\star\star\star}$\\
Young Child with Preadolescent & $9.10 \; (1.52)^{\star\star\star}$\\
Preadolescent with Preadolescent & $8.17 \; (1.49)^{\star\star\star}$\\
Adolescent with Adolescent & $7.69 \; (1.47)^{\star\star\star}$\\
Young Child with Young Adult & $9.65 \; (1.82)^{\star\star\star}$\\
Preadolescent with Young Adult & $7.22 \; (1.50)^{\star\star\star}$\\
Adolescent with Young Adult & $7.73 \; (1.49)^{\star\star\star}$\\
Young Adult with Young Adult & $7.65 \; (1.48)^{\star\star\star}$\\
Young Child with Older Female Adult & $10.24 \; (1.50)^{\star\star\star}$\\
Preadolescent with Older Female Adult & $9.67 \; (1.48)^{\star\star\star}$\\
Adolescent with Older Female Adult & $8.90 \; (1.47)^{\star\star\star}$\\
Older Female Adult with Older Female Adult & $7.44 \; (1.51)^{\star\star\star}$\\
Young Child with Older Male Adult & $9.09 \; (1.48)^{\star\star\star}$\\
Preadolescent with Older Male Adult & $8.76 \; (1.47)^{\star\star\star}$\\
Adolescent with Older Male Adult & $8.20 \; (1.47)^{\star\star\star}$\\
Older Female Adult with Older Male Adult & $10.11 \; (1.48)^{\star\star\star}$\\
\quad   if child absent & $-1.21 \; (0.30)^{\star\star\star}$\\
Older Male Adult with Older Male Adult & $6.57 \; (1.49)^{\star\star\star}$\\
Older Female Adult with Senior & $8.13 \; (1.46)^{\star\star\star}$\\
Older Male Adult with Senior & $7.51 \; (1.49)^{\star\star\star}$\\
Senior with Senior & $7.82 \; (1.45)^{\star\star\star}$\\
Adolescent with Young Child or Preadolescent & $8.07 \; (1.47)^{\star\star\star}$\\
Young Adult with Older Adult & $8.01 \; (1.47)^{\star\star\star}$\\
Young Child or Preadolescent with Senior & $8.29 \; (1.56)^{\star\star\star}$\\
Adolescent or Young Adult with Senior & $9.91 \; (1.75)^{\star\star\star}$\\
\bottomrule
\end{tabular}

Significance: $^{\star\star\star}\le 0.001<^{\star\star}\le 0.01< ^\star \le 0.05$
\end{center}
\end{table}
\begin{table}[H]
  \caption{\label{tab:omnibus-m2p} Omnibus tests for selected groups of effects in \Model{\MIIp}. Effects are net of the rest of the model.}
  \footnotesize
  \begin{center}

\begin{tabular}{lrr}
\toprule
Effects & Wald $\chi^2$ (df) & $\pval$\\
\midrule
any 2-star & 46.1 (3) & \ensuremath{<0.001}\\
any triangle & 100.9 (3) & \ensuremath{<0.001}\\
any $\log(\nactors_\sampidx)$ or $\log^2(\nactors_\sampidx)$ & 2896.6 (6) & \ensuremath{<0.001}\\
any $\log^2(\nactors_\sampidx)$ & 54.2 (3) & \ensuremath{<0.001}\\
2-star or triangle $\log^2(\nactors_\sampidx)$ & 21.8 (2) & \ensuremath{<0.001}\\
\bottomrule
\end{tabular}

\end{center}
\end{table}

\begin{figure}[H]

{\centering \subfloat[Edges\label{fig:unnamed-chunk-147-1}]{\includegraphics[width=\maxwidth]{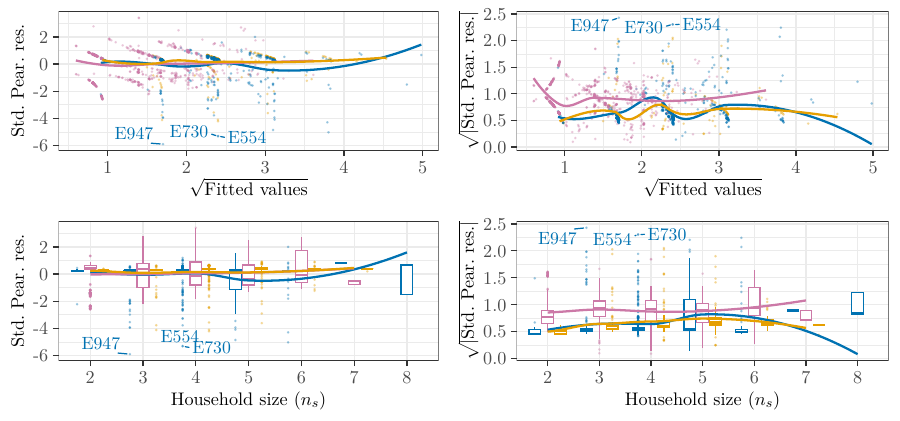} }\newline
\subfloat[2-stars\label{fig:unnamed-chunk-147-2}]{\includegraphics[width=\maxwidth]{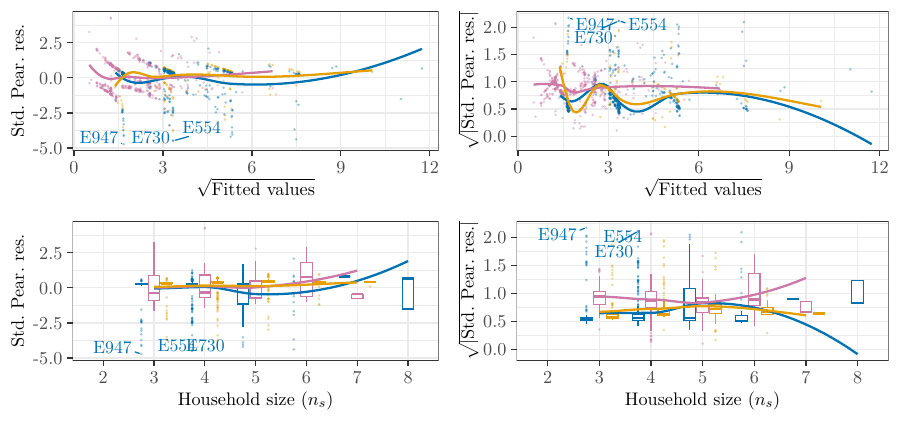} }\newline
\subfloat[Triangles\label{fig:unnamed-chunk-147-3}]{\includegraphics[width=\maxwidth]{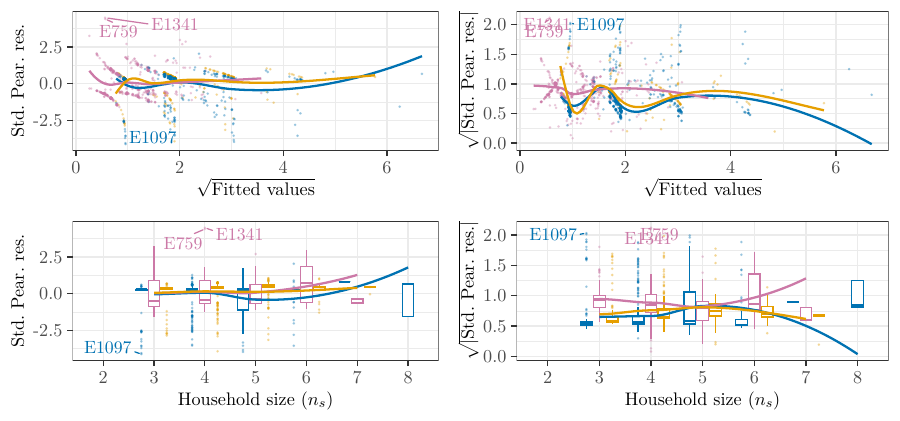} }

}

\caption{\label{fig:resid-plots-m2p}Residual plots of network statistics against fitted values and network size for \Model{\MIIp}. \subdataleg}\label{fig:unnamed-chunk-147}
\end{figure}

\begin{figure}[H]

{\centering \includegraphics[width=\maxwidth]{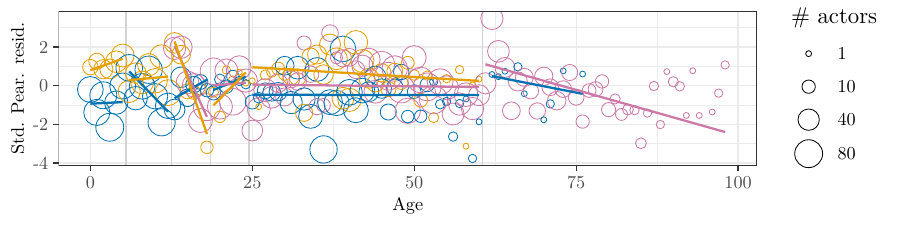} 

}

\caption{\label{fig:age-resid-m2p}Residual plot for \Model{\MIIp} of edges incident on actors of a given age against age. \subdataleg}\label{fig:unnamed-chunk-148}
\end{figure}

\begin{figure}[H]

{\centering \includegraphics[width=\maxwidth,height=1in]{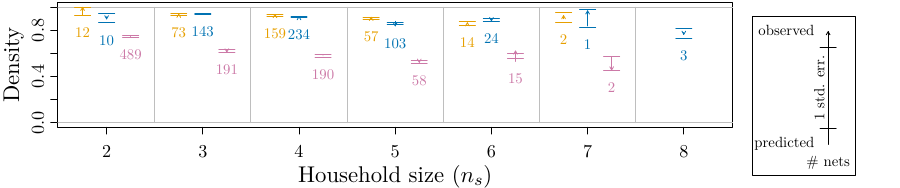} 

}

\caption{\label{fig:density-error-m2p}Average prediction errors of density in \Model{\MIIp}. Values are averaged over the networks grouped by size and subset. \subdataleg}\label{fig:unnamed-chunk-149}
\end{figure}

\begin{table}[H]
  \caption{\label{tab:nns-anova-m2p}Analyses of variance for fitting residuals for \Model{\MIIp} against household size (represented as a categorical predictor, with a dummy variable for each size).}
  \centering
  \footnotesize

\begin{tabular}{lrrrrr}
\toprule
\quad{}Source & df & Sum Sq. & Mean Sq. & $F$ & \pval\\
\midrule
edges &  &  &  &  & \\
\quad{}$\nactors_{\sampidx}$ (categorical) & 7 & 9.5 & 1.36 & 1.4 & \ensuremath{0.218}\\
\quad{}Residuals & 1773 & 1771.8 & 1.00 &  & \\
\addlinespace
2-stars &  &  &  &  & \\
\quad{}$\nactors_{\sampidx}$ (categorical) & 6 & 8.6 & 1.44 & 1.5 & \ensuremath{0.187}\\
\quad{}Residuals & 1263 & 1241.2 & 0.98 &  & \\
\addlinespace
triangles &  &  &  &  & \\
\quad{}$\nactors_{\sampidx}$ (categorical) & 6 & 7.8 & 1.29 & 1.3 & \ensuremath{0.232}\\
\quad{}Residuals & 1263 & 1209.5 & 0.96 &  & \\
\bottomrule
\end{tabular}

\end{table}

\begin{table}[H]
  \caption{\label{tab:reg-res-m2p} Regression of residuals of \Model{\MIIp} for each network structural property on each candidate predictor.}
  \footnotesize
  \begin{center}

\begin{tabular}{lrr}
\toprule
Relationship Effect\\\quad{}$\times$ Network-Level Effect & Wald $\chi^2$ (df) & $\pval$\\
\midrule
edges &  & \\
$\quad\times\log(\text{pop.\ dens.\ in post code})$ (linear) & 1.9 (1) & \ensuremath{0.171}\\
\quad$\times\log(\text{pop.\ dens.\ in post code})$ (quadratic) & 2.6 (2) & \ensuremath{0.277}\\
\quad if city post code & 0.6 (1) & \ensuremath{0.458}\\
\quad if Brussels post code & 0.3 (1) & \ensuremath{0.594}\\
\quad if weekend & 1.0 (1) & \ensuremath{0.328}\\
2-stars &  & \\
$\quad\times\log(\text{pop.\ dens.\ in post code})$ (linear) & 0.3 (1) & \ensuremath{0.587}\\
\quad$\times\log(\text{pop.\ dens.\ in post code})$ (quadratic) & 0.3 (2) & \ensuremath{0.855}\\
\quad if city post code & 0.9 (1) & \ensuremath{0.330}\\
\quad if Brussels post code & 0.0 (1) & \ensuremath{0.876}\\
\quad if weekend & 1.0 (1) & \ensuremath{0.321}\\
triangles &  & \\
$\quad\times\log(\text{pop.\ dens.\ in post code})$ (linear) & 0.3 (1) & \ensuremath{0.594}\\
\quad$\times\log(\text{pop.\ dens.\ in post code})$ (quadratic) & 0.4 (2) & \ensuremath{0.828}\\
\quad if city post code & 0.9 (1) & \ensuremath{0.330}\\
\quad if Brussels post code & 0.1 (1) & \ensuremath{0.743}\\
\quad if weekend & 0.7 (1) & \ensuremath{0.420}\\
\bottomrule
\end{tabular}

\end{center}
\end{table}
\begin{table}[H]
  \caption{\label{tab:dataset-m2p} Tests of the null hypothesis of no effect of dataset $H$ on the specified network statistic over and above \Model{\MIIp}.}
  \footnotesize
  \begin{center}

\begin{tabular}{lrr}
\toprule
Statistic & Score $\chi^2$ (df) & $\pval$\\
\midrule
Omnibus & 3.0 (2) & \ensuremath{0.2183}\\
\quad edges & nonparam. & \ensuremath{0.0836}\\
\quad 2-stars & nonparam. & \ensuremath{0.1365}\\
\bottomrule
\end{tabular}

\end{center}
\end{table}

\begin{table}[H]
  \caption{\label{tab:pearson-sd-m2p}Sample standard deviations of Pearson residuals for edge, two-star, and triangle counts in \Model{\MIIp} for the dataset and its subsets. Values substantially greater than 1 indicate overdispersion relative to the model.}
  \footnotesize
\begin{center}

\begin{tabular}{lrrrr}
\toprule
Statistic & Overall & $H$ & \Ewc & \Enc\\
\midrule
edges & 1.001 & 0.822 & 1.07 & 1.01\\
2-stars & 0.993 & 0.852 & 1.04 & 1.02\\
triangles & 0.980 & 0.879 & 1.01 & 1.01\\
\bottomrule
\end{tabular}

\end{center}
\end{table}

\begin{table}[H]
  \caption{\label{tab:pearson-mix-m2p}Pearson residuals for edge, two-star, triangle, and mixing counts in \Model{\MIIp} for the dataset and its subsets. Extreme positive or negative residual values indicate that the model poorly accounts for effect of the presence of a child (the main selection criterion for the $H$ dataset) and/or that cells in the mixing model should not have been merged. Mixing effects for which a child effect is inherent (i.e., involving a young child and/or a preadolescent but no other on one side of the relation) or which are not found in the $H$ dataset (i.e., seniors) are also excluded.}
  \footnotesize
\begin{center}

\begin{tabular}{lrrrr}
\toprule
Effect & Overall & $H$ & \Ewc & \Enc\\
\midrule
edges & $ 0.12$ & $ 1.78$ & $-1.61$ & $ 0.19$\\
2-stars & $ 0.21$ & $ 1.57$ & $-1.37$ & $ 0.47$\\
triangles & $ 0.20$ & $ 1.52$ & $-1.33$ & $ 0.45$\\
Adolescent with Adolescent & $ 0.12$ & $-1.32$ & $ 0.22$ & $ 0.58$\\
Adolescent with Young Adult & $ 0.14$ & $-1.55$ & $ 0.42$ & $ 0.38$\\
Young Adult with Young Adult & $ 0.13$ & $ 1.76$ & $-0.85$ & $-0.10$\\
Adolescent with Older Female Adult & $ 0.12$ & $ 1.73$ & $-1.01$ & $-0.21$\\
Young Adult with Older Female Adult & $ 0.60$ & $-0.35$ & $-0.19$ & $ 0.78$\\
Older Female Adult with Older Female Adult & $-0.03$ & $ 0.54$ & $ 0.09$ & $-0.11$\\
Adolescent with Older Male Adult & $ 0.10$ & $ 1.56$ & $-1.24$ & $-0.09$\\
Young Adult with Older Male Adult & $-0.49$ & $-1.01$ & $ 0.62$ & $-0.46$\\
Older Female Adult with Older Male Adult & $ 0.03$ & $ 0.14$ & $-0.07$ & $ 0.01$\\
Older Male Adult with Older Male Adult & $-0.07$ & $-1.38$ & $-2.02$ & $ 0.48$\\
\bottomrule
\end{tabular}

\end{center}
\end{table}

\end{document}